**Minimizing the Probability of Ruin in Retirement**

CHRISTOPHER J. ROOK*

**ABSTRACT**

Retirees who exhaust their savings while still alive are said to experience financial ruin.  The savings are typically grown during life's accumulation phase then spent during the retirement decumulation phase.  Extensive research into invest-and-harvest decumulation strategies has been conducted, but recommendations differ markedly.  This has likely been a source of concern and confusion for the retiree.  Our goal is to find what has heretofore been elusive, namely an optimal decumulation strategy.  Optimality implies that no alternate strategy exists or can be constructed that delivers a lower probability of ruin, given a fixed inflation-adjusted withdrawal rate.

* The author is a consultant statistical programmer and studies in the Department of Systems Engineering at Stevens Institute of Technology.  All appendices referred to in the text can be found in the Internet Appendix document that accompanies this research.

The retirement literature comprises a vast body of research that has evolved disparately with financial economists optimizing portfolio-choice or life-cycle models and practitioners adhering to rules-of-thumb (Milevsky and Huang (2010)). Economists attempt to smooth a person's utility of consumption over their earning and retirement years and have criticized practitioner heuristics as being ad-hoc, simplistic, inefficient, misguided, far from optimal, and potentially unsafe (see Kotlikoff (2008); Scott, Sharpe, and Watson (2009); Irlam and Tomlinson (2014); Finke, Pfau, and Blanchett (2013)). Practitioners counter that they consider risk tolerance when forming and assessing a client's financial plan (Mitchell and Smetters (2013)), and assert that most clients seek constant, not variable spending and consumption during retirement. Another practitioner concern is that utility functions are difficult to estimate accurately and if misspecified results in maximization of the wrong expression (Kitces (2012)).

Our focus here is on the most ubiquitous of practitioner heuristics, namely the safe withdrawal rate ($W_R$). This heuristic posits that the retiree may withdraw $W_R$-% from their portfolio during year one and adjust it for inflation each year thereafter. The popularity of this heuristic among practitioners is irrefutable. Scott et al. note that "retail brokerage firms, mutual fund companies, retirement groups, investor groups, financial websites, and the popular financial press all recommend it." Milevsky and Huang (2010) add that "it is hard to overstate the influence" this heuristic has had on modern retirement planning, noting that its results have been reported "thousands of times in the last two decades" making it "destined for the same immortality enjoyed by other rules-of-thumb".

The effectiveness of a safe $W_R$ strategy is measured by its success rate which is the percentage of times the portfolio lasts a fixed number of years, or outlives the retiree. Letting P(Ruin) denote the probability of ruin, the success rate is 100%∗(1-P(Ruin)). Maximizing the success rate and minimizing P(Ruin) are thus equivalent optimization problems. Retirees using a safe $W_R$ strategy face the difficult decision of setting its initial value and determining an asset allocation that is appropriate. Withdraw too little and face a reduced standard of living; withdraw too much and face financial ruin. Likewise, investing the wrong proportion in stocks or bonds can lead to ruin from excessive or insufficient risk. Unfortunately there is no consensus regarding a precise "safe" initial $W_R$ or the optimal asset allocation for retirees. Generally a $W_R$ between 3% and 6% is suggested but success rates vary considerably. The term glide-path refers to asset allocations that either shed or acquire equities systematically over time. Some advocate for a declining, and others for a rising glide-path during retirement. With such a large cohort (> 76 million U.S. Baby Boomers) on the precipice of spending their savings, finding optimal decumulation strategies is increasingly important. The first Baby Boomers to retire at age 65 began doing so in 2011, and these retirements will continue at a rate of 10,000/day until 2030.[1]

The general approach to retirement planning by practitioners has been to develop intuition-based heuristics and test them using historical data, bootstrapping, or simulation. These heuristics usually apply to withdrawals or to the glide-path. Our goal here is to introduce a theoretical framework for the safe $W_R$ heuristic into the literature. In addition, we optimize the glide-path to achieve a minimum P(Ruin) for retirees. We show that, when used optimally, the safe $W_R$ is not merely a heuristic but an academically sound approach to retirement planning based on theory spanning a wide array of disciplines. At each time point during decumulation there exists a minimum P(Ruin) based on an optimal asset allocation and we seek to find these values. We formulate a model that reveals precisely these measures as a function of both time and a quantity we term the *ruin factor*. Our model is optimal in the sense that no alternate $W_R$ strategy can deliver a lower P(Ruin), but it is also intractable under common market assumptions. We use numerical techniques to approximate the optimal solution and the user can achieve any desired degree of accuracy by adjusting the discretization precision.

This research is organized as follows. In Section I the literature of practitioner-based decumulation strategies is reviewed and in Section II we formulate and optimize models for retirement horizons that are fixed or random in length. In Section III we implement these models using historical market returns to estimate randomness, and we extend the random horizon model to a group of retirees. In Section IV we conclude with a non-technical summary and provide source code from a full implementation in the Appendix. Throughout the discussion we offer suggestions for extending these models in various dimensions, for example by adding asset classes, tailoring the expense ratio, and using forecasts to vary the return distribution across time.

# I. Literature Review

Bengen (1994) determined the safe initial $W_R$ for retirees by analyzing U.S. historical return sequences on portfolios of stocks and bonds. He concluded that a 4.15% inflation-adjusted $W_R$ would not have been exhausted over any prior 30-year period using 50% stocks, and therefore considered it safe. Bengen (1994) recommended a fixed equity allocation between 50% and 75% which would only change if the retiree's objectives changed. This research led to what is commonly referred to as the "4% rule". Bengen (2006) updated this research to include small cap stocks in the portfolio and used the term SAFEMAX to refer to the highest $W_R$ that yields a >$0 terminal balance after 30 years for all historical sequences analyzed. When at least 20% of the portfolio's stock was invested in small caps, the SAFEMAX grew to 4.42%.

Cooley, Hubbard, and Walz (1998) also applied historical return sequences to retirement portfolios of stocks and bonds. They concluded that a 4% $W_R$ on accounts with 50% and 75% stocks had success rates (>$0 terminal balance) of 95% and 98%, respectively over 30-year horizons. In 2011, Cooley et al. updated the analysis using 14 years of new data and found that portfolios with 50% and 75% stocks had success rates of 96% and >99%, respectively.

Pye (2000) analyzed the sustainable $W_R$ by making distributional assumptions about investment returns and tested various withdrawal strategies using simulation. Portfolios were constructed using stocks and TIPS and withdrawals were reduced once thresholds became violated. If the portfolio approaches ruin withdrawals are restricted, reducing the probability of a shortfall. Instead of measuring the probability of ruin, Pye (2000) measured the probability of hitting the limits and facing a reduced standard of living (success=no reduction). A 4% $W_R$ using 100% stocks and a 4.5% $W_R$ using 40% stocks both had 35-year success rates of 81%.

Guyton (2004) developed decision rules for making withdrawals and tested them on the 30-year period between 1973 and 2003, considered challenging due to high inflation and two severe market downturns. Fixed portfolios having either 65% or 80% equities were used and the rules limited inflation adjustments on $W_R$ to years when the account balance did not decline. Withdrawals were also taken from stocks and bonds in a systematic manner that drew first from accounts with outsized prior year gains. Using these rules, Guyton (2004) found that portfolio equity ratios of 65% and 80% could sustain an initial $W_R$ of 5.4% and 5.8%, respectively for the 30-year period studied plus 10 additional years. The foregoing of an inflation-adjustment at predetermined times thus results in a higher $W_R$ when compared to most other research.

Milevsky and Robinson (2005) introduced the stochastic present value (SPV) to derive the ruin PDF over a retiree's life expectancy which was approximated by an exponential RV. The SPV is the present value of the retiree's future spending discounted using random market returns. Ruin occurs if the SPV exceeds the retiree's starting account balance. Milevsky and Robinson (2005) found that under the assumption of lognormal portfolio returns the SPV RV follows a reciprocal gamma distribution with parameters directly related to the parameters of the lognormal returns (mean and variance) and the exponential life expectancy (rate). The solution is closed form in nature, but serves as an approximation to the true probability of ruin when exponential lifetimes are used. Milevsky and Robinson (2005) found that equity ratios of 50% and 100% yield success rates of 91.0% and 87.7%, respectively using a 4% $W_R$.

Blanchett (2007) analyzed the safe $W_R$ by bootstrap sampling historical returns on portfolios of stocks and bonds. Fixing the success rate at 95%, Blanchett (2007) found that a portfolio using 40% equities had the highest $W_R$ of 4.2% over a 30-year horizon. Blanchett (2007) also analyzed 4 downward sloping (linear, stair, concave, convex) equity glide-paths along with the constant allocation (no glide-path). Each was tested with different starting equity percentages and resulted in 43 separate paths analyzed on horizons of 20 to 40 years. Blanchett (2007) found that a constant allocation performed best. Specifically, over 30-years a 4% $W_R$ with constant allocation of 30% stocks would generate the lowest failure rate of 2.06%.

Stout (2008) formulated retirement ruin as the objective of a stochastic optimization problem and solved it using simulation. Fixed allocation percentages to stocks and bonds were varied until the minimum probability of ruin was found. The objective function thus depends on random quantities such as market returns and age of death using survival probabilities from the U.S. Centers for Disease Control. Stout (2008) found that a 65-year old retiree using a 4.25% $W_R$ would achieve a success rate of 96.75%, assuming an optimal allocation of 50.25% in stocks. Further, as $W_R$ increases the equity ratio must also increase to maintain optimality.

Target-date (TD) funds are financial products that implement declining equity glide-paths based on the retirement date. Some cease at retirement, while others continue reallocating through retirement. Morningstar, Inc. (2009) surveyed financial firms and summarized the glide-paths across a myriad TD funds. They found that retirees from 1995, 2000, 2010, and 2015 currently have on average 20%, 34.5%, 50.2%, and 62.1% in equities, respectively. This survey provides a great deal of insight into the glide-paths used by financial firms in their TD funds. On average the retiree begins with about 62% equities which decreases to 20% after two decades.

There is not widespread agreement that declining equity glide-paths outperform fixed allocations, however. Similar to Blanchett (2007), Cohen, Gardner, and Fan (2010) found that a constant equity mix will outperform a declining equity glide-path almost universally, raising doubts about TD fund strategies. Cohen et al. (2010) used a fixed retirement horizon and found that retirees should have fewer equities than is often recommended. Namely, equity percentages of 32%, 60%, and 80% yielded success rates of 94%, 88%, and 84%, respectively using a 4% $W_R$. Their rational for a lower equity ratio is to prevent against sequence of return (SOR) risk, which is the risk of losses early in retirement. When coupled with inflation-adjusted withdrawals, early losses increase the probability of experiencing financial ruin.

Fan, Murray, and Pittman (2013) developed an adaptive allocation model that tracks the retiree's account balance over time and adjusts the asset allocation to maximize expected wealth over 20 years, prior to buying an annuity. A decision tree is built using combinations of market returns and their impact on different asset allocations between time points. By folding back the tree an optimal policy reveals itself and by eliminating sub-optimal nodes it serves as a guide for the retiree over time. The allocation is optimized at each node in the tree using an objective function that punishes shortfalls quadratically. The adaptive model results in higher expected surpluses and lower expected losses compared to fixed strategies. A feature of this model is that it becomes most conservative when current assets match discounted future spending. When the retiree experiences strong returns the shortfall penalty approaches zero revealing a feedback mechanism that encourages the building of more wealth as wealth grows. Therefore the adaptive model advocates for a higher equity ratio in both underperforming and outperforming markets.

Pfau and Kitces (2014) researched various equity glide-paths for periods of typical and atypical (sub-standard) returns using simulation.  For 2 of the 3 periods tested, a 4% $W_R$ yielded the highest success rate using equity glide-paths that start between 20% and 40%, and end between 60% and 80%.  These glide-paths also led to the smallest shortfall magnitude for all periods analyzed.  Rising equity glide-paths thus outperformed both static and declining glide-paths.  Pfau and Kitces (2014) offer the intuition behind these findings, namely that the portfolio assumes greatest SOR risk at the beginning when it is largest, and the retiree should limit their equity exposure.  Over time this risk lessens and adding equities is warranted.  Pfau and Kitces (2014) note that a rising glide-path during a poor market allows the retiree to accumulate stocks when prices are low, reaping the benefits of a market upturn.  Accumulating equities during a rising market is viewed as an opportunity to build a legacy for heirs.

## II.   Formulating and Solving the Model

We formulate an optimal retirement decumulation model in Section II and implement it in Section III.  The solution can serve as a roadmap for decumulating retirees.  Most conclusions are justified in appendices and the reader is encouraged to review those sections as well.

### A.  Definitions and Notation

Assume the retiree owns an investment account consisting of stocks and bonds and that financial ruin occurs *iff* the retiree empties their account while still alive.  Define:

- $t$ = time point ($t$=0, 1, …, $T_D$), with retirement starting at $t$=0
- $T_D$ = time of last withdrawal prior to death (can be either fixed or random)
- $T_D$-$k$ = $k$ time points before $T_D$ after making the withdrawal  (there are k withdrawals remaining at time $t$=$T_D$-$k$)
- Ruin($t$) = the event of ruin is experienced at time t
- Ruin$^C$($t$) = the event of ruin is not experienced at time t
- \$A = value of retirement account at time $t$=0
- $\alpha$ = proportion of retirement account in stocks ($0 \leq \alpha \leq 1$)
- $(1-\alpha)$ = proportion of retirement account in bonds

· $W_R$ = initial withdrawal rate adjusted for inflation at each time point

· $I_t$ = inflation rate between times t-1 and t

· $R_{(t,\alpha)}$ = total rate of return (RoR) for the retirement account with $100*\alpha$% stocks and $100*(1-\alpha)$% bonds between times t-1 and t

· $r_{(t,\alpha)} \mid I_t, R_{(t,\alpha)}$ = inflation-adjusted RoR for the retirement account with $100*\alpha$% stocks and $100*(1-\alpha)$% bonds between times t-1 and t, given $I_t$ and $R_{(t,\alpha)}$

· $E_R$ = expense ratio charged by the financial institution per time t (fixed or $\alpha*E_{R(s)} + (1-\alpha)*E_{R(b)}$ where $E_{R(s)}$, $E_{R(b)}$ are stock and bond expense ratios)

· $\hat{r}_{(t,\alpha)} = (1-E_R)*(1+r_{(t,\alpha)} \mid I_t, R_{(t,\alpha)})$ inflation/expense-adjusted compounded return on the account with $100*\alpha$% stocks and $100*(1-\alpha)$% bonds between times t-1 and t

The quantities $I_t$, $R_{(t,\alpha)}$, $r_{(t,\alpha)} \mid I_t, R_{(t,\alpha)}$, and $\hat{r}_{(t,\alpha)}$ are continuous RVs and $T_D$ can be fixed or (discrete) random. $W_R$ is based on \$A and is withdrawn at time t after adjusting for inflation. The real RoR for an $\alpha$-based portfolio at time t is $r_{(t,\alpha)} \mid I_t, R_{(t,\alpha)}$, which is calculated after $R_{(t,\alpha)}$ and $I_t$ are observed, therefore it is defined conditionally. The quantity $\hat{r}_{(t,\alpha)}$ reflects the inflation/ expense-adjusted compounded return for an $\alpha$-based portfolio at time t. The RV $\hat{r}_{(t,\alpha)}$ thus depends on 5 quantities: t, $I_t$, $R_{(t,\alpha)}$, $\alpha$, $E_R$, where t, $\alpha$, and $E_R$ are deterministic and $I_t$, $R_{(t,\alpha)}$ random. The real balance at time t-1 multiplied by $\hat{r}_{(t,\alpha)}$ is the real balance at time t. Finally, Ruin(t) and Ruin$^C$(t) are random events that occur with some probability.

The event Ruin(t) occurs *iff* the account balance cannot sustain the withdrawal at time t. It can only occur once. When expressing P(Ruin(t)) we condition on no ruin occurring before time t. For example, the probability of ruin at time t=3 is P(Ruin(3)|Ruin$^C$(1)∩Ruin$^C$(2)). The conditioning component can be expressed as P(Ruin$^C$(1) ∩ Ruin$^C$(2)) = P(Ruin$^C$(1))*P(Ruin$^C$(2) | Ruin$^C$(1)). The probability of experiencing Ruin(3) is then expressed as:

$$P(\text{Ruin}(3) \mid \text{Ruin}^C(1) \cap \text{Ruin}^C(2)) = \frac{P(\text{Ruin}^C(1) \cap \text{Ruin}^C(2) \cap \text{Ruin}(3))}{P(\text{Ruin}^C(1))*P(\text{Ruin}^C(2) \mid \text{Ruin}^C(1))} \quad (1)$$

The probability of ruin, P(Ruin), changes over time. If the retiree using a fixed $W_R$ owns an $\alpha$-based portfolio whose real account balance doubles between times t=0 and t=1 then a very different P(Ruin) exists at time t=1, than did at time t=0. We assume the retiree's objective is to minimize P(Ruin), not build wealth, given $W_R$. Ruin probabilities can thus perish quickly and should be reevaluated at regular intervals by the retiree.

## B. The Retirement Sequence

We assume the retiree starts time t=0 with an account balance of $A. The withdrawal rate $W_R$ is based on $A and adjusted for inflation at each time point. At time t=1 the retiree attempts their first withdrawal of $(\$A)*(W_R)*(1+I_1)$ from the account which has compounded returns for one time point. If the account cannot sustain the withdrawal Ruin(1) occurs, otherwise $\text{Ruin}^C(1)$ occurs. The pre-withdrawal account balance is $(\$A)*(1+R_{(1,\alpha)})*(1-E_R)$, where $R_{(1,\alpha)}$ is the portfolio's total RoR between times t=0 and t=1. Under an optimal strategy, assuming $\text{Ruin}^C(1)$, the retiree should reassess P(Ruin) at any future time point (t=2, 3, …, $T_D$) and reallocate the portfolio's assets optimally.

This scenario assumes no withdrawal is made at time t=0. We base the model on this assumption and refer to it as *standard form*. In standard form, the first asset allocation decision occurs at time t=0 and the first withdrawal occurs at time t=1. If a retiree presents differently they will be converted to standard form before applying the model. For example, a retiree with time t=0 balance of $B that requires funding between times t=0 and t=1 can be converted to standard form by using a time t=0 withdrawal rate of $W_0=W_R/(1+W_R)$ applied to $B. The withdrawal at time t=0 is $(W_0)*(\$B)$ and the remaining balance is $A=\$B*(1-W_0)$. The standard form model is then applied using $A and $W_R$. Note that this conversion preserves the retiree's purchasing power from time t=0 to time t=1 since $(W_0)*(\$B)*(1+I_1) = (WR)*(\$A)*(1+I_1)$.[2]

## C. The Ruin Factor

We now define a quantity that encapsulates much information needed to assess P(Ruin). It should be updated by the retiree at each time point and is actionable in the sense that it helps reveal the precise asset allocation that minimizes P(Ruin) for the remainder of retirement. Let $r_{(t,\alpha)}$ denote $r_{(t,\alpha)} | I_t, R_{(t,\alpha)}$. Define the *ruin factor*, RF(t) as $W_R$ for t = 0 and:

$$RF(t) \quad = RF(t-1) / [(1+r_{(t,\alpha)})(1-E_R) - RF(t-1)] \qquad (2a)$$

$$= RF(t-1) / [\bar{r}_{(t,\alpha)} - RF(t-1)], \quad \text{for } t = 1, 2, …, T_D. \qquad (2b)$$

Given no ruin prior to time t, we show that the event Ruin(t) occurs as:

$$\text{Ruin(t)} \mid \text{Ruin}^C(1) \cap \ldots \cap \text{Ruin}^C(t\text{-}1) \;\leftrightarrow\; (1 + r_{(t,\alpha)}) * (1 - E_R) \;\leq\; RF(t\text{-}1) \qquad (3a)$$

$$\leftrightarrow\; \hat{r}_{(t,\alpha)} \;\leq\; RF(t\text{-}1) \qquad (3b)$$

(See Appendix B.)  At time t=t-1, RF(t-1) is known and Ruin(t) occurs *iff* $\hat{r}_{(t,\alpha)} \leq RF(t\text{-}1)$. Assuming the CDF of $\hat{r}_{(t,\alpha)}$ is known or estimated, RF(t-1) can be calculated and P(Ruin(t)) evaluated across various α at time t=t-1.  RF(t) can be defined using any unit of time assuming $\hat{r}_{(t,\alpha)}$ is the appropriate return.  Note that RF(t) is positive until ruin occurs (if it occurs) when it becomes negative (or ∞).  To see this note that at any time t, $\text{Ruin}^C(t)$ occurs ↔:

$$\hat{r}_{(t,\alpha)} > RF(t\text{-}1) \qquad (4)$$

This condition states that the upcoming time point's return $\hat{r}_{(t,\alpha)}$, must exceed RF(t-1) to avoid ruin.  If this fails to hold, $\hat{r}_{(t,\alpha)} \leq RF(t\text{-}1)$ and RF(t) is either negative (when $\hat{r}_{(t,\alpha)} < RF(t\text{-}1)$) or undefined (when $\hat{r}_{(t,\alpha)} = RF(t\text{-}1)$).  If RF(t) declines between time points the retiree has improved their position with respect to ruin, and vice versa.  RF(t) declines when the denominator exceeds 1,  namely $\hat{r}_{(t,\alpha)} > 1 + RF(t\text{-}1)$.  If the retiree generates returns ($r_{(t,\alpha)} - E_R - r_{(t,\alpha)} * E_R$) that exactly equal $W_R = RF(0)$, then RF(t) is constant and equals $W_R$ over time.

If $\text{Ruin}^C(t)$ occurs the real account balance at time t is $(\$A) * RF(0)/RF(t)$.  (See Appendix A.)  This reveals that the reciprocal of RF(t) equals the number of real withdrawals remaining at time t.  At time t, RF(t) is computed and known.  Future ruin factors are unknown since they are functions of unobserved market returns, thus they are RVs.  At the start of retirement RF(0)=$W_R$ is the only known ruin factor.  The retiree exercises some control over future RF(t) values via the choice of α since RF(t)=RF(t-1)/[$\hat{r}_{(t,\alpha)} - RF(t\text{-}1)$].  If a low-α portfolio is used between times t-1 and t, the retiree is more confident of $\hat{r}_{(t,\alpha)}$, and thus RF(t).  Conversely, if a high-α portfolio is used $\hat{r}_{(t,\alpha)}$ has greater uncertainty and so too does RF(t).

*D.  Example of Two Retirees Facing Different SORs*

Consider two retirees with rebalanced α=0.5 portfolios.  Let the unit of time be years with $T_D$=30.  Both begin time t=0 with account balances of $A use $W_R$=4% with $E_R$=0.5%.  To avoid ruin, the retirees must have >$0 account balances after making the time $T_D$=30 withdrawal.  At the start of each year (t ≥ 1) RF(t) is calculated and the required withdrawal is made. Figure 1 tracks RF(t) over the retirement horizon for both retirees (lines are smoothed) and indicates that Retiree 1 experienced the event Ruin(20) when funds were insufficient to make the time t=20 withdrawal. Retiree 2 did not experience a ruin event, making 30 successful withdrawals for spending. Note that at year 10 Retiree 1 was in a better financial state than Retiree 2 (RF(10) of 0.067 vs 0.077). A Question to consider: Was an α=0.5 portfolio appropriate for Retiree 1 at all time points during decumulation?

**Figure 1**
**The Ruin Paths of Two Retirees**

This figure details 2 ruin paths retirees may experience during decumulation and shows how the ruin factor, RF(t), is tracked over time. We know from (3b) that, for a given α, P(Ruin(t)) is an increasing function of RF(t-1). For a fixed t, larger RF(t) correspond to higher P(Ruin(> t)). Retiree 1 indicates it is extremely difficult to recover once RF(t) begins a parabolic climb. RF(t) has initial value of RF(0)=$W_R$ and we seek to keep it depressed over time. If ruin occurs RF(t) becomes negative or ∞ (not shown).

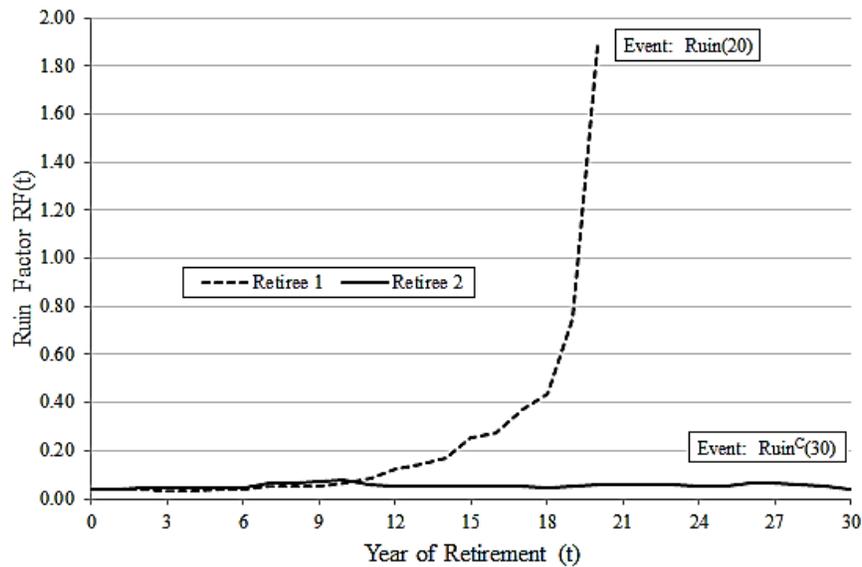

### E. The Event of Financial Ruin

We define Ruin(t) as the event of financial ruin at time t. Let Ruin(≤ t) represent the event of financial ruin occurring at or before time t and let Ruin(> t) represent the event of financial ruin occurring after time t for t=1, 2, …, $T_D$. (See Figure 2.)

**Figure 2**
**Venn Diagram of Ruin Event**

This figure expresses financial ruin as an event and demonstrates the relationship between ruin events at different time points. For example, if Ruin(1) occurs then Ruin($\leq 2$) occurs since Ruin(1) $\subseteq$ Ruin($\leq 2$). However, Ruin($\leq 2$) can occur without Ruin(1) occurring. Note that a ring of the oval represents the event of ruin occurring at that time point. This visual helps us formulate probability statements and also assists during the upcoming induction process when we are faced with restricted sample spaces.

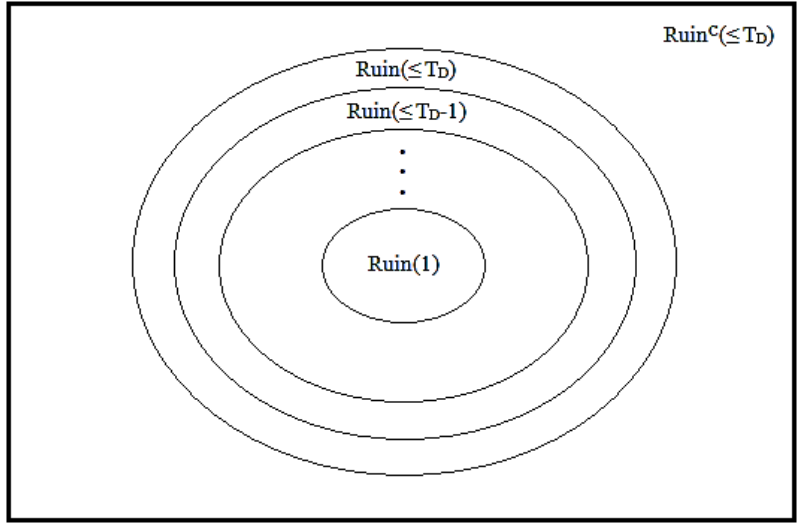

It follows that,

$$\text{Ruin}(t) \quad \equiv \quad \text{Ruin}(\leq t) \ \cap \ \text{Ruin}^C(\leq t\text{-}1), \tag{5}$$

which represents a single ring of the oval in Figure 2. Note that P(Ruin) can be expressed as:

$$P(\text{Ruin}(> 0)) \quad = \quad 1 - P(\text{Ruin}^C(\leq T_D)) \tag{6a}$$

$$= \quad 1 - P(\text{Ruin}^C(1) \cap \text{Ruin}^C(2) \cap \ldots \cap \text{Ruin}^C(T_D)) \tag{6b}$$

$$= \quad 1 - P(\text{Ruin}^C(1))*P(\text{Ruin}^C(2) \cap \ldots \cap \text{Ruin}^C(T_D) \mid \text{Ruin}^C(1)) \tag{6c}$$

where the last expression holds since, for any events A, B, and C:

$$P(A \cap B \cap C) \quad = \quad P(A)*P(B \cap C \mid A). \tag{7}$$

The retiree's objective is to minimize $P(\text{Ruin}) = P(\text{Ruin}(\leq T_D)) = P(\text{Ruin}(> 0))$, given $W_R$. As noted, the debate regarding the true value of P(Ruin) is unsettled. Further, the preferred glide-path is subject to much disagreement. It would be helpful for the retiree to know the exact optimal (minimum) P(Ruin) for each possible $W_R$ at every time t, along with the $\alpha$ required to achieve this optimum. The retiree would benefit from a roadmap that informs them how their P(Ruin) changes over time and the actions needed to maintain optimality. Ultimately such a roadmap should be available at time t=0, guiding the retiree through decumulation.

*F. Minimizing P(Ruin) at the Next Time Point*

The retiree's objective is to minimize P(Ruin) during decumulation and the choice of α at each time point must consider the remaining horizon. A relevant sub-problem is minimizing P(Ruin) at the next time point $t \leq T_D$ (fixed). Assume the withdrawal at time t-1 has been made and RF(t-1) > 0 calculated. The α that minimizes P(Ruin(t)) is found by inspecting the tail of all $\hat{r}_{(t,\alpha)}$ PDFs where Ruin(t) ↔ $\hat{r}_{(t,\alpha)} \leq$ RF(t-1). Figure 3 displays a collection of such PDFs.

**Figure 3**
**Generic $\hat{r}_{(t,\alpha)}$ PDFs for Various Asset Allocations (α)**

This figure demonstrates how we evaluate a collection of PDFs and select one that minimizes P(Ruin(t)) recalling that Ruin(t) ↔ $\hat{r}_{(t,\alpha)} \leq$ RF(t-1). As we increase α the PDF of $\hat{r}_{(t,\alpha)}$ shifts right and the variance increases resulting in a shorter and wider distribution. Note that P(Ruin(t)) = $F_{\hat{r}(t,\alpha)}$(RF(t-1)), where $F_x(\cdot)$ represents the CDF of the RV x. The probability of ruin is thus a left tail area under the PDF of $\hat{r}_{(t,\alpha)}$.

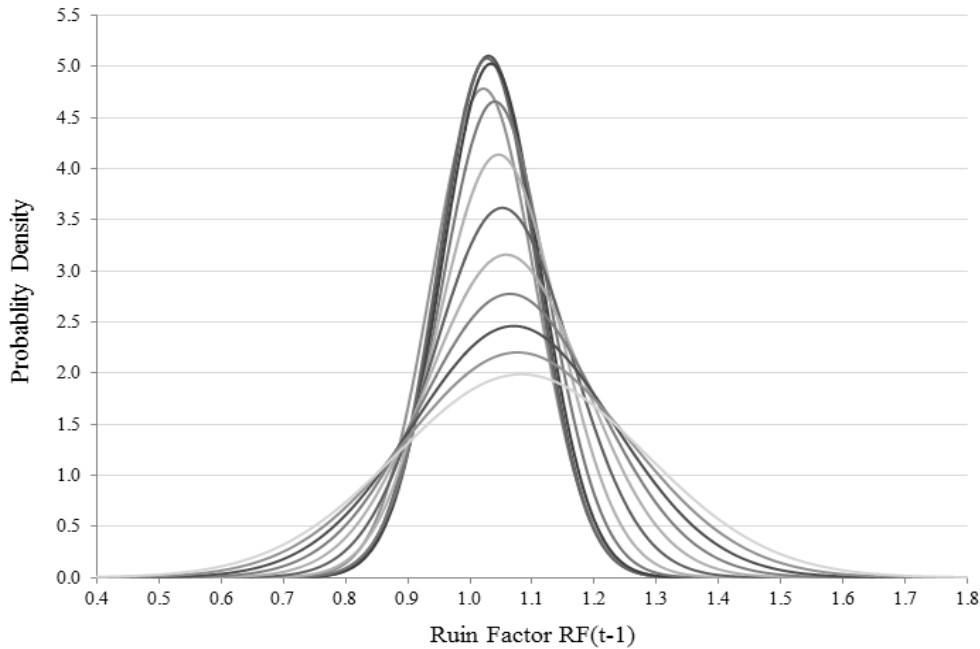

In Figure 3, the density with low peak and fat tail may represent an α=1 portfolio, and the density with high peak and smallest variance may represent the minimum volatility portfolio described by Markowitz (1952). Given $Ruin^C(t-1)$, the $\alpha$ that minimizes $P(Ruin(t))$ can be determined by choosing the PDF with smallest tail area to the left of $RF(t-1)$. This tail area is defined by $F_{\hat{r}_{(t,\alpha)}}(RF(t-1))$ where $F_x(x) = P(x \leq x)$ denotes the CDF of the RV x.

## G. Minimizing P(Ruin) at any Time Point for Fixed $T_D$

To minimize $P(Ruin)$ at any time during decumulation, we use backward induction. The induction process begins at time $t=T_D$ and each step unfolds similarly employing the expressions in (6). Namely, we assume the retiree arrives at that time point and successfully makes their withdrawal then recalculates $RF(t)$ which is $> 0$ since $Ruin^C(t)$ occurs. At $t=0$ the process ends and reveals the exact optimal (minimum) $P(Ruin)$ for all $W_R$, along with the dynamic asset allocation required to maintain optimality throughout decumulation. Let,

$V(t, RF(t))$ = Optimal (minimum) probability of ruin after time t given $RF(t) > 0$

$\alpha(t, RF(t))$ = Value of $\alpha$ required to achieve $V(t, RF(t))$.

In terms of ruin events, the value function $V(t, RF(t))$ is defined using (6c) as:

$$V(t, RF(t)) = Min_{(\alpha)}\{ P(Ruin(> t)) \} \tag{8a}$$

$$= Min_{(\alpha)}\{ P(Ruin(t+1) \cup Ruin(t+2) \cup \ldots \cup Ruin(T_D)) \} \tag{8b}$$

$$= Min_{(\alpha)}\{ 1 - P(Ruin^C(t+1)*P(Ruin^C(t+2) \cap \ldots \cap Ruin^C(T_D) \mid Ruin^C(t+1)) \}, \tag{8c}$$

We assume that returns from $\alpha$-based portfolios at different time points are independent, consistent with the theory of efficient markets. That is, any predictive capacity contained in past patterns of returns is deemed already accounted for, rendering it valueless to the retiree investor.

### G.1 The Induction Process for Fixed $T_D$

Assume the retiree arrives at time $t=T_D$ and makes their final withdrawal. The restricted sample space $S=\{Ruin^C(\leq T_D)\}$ includes a single event. $RF(T_D)$ ($> 0$) need not be computed as there are no future withdrawals. At $t=T_D$, $P(Ruin(>T_D))=0$, and $V(T_D, RF(T_D))=0$, $\forall RF(T_D) > 0$ serves as a boundary condition (B.C.). At any other time point t, assume the retiree makes their withdrawal and updates $RF(t)$ ($> 0$) based on the return just observed, $\hat{r}_{(t,\alpha)}$. The retiree faces the restricted sample space $S=\{Ruin(t+1), Ruin(t+2), \ldots, Ruin(T_D), Ruin^C(\leq T_D)\}$ and seeks $\alpha$ to minimize $P(Ruin(> t))$. Using (8c), this is equivalent to: (See Appendix C.)

$$V(t, RF(t)) = \begin{array}{l} \left. \text{Min}_{(0 \le \alpha \le 1)} \left\{ \begin{array}{l} 1 - (1 - F_{\hat{r}_{(t+1,\alpha)}}(RF(t)))* \\ \\ (1 - E_{\hat{r}_{(t+1,\alpha)}^+}\left[V(t+1, \frac{RF(t)}{\hat{r}_{(t+1,\alpha)} - RF(t)})\right]) \end{array} \right\} \right. \\ \text{For: } 0 \le t \le T_D\text{-}1, \ RF(t) > 0, \ V(T_D, RF(T_D)) = 0. \end{array} \quad (9)$$

Optimality is achieved at $\alpha(t, RF(t))$ and the expectation is wrt $\hat{r}_{(t+1,\alpha)}^+ = \hat{r}_{(t+1,\alpha)} \mid \hat{r}_{(t+1,\alpha)} > RF(t)$. Assuming $f(\hat{r}_{(t+1,\alpha)})$ is the PDF of $\hat{r}_{(t+1,\alpha)}$ and $RF(t)$ is known, $\hat{r}_{(t+1,\alpha)}^+$ has PDF:

$$\hat{r}_{(t+1,\alpha)}^+ = (\hat{r}_{(t+1,\alpha)} \mid \hat{r}_{(t+1,\alpha)} > RF(t)) \sim \begin{cases} \frac{f(\hat{r}_{(t+1,\alpha)})}{\int_{RF(t)}^{\infty} f(\hat{r}_{(t+1,\alpha)}) \, d\hat{r}_{(t+1,\alpha)}} & , \text{ for } \hat{r}_{(t+1,\alpha)} > RF(t) \\ 0 & , \text{O.W.} \end{cases} \quad (10)$$

The choice of $\alpha$ in (9) balances minimizing P(Ruin) at the next time point, and after the next time point. Selecting a low volatility portfolio early in retirement to avoid ruin at the next time point comes with a price because that portfolio results in a higher expected ruin factor, which increases P(Ruin) after the next time point. A balance is struck at $\alpha(t, RF(t))$.

## H. Minimizing P(Ruin) at any Time Point for Random $T_D$

For random $T_D$, select $S_{Max}$ such that $P(T_D > S_{Max}) = 0$ and build discrete time hazard probabilities $P(T_D = t \mid T_D \ge t)$ for t=0, 1, 2, …, $S_{Max}$. Induction begins at t=$S_{Max}$ but now must recognize that any time t may represent $T_D$. The random $T_D$ value function is defined as:

$V_R(t, RF(t))$ = Optimal (minimum) probability of ruin after time t given $RF(t) > 0$

$\alpha_R(t, RF(t))$ = Value of $\alpha$ required to achieve $V_R(t, RF(t))$.

Given $T_D \ge t$, Ruin$^C$(t) occurs if the withdrawal at time t is successful and Ruin$^C(\le S_{Max})$ occurs at time t if the withdrawal made is the last attempted. The sub-problem presented in Section II-F must change to reflect this. We assume that $T_D$ and $\hat{r}_{(t,\alpha)}$ are independent RVs $\forall$ t=1, …, $S_{Max}$.

## H.1 The Induction Process for Random $T_D$

Assume the retiree arrives at time t=$S_{Max}$ and makes their withdrawal. $RF(S_{Max})$ (> 0) need not be calculated since there are no more withdrawals and Ruin$^C(\le S_{Max})$ has occurred. At t=$S_{Max}$, P(Ruin(>$S_{Max}$))=0 and a B.C. is $V_R(S_{Max}, RF(S_{Max}))$=0, $\forall$ $RF(S_{Max}) > 0$. For any other time t, assume the retiree arrives at that time point, makes their withdrawal and updates $RF(t) > 0$.

The value function $V_R(t, RF(t))$ is defined as:

$$V_R(t, RF(t)) = \text{Min}_{(\alpha)}\left\{P(\text{Ruin}(>t))\right\} \tag{11a}$$

$$= \text{Min}_{(\alpha)}\left\{1 - P(\text{Ruin}^C(t+1) \cap P(\text{Ruin}^C(t+2) \cap \dots \cap \text{Ruin}^C(S_{Max}))\right\} \tag{11b}$$

$$= \text{Min}_{(\alpha)}\left\{\begin{array}{l} P(T_D > t \mid T_D \geq t) * [1 - \{P(\hat{r}_{(t+1,\alpha)} > RF(t)) * \\ \qquad P(\text{Ruin}^C(t+2) \cap \dots \cap \text{Ruin}^C(S_{Max}) \mid \hat{r}_{(t+1,\alpha)} > RF(t))\}] \end{array}\right\} \tag{11c}$$

We show in Appendix D that this minimization over $(0 \leq \alpha \leq 1)$ is equivalent to:

$$V_R(t, RF(t)) = \left[\begin{array}{l} \text{Min}_{(0 \leq \alpha \leq 1)}\left\{\begin{array}{l} P(T_D > t \mid T_D \geq t) * \{1 - (1 - F_{\hat{r}(t+1,\alpha)}(RF(t))) * \\ \qquad (1 - E_{\hat{r}(t+1,\alpha)^+}\left[V_R(t+1, \frac{RF(t)}{\hat{r}_{(t+1,\alpha)} - RF(t)})\right])\} \end{array}\right\} \\ \text{For: } 0 \leq t \leq S_{Max}\text{-}1, \ RF(t) > 0, \ V_R(S_{Max}, RF(S_{Max})) = 0. \end{array}\right. \tag{12}$$

Optimality is achieved at $\alpha_R(t, RF(t))$ and the expectation is over the conditional RV $\hat{r}_{(t+1,\alpha)}{}^+$. $V_R(\cdot)$ in (12) resembles $V(\cdot)$ in (9) except for the factor $P(T_D > t \mid T_D \geq t)$, which bounds $P(\text{Ruin})$ from above when $T_D$ is random. This is intuitive since $P(\text{Ruin})$ at any future time point cannot exceed the probability of living to attempt another withdrawal. We see that the fixed $T_D$ model in (9) is a special case of (12) with $P(T_D = S_{Max})=1$ or $P(T_D > t \mid T_D \geq t)=1$, t=0,1,…,$S_{Max}$-1 and $P(T_D > S_{Max} \mid T_D \geq S_{Max})=0$. $V(t, RF(t))$ and $V_R(t, RF(t))$ represent value functions of dynamic programs (DPs) with $V(0, RF(0))$ and $V_R(0, RF(0))$ being the optimal (minimum) $P(\text{Ruin})$ at any time point given $RF(0)=W_R$. Time t is considered the DP's *stage*, and $RF(t)$ the *state*. Since the solution finds the minimum $P(\text{Ruin}) \ \forall \ (t, RF(t))$ it is also a roadmap for maintaining optimality over time. Both $V(\cdot)$ and $V_R(\cdot)$ are exact, not simulated, and optimal meaning it is not possible to achieve a lower $P(\text{Ruin})$ under any other strategy using $W_R=RF(0)$ and the same stocks/bonds.

The challenge we face is solving these DPs for all $RF(t) > 0$, at each t. The nature of $V(\cdot)$ and $V_R(\cdot)$ will depend on the distributional assumptions made for inflation/expense-adjusted returns. If the expression being minimized has known form then finding $\alpha(\cdot)$ is handled with calculus. Closed form expressions for $V(\cdot)$ and $V_R(\cdot)$ are unlikely however. To overcome this we discretize the $\alpha$ and $RF(t)$ dimensions for $0 \leq \alpha \leq 1$ and $RF(t) > 0$, generating a numerical approximation to the optimal solution not based on simulation. The user can obtain an approximation to any desired degree of accuracy by increasing the discretization precision.

## I. Discretizing Along α and RF(t) Dimensions

Define $\alpha_{\{\cdot\}}$ and $R_{\{\cdot\}}$ as the sets that result from discretization precisions $P_\alpha$ and $P_R$ along the α and RF(t) dimensions, respectively with maximum RF(t) = $RF_{Max}$ ∀ t. That is,

$$\alpha_{\{\cdot\}} = \{0, 1/P_\alpha, 2/P_\alpha, \ldots, 1\text{-}(1/P_\alpha), 1\} \tag{13a}$$

$$R_{\{\cdot\}} = \{1/P_R, 2/P_R, 3/P_R, \ldots, RF_{Max}\text{-}(1/P_R), RF_{Max}\}. \tag{13b}$$

Along the ruin factor dimension each value (> $1/P_R$) in the set $R_{\{\cdot\}}$ will serve as the midpoint of a bucket constructed around it. Every RF(t) > 0 maps to one of the $(P_R)*RF_{Max}+1$ buckets, with the last bucket containing all ruin factors > $RF_{Max}+1/(2P_R)$. (See Figure 4.)

### Figure 4
### Discretization of the Positive Ruin Factor Dimension

The DPs proposed in (9) and (12) are unlikely to exhibit closed form solutions since CDFs of RVs used to model market returns typically do not have closed functional forms. To implement these DPs we discretize the continuous dimensions of α and RF(t). Discretizing α is trivial, but discretizing RF(t) is more involved. In this figure we show the RF(t) discretization strategy given ruin factor precision $P_R$ and maximum ruin factor $RF_{Max}$. The discretization is contiguous (no gaps).

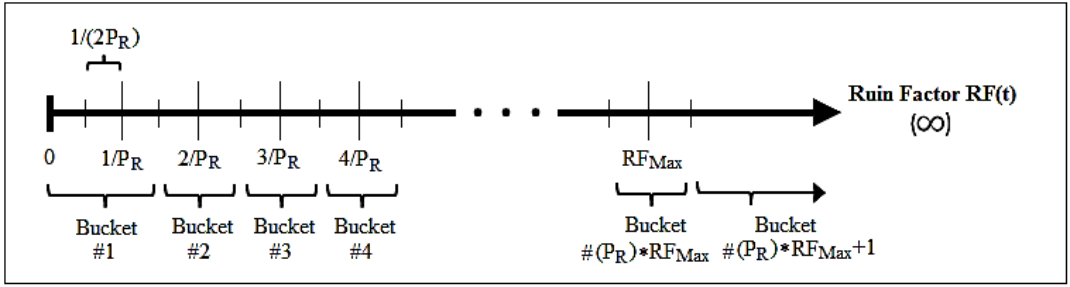

We will use the bucket midpoints when the discrete DP implementation requires a single RF(t) value. Since the buckets are contiguous (no gaps), the solution provides coverage for all RF(t) > 0. The value functions V(·) and $V_R$(·) are derived for (t, i/$P_R$), i=1, 2, 3, …, $(P_R)*RF_{Max}$, and V(t, RF(t)) = 1 for RF(t) > $RF_{Max}$ + 1/(2$P_R$). As $P_R$→∞ the buckets collapse to the midpoint and the discretization along the RF(t) dimension approaches the continuous solution. We handle the conditional expectations by discretizing the probability distribution of $\hat{r}_{(t+1,\alpha)}^{+}$ in a manner that coincides with the bucket boundaries. At time t the probability that RF(t+1) falls in bucket i, i = 1, 2, …, $(P_R)*RF_{Max}$, $(P_R)*RF_{Max}+1$ given that RF(t+1) > 0 is completely determined by the CDF of $\hat{r}_{(t+1,\alpha)}$. The probability of falling into any bucket is precisely P(Ruin$^C$(t+1)) = P(RF(t+1) > 0) which is also determined by the CDF of $\hat{r}_{(t+1,\alpha)}$. The expectations over $\hat{r}_{(t+1,\alpha)}^{+}$ in (9) and (12) become sums over the values of $V(t+1,\ RF(t+1))$ and $V_R(t+1,\ RF(t+1))$ multiplied by the appropriate discrete conditional probability (see Appendix E). Let $V_d(t,\ RF(t))$ and $\alpha_d(t,\ RF(t))$ denote the discretized versions of $V(\cdot)$ and $\alpha(\cdot)$ from (9), respectively. Then,

$$V_d(t,\ i/P_R) = \begin{cases} \text{Min}_{(\alpha\,\in\,\alpha_{[\cdot]})} \left\{ 1 - (1 - F\hat{r}_{(t+1,\alpha)}(i/P_R))* \right. \\ \qquad\qquad \left. (1 - \textstyle\sum_{j=1}^{(P_R)*RF_{Max}+1}\{P(j)*[V_d(t+1,\ j/P_R)]\}) \right\} \\ \text{For } 0 \leq t \leq T_D\text{-1, } i=1, 2, 3, \ldots, (P_R)*RF_{Max}, \\ \text{w/B.C.'s } V_d(T_D,\ RF(T_D)) = 0 \ \ \forall\ RF(T_D) > 0, \\ \qquad\qquad V_d(t,\ RF(t)) = 1 \text{ for } RF(t) > RF_{Max}+1/(2P_R). \end{cases} \qquad (14)$$

Here, $P(j)$ denotes $P(RF(t+1)$ in Bucket # $j \mid \hat{r}_{(t+1,\alpha)} > RF(t))$ with optimal $\tilde{\alpha}=\alpha_d(t,\ RF(t))$. This DP can be solved in any programming language with the user's choice of $P_\alpha$ and $P_R$. The code operates on a grid over the $(t,\ RF(t))$ dimensions (see Figure 5). A basic (but inefficient) strategy for coding this DP is to start at the diamond, follow the arrows, and end at the square. Values of $V_d(\cdot)$ and $\alpha_d(\cdot)$ are derived and retained at each cell in the grid. As indicated, $V(\cdot)$ from (9) exhibits structure across the t and $RF(t)$ dimensions. Given $RF(t)$, $V(\cdot)$ decreases with t, and given t, $V(\cdot)$ increases with $RF(t)$. Justifications are by indirection. Assume that one unit of time elapses and $RF(t)$ does not change yet $V(t,\ RF(t))$ increases. As noted, the real account balance at time t is $(\$A)*RF(0)/RF(t)$. Therefore, the real account balance at time t+1 equals the real account balance at time t, but P(Ruin) has increased. This is a contradiction. Similarly, assume that at time t, $RF_1(t) < RF_2(t)$ but $V(t,\ RF_1(t)) > V(t,\ RF_2(t))$. This is also a contradiction since $(\$A)*RF(0)/RF_1(t) > (\$A)*RF(0)/RF_2(t)$ and $W_R=RF(0)$ for both portfolios.

$V(\cdot)$ and $V_R(\cdot)$ are optimal in the sense that no alternate strategy can produce a lower P(Ruin) using the same stocks/bonds, given $W_R$. The discrete DP is a numerical approximation of the optimal strategy and its implementation is CPU intensive, especially for large $P_\alpha$ and $P_R$. In general, larger $P_\alpha$ decrease P(Ruin) while larger $P_R$ improve the approximation's accuracy.

**Figure 5**
**Discretized DP Coded as a Time by Ruin Factor Grid**

The discretized version of the DP in (9) is given in (14) and it is directly extendable to the random TD model in (12). We can solve these DPs using many strategies but generally favor solutions with shorter processing times. One basic but inefficient strategy is shown in this figure. Here, our code would iterate over both the time and RF(t) dimensions starting at the diamond and ending at the square. In this figure $T_D$ is the time of final withdrawal, $P_R$ the ruin factor precision and V(·) the DP's value function.

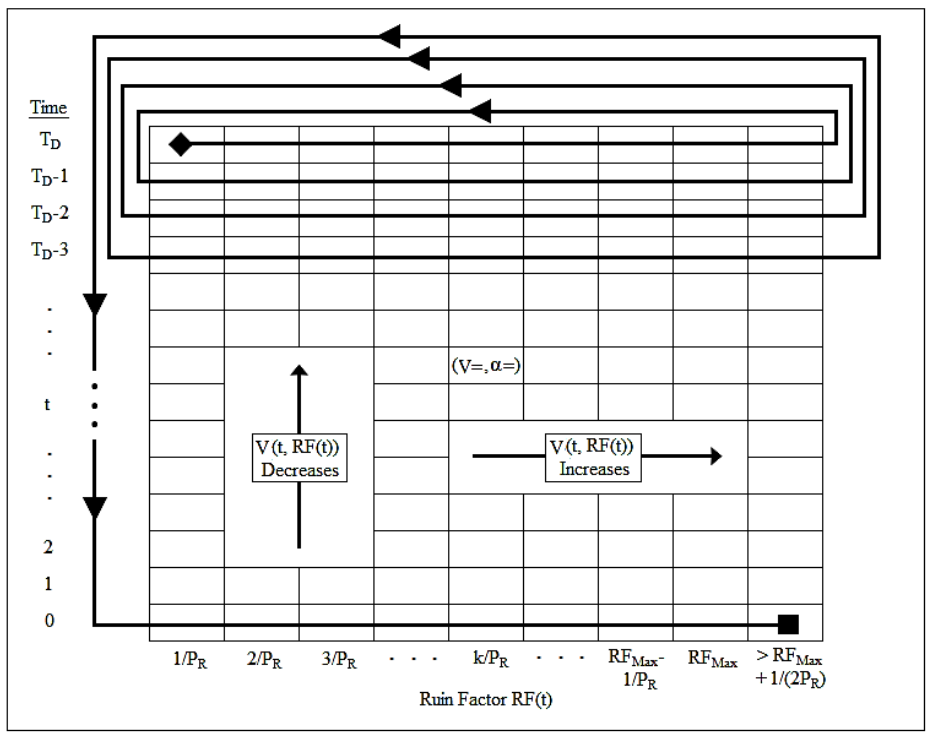

## III. Implementations for Fixed and Random TD

We use the technique in (14) to code the DP for fixed $T_D$ in (9) with $P_\alpha$=1,000, $P_R$=5,000, $RF_{Max}$=2.75, and $E_R$={0.5%, 0.0%}. Under this discretization 13,751 RF(t) buckets exist at each time t. We set the unit of time to years with $T_D$=30 resulting in a Figure 5 grid of 426,281 cells populated with (V, α). Each year the retiree calculates RF(t), slots it to a bucket and consults the grid. The value of V(·) reflects the optimal (minimum) P(Ruin) at any future time point up to t=$T_D$ using asset allocation α. Note that DPs require an optimal policy be followed at the current, and all future time points reflecting the *principle of optimality*. To facilitate DP coding we must make distributional assumptions for stock/bond returns.

*A. Distributional Assumptions*

We use historical annual S&P 500 total returns, 10-year Treasury Bond total returns, and the CPI-U from 1928 − 2013 to represent stock returns, bond returns, and the inflation rate, respectively in this analysis. We implicitly assume that future stock/bond returns originate from the same underlying probability distributions as past returns and N=86 years of historical returns are considered. After adjusting for inflation we find that real total stock and bond returns originate as random samples from underlying normal distributions with ($\mu$, $\sigma$) of (0.0825, 0.2007) and (0.0214, 0.0834), respectively. Further, a small positive correlation of $\rho$=0.04387 exists between these returns at each time t. (See Appendix F.)

The models in (9) and (12) make no assumption regarding the underlying distributions of asset class returns. The only requirement is that tail probabilities for convex combinations of correlated stock/bond returns are known or can be simulated/approximated at the bucket boundaries over the set $\alpha_{\{\cdot\}}$. The bucket boundaries are fixed and known once $P_R$ has been selected, therefore the tail probabilities can be preprocessed. Array references would replace CDF function calls in the code. Further, the CDF call at the bucket midpoint can be replaced by the average boundary CDF probabilities without impacting convergence of the discrete DP to its continuous counterpart.

In this implementation we assume stock/bond returns are *iid* over time, but the *identical distribution* assumption is not required. Users may apply forecasts where means and variances change as a function of time. Further, our models assume 2 asset classes but this too can be relaxed if the CDF of linear combinations of (possibly correlated) returns across multiple asset classes is obtainable. Clever coding may be required to keep the runtime reasonable since the optimization would be over more than just $\alpha$ at each (t, RF(t)).

*B. Discretized DP Results for Fixed $T_D$*

The discrete DP implementation for fixed $T_D$ described in Section III populates (V, $\alpha$) for a grid of 426,281 cells and 46 are shown in Figure 6. Here, V is the minimum P(Ruin) at any future time point, and $\alpha$ is the required stock proportion. Since RF(0)=$W_R$, the time t=0 row reveals the optimal P(Ruin) for initial $W_R$ between 3.5% and 4.5%, along with the α required for optimality during the first year. Noteworthy is that a 4% inflation-adjusted $W_R$ with $E_R$=0.5% yields an optimal P(Ruin) of 0.065 (success rate = 93.5%) and starting α of 41.4% in stocks.

**Figure 6**
**Select Discrete Implementation Cells using Expense Ratio $E_R$=0.5%**

This figure displays a sample of 46 cells from the 426,281 cell optimal solution for (fixed) $T_D$=30 years, $P_\alpha$=1,000, $P_R$=5,000, and $E_R$=0.5%. We start the optimal strategy at time t=0 using the asset allocation (α) from the cell where $W_R$=RF(0). For a retiree in standard form, the first withdrawal occurs at time t=1 and the last at time t=$T_D$=30. At each time t the retiree attempts their withdrawal and if successful updates RF(t) then consults the grid for the new optimal (α). At time t, the optimal P(Ruin(>t)) is V which changes over time. Three paths are shaded and indicate possible routes through the grid (i.e., glide-paths). When our portfolio performs well RF(t) decreases and we track leftward. When it performs poorly RF(t) increases and we track rightward. When $T_D$=30, the first and last α decisions are at time t=0 and t=29, respectively. The first and last withdrawals are at times t=1 and t=30, respectively.

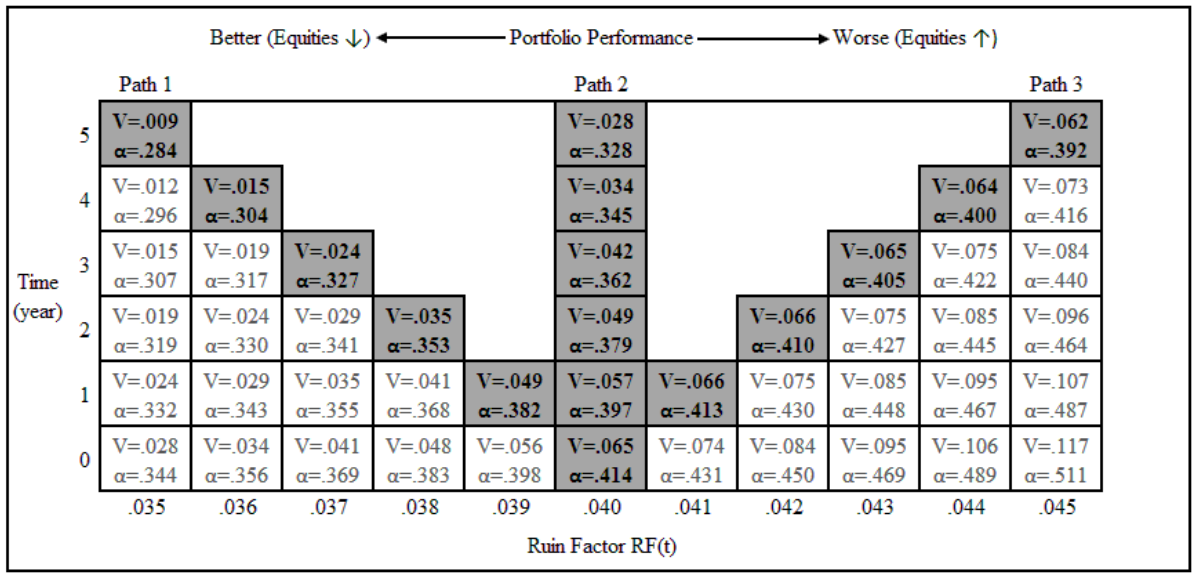

Three paths are shaded within Figure 6 for illustration. If the retiree uses an initial 4% $W_R$ and experiences favorable returns RF(t) decreases over time and the retiree will trace a leftward path perhaps similar to Path 1. Stable returns close to $W_R$ result in a constant RF(t) and the retiree's path could resemble Path 2. Unfavorable returns increase RF(t) and the retiree could track rightward along Path 3. As shown in Paths 1 and 2, P(Ruin) decreases in a favorable/stable market and α declines. Thus Paths 1 and 2 roughly follow the glide-path strategy used in TD funds that extend through the retirement date. Path 3, however reveals that the optimal equity allocation roughly remains constant or increases in an unfavorable market. This is consistent with findings by researchers who warn of poor sequences of returns early in retirement. See Cohen et al. (2010), and Pfau and Kitces (2014). Note that a year 1 inflation/expense-adjusted return of 1.799% would result in a rounded RF(1)=0.041 when RF(0)=0.040.

Figure 7 displays the results for $E_R$=0.0%. The optimal P(Ruin) using a 4% initial $W_R$ is 0.043 (success rate = 95.7%) with starting α of 36.8% in equities. The same pattern of equities decreasing during a favorable market and increasing in an unfavorable market holds true. The 95.7% success rate applies to any retiree who agrees to act optimally at all time points.

**Figure 7**
**Select Discrete Implementation Cells using No Expense Ratio (E$_R$=0.0%)**

This figure displays the same information as Figure 6, except it is for the solution with $E_R$=0.0%. With no expense ratio the time t=0 ruin probabilities (V) are lower as is the asset allocation (α), which is expected.

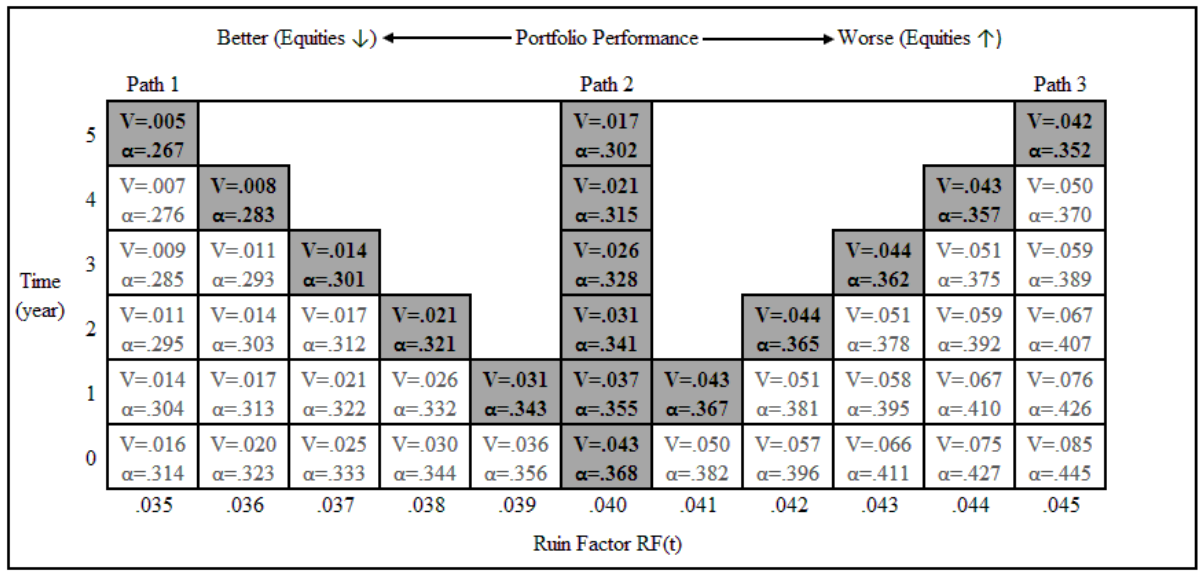

The grids in Figures 6 and 7 apply to all retirees using any initial $W_R$. If two retirees begin time t=0 with different $W_R$ and end up with equal RF(t) at time t then their returns as well as (V, α) will be identical for the remainder of retirement. This also means they will chart the exact same course through the grid after the point of intersection, and $W_R$ plays no role in the analysis. The intuition behind this conclusion is that the retiree with higher $W_R$ must experience favorable returns and chart a leftward path prior to the point of intersection. While charting this leftward path the retiree accumulates enough wealth to support larger withdrawals. We make no assessment here regarding the likelihood of such intersections occurring, however.

## B.1 Numerical Accuracy of the Solution

We assess the numerical approximation's accuracy by doubling $P_R$ from 5,000 to 10,000. This discretization creates 27,501 buckets at each time t and 852,531 total cells in the Figure 5 grid. The solution is directly comparable to Figure 7 ($E_R$=0.0%). The larger $P_R$ changes P(Ruin) and $\alpha$ modestly, but from a practical sense the results using $P_R$=5,000 are more than adequate when $T_D$=30. (See Table I.)

### Table I
### Comparison of V and α for $P_R$=5,000 vs. $P_R$=10,000 with $E_R$=0.0%

This table compares the minimum probability of ruin $V_d(0, RF(0))$ and the optimal asset allocation $\alpha_d(0, RF(0))$ at t=0 for models with $P_R$=5,000 vs. $P_R$=10,000 using 3 withdrawal rates ($W_R$). Here, $P_R$ is the RF(t) discretization precision. Higher values of $P_R$ lead to a more accurate numerical approximation when we discretize the DPs in (9) and (12), but also lead to longer run times. We show that very little is gained by doubling $P_R$ from 5,000 to 10,000. The optimal solution using $P_R$=5,000 has 426,281 total cells and the optimal solution using $P_R$=10,000 has 852,531 total cells when the alpha precision is $P_\alpha$=1,000.

| Time | RF(t) | $V_d(0, RF(0))$  [$\alpha_d(0, RF(0))$] | |
|---|---|---|---|
| (t) | (=$W_R$) | $P_R$=5,000[a] | $P_R$=10,000 |
| 0 | 0.035 | 0.01638 [0.314] | 0.01637 [0.314] |
| 0 | 0.040 | 0.04252 [0.368] | 0.04251 [0.368] |
| 0 | 0.045 | 0.08455 [0.445] | 0.08454 [0.445] |

[a] From Figure 7, row t=0.

## B.2 Consequences of Suboptimal Strategies

We assess the practical worth of the model in (9) and its implementation in Section III-B by comparing it with suboptimal strategies. Fixed portfolios having $\alpha \in$ {0.00, 0.25, 0.50, 0.75, 1.00} are considered under the same assumptions set forth in Section III-A, with $T_D$=30, $E_R$=0.0%, and $W_R$=4%. When $\alpha$ is fixed, simulation can be used to estimate P(Ruin) or the DP can be solved separately for each $\alpha$ with $|\alpha_{\{\cdot\}}|=1$. The minimum is over one $\alpha$ at each cell and V(0, 0.04) represents a numerical approximation to P(Ruin) using that strategy. The results are evaluated against V(0, 0.04) in Figure 7. An $\alpha$=0.50 portfolio generates the lowest P(Ruin) having a success rate of 91.5%, compared with 95.7% under the optimal strategy. (See Table II.)

**Table II**

**Comparable P(Ruin) using Fixed α Strategies**

This table lists the P(Ruin) values using $T_D$=30 years, $E_R$=0.0%, and $W_R$=4% for various fixed asset allocation (α) strategies. These ruin probabilities are directly comparable to those in Figure 7. We can generate these probabilities using simulation or the DP with a single α. Both methods were used and the results are shown to be virtually identical. Our point is to show the difference in P(Ruin) between the best fixed α strategy and the optimal strategy. The best fixed α is α=0.5 with corresponding P(Ruin) = 0.0850 (success rate = 91.5%). We saw in Figure 7 the optimal strategy yields a success rate of 95.7%. Thus, the consequence of using a suboptimal fixed α strategy is to accept a doubling of P(Ruin).

| Fixed α | Simulation[a] | DP Solved for $\alpha_{\{1..\}}$=$\{\alpha\}$[b] |
|---------|---------------|----------------------------|
| 0.00 | 0.3851 | 0.3850 |
| 0.25 | 0.1204 | 0.1204 |
| 0.50 | 0.0850 | 0.0851 |
| 0.75 | 0.1070 | 0.1070 |
| 1.00 | 0.1461 | 0.1460 |

[a] N=2.5 million/simulation
[b] $P_R$=5,000

The magnitude of difference between optimal and suboptimal P(Ruin) values increases with $T_D$. For example, using $T_D$=50, $E_R$=0.0%, and $W_R$=4% a fixed α=0.75 generates the lowest P(Ruin) of 0.217 (success rate = 78.3%) whereas the optimal strategy derived using $P_R$=10,000 and $P_\alpha$=1,000 generates a P(Ruin) of 0.137 (success rate = 86.3%).

*B.3 Algorithm Scalability*

Let k=$T_D$, M=$P_\alpha$, and N=$RF_{Max}*P_R$. The runtime for the discrete DP algorithm in (14) scales at O(k*M*N^2). This is because each cell in a row of the Figure 5 grid requires an expected value computation that accesses N cells at the next time point, which occurs for M asset allocations over all N cells in each of k rows. Therefore without pruning or parallel processing the implementation from Section III-B.1 has a 4X longer runtime than the implementation from Section III-B (since k and M did not change, but N is 2X larger). We can achieve significant and scalable efficiencies in the code, however (see Appendix H).

*C. Random $T_D$ Model for Multiple Retirees*

Define a multi-person unit (MPU) as a group that pools retirement funds and distributes systematic withdrawals amongst surviving members at each time point. The MPU makes withdrawals while it is alive and it is alive when at least one member is alive. Since systematic withdrawals are used ruin can occur. The general MPU consists of K females and L males. Let $F_i$ and $M_j$ represent the remaining lifetimes for female i and male j, respectively. At MPU retirement start $F_i$ and $M_j$ are independent discrete RVs with know PMF and:

$$T_D = \max\{F_1, \ldots, F_i, \ldots, F_K, M_1, \ldots, M_j, \ldots, M_L\} \tag{15}$$

Let $F_{T_D}(\cdot)$ and $f_{T_D}(\cdot)$ represent the CDF and PMF of $T_D$, respectively. The discrete time hazard probabilities $P(T_D=t \mid T_D \geq t)$ are derived as $f_{T_D}(t) / [1 - F_{T_D}(t-1)]$ for t=0, 1, 2, …, $S_{Max}$ with $F_{T_D}(-1)$ = 0. The PMF is constructed as $f_{T_D}(t) = F_{T_D}(t) - F_{T_D}(t-1)$ and the CDF is built by recognizing that the maximum of a set is less than or equal to a given value *iff* all members of the set are less than or equal to that value. Namely,

$$T_D \leq t \quad \leftrightarrow \quad \max\{F_1, \ldots, F_i, \ldots, F_K, M_1, \ldots, M_j, \ldots, M_L\} \leq t \tag{16a}$$

$$\leftrightarrow \quad (F_1 \leq t) \cap \ldots \cap (F_K \leq t) \cap (M_1 \leq t) \cap \ldots \cap (M_K \leq t) \tag{16b}$$

$$\rightarrow \quad F_{T_D}(t) \;=\; P(T_D \leq t) \;=\; P(F_1 \leq t)*\ldots*P(F_K \leq t)*P(M_1 \leq t)*\ldots*P(M_K \leq t). \tag{16c}$$

The individual RHS probabilities in (16c) are derived from published SSA life tables and the random $T_D$ model is solved by treating the MPU as an individual with known hazard probabilities. The solution is an optimal decumulation strategy and the MPU can select $W_R$ by first determining their desired success rate. Note that the same age M/F couple analysis in Section III-C.1 is a special-case MPU having K+L=2 (K=L=1).

A risk with the random $T_D$ models we propose here is that the hazard probabilities can become dated, rendering the solution obsolete. Suppose an MPU is formed and several members perish unexpectedly in year 1. This MPU will have remaining hazard probabilities that differ markedly from those used to construct the optimal strategy and the optimization should be rerun. The same applies to a couple where one member dies early in retirement. As K+L increases the risk lessens, meaning that couples are most exposed. They are also the easiest to adjust. Simply transfer the living member to the random $T_D$ model solved for single retirees of that gender starting at the appropriate time point. Random $T_D$ MPU models should thus be monitored over time and updated when realized hazard probabilities differ from those used in the optimization.

*C.1 Discretized DP Results for Random $T_D$*

We discretize the random $T_D$ model in (12) using the method proposed in Section III-C for retiring same-age male/female (M/F) couples (MPU with K=L=1). Hazard probabilities are derived from SSA life tables with time t=0 reflecting age 65. We assume the MPU is in standard form. (Refer to Section II-B.) The first withdrawal attempt, if necessary, would occur at time t=1 and the last at time t=$S_{Max}$=48 (years). We use precisions $P_\alpha$=1,000 and $P_R$=5,000 with $E_R$=0.0% and $W_R$={4%, 5%, 6%}. The results are compared against the best performing fixed $\alpha$ strategy found using simulation (N=2.5 million per $W_R$). All fixed $\alpha$ values between 0.0 and 1.0 were assessed in increments of 0.05. The best performing fixed $\alpha$ strategy using $W_R$=4% is $\alpha$=0.45 and the corresponding P(Ruin) is 0.0421 (success rate = 95.8). The optimal strategy starts time t=0 with $\alpha$=0.356 and yields a minimum P(Ruin) of 0.0287 (success rate = 97.1%) which is a 31.8% improvement. Comparisons for $W_R$={4%, 5%, 6%} are shown in Table III.

**Table III**
**P(Ruin) for Random Same-Age (65) M/F Couple $T_D$: Optimal vs. Fixed $\alpha$ Strategies**

In this table we present a comparison of the optimal random $T_D$ model vs. the best performing fixed $\alpha$ strategy using withdrawal rates of $W_R$={4%, 5%, 6%} for a same-age male/female couple. We assume the retirees are in standard form with time t=0 reflecting age 65. The first withdrawal attempt occurs at time t=1 (if $T_D \geq 1$) and the last at t=$S_{Max}$=48 (if $T_D$=48). The discrete time hazard probabilities P($T_D$=t | $T_D \geq$ t) were derived from life-tables at SSA.gov for t=0, 1, …, 48. The model was discretized using precisions $P_\alpha$=1,000 and $P_R$=5,000 with $E_R$=0.0%. The best performing fixed $\alpha$ solutions were found using simulation (N=2.5 million/$W_R$) with the test set {$\alpha$: $\alpha$=0.00 to 1.00 by 0.05}.

| Demo | $W_R$ | Suboptimal | Optimal | % Decrease in |
|------|-------|------------|---------|---------------|
| | | P(Ruin) [Best Fixed $\alpha$] | P(Ruin) [$\alpha$ at t=0] | P(Ruin) |
| M/F Couple | 4% | 0.0421 [0.45] | 0.0287 [0.356] | 31.8% |
| | 5% | 0.1349 [0.60] | 0.0978 [0.481] | 27.5% |
| | 6% | 0.2523 [0.80] | 0.2009 [0.672] | 20.4% |

The corresponding optimal solution grid for the random $T_D$ model above is shown in Figure 8 and reveals the familiar pattern of $\alpha$ decreasing when returns are favorable, and increasing when returns are unfavorable. As noted in Section III-C, this model is exposed to dated hazard risk and requires close monitoring with possible reoptimization over time.

**Figure 8**
**Select Discrete Implementation Cells for Random $T_D$:  M/F Couple w/$E_R$=0.0%**

This figure displays a sample of 46 cells from the 673,799 cell optimal solution for the random $T_D$ model with $P_\alpha$=1,000, $P_R$=5,000, $E_R$=0.0%, and $S_{Max}$=48.  The first asset allocation decision for an MPU in standard form is at time t=0 and the last at time t=47 ($S_{Max}$-1).  The first withdrawal is attempted at time t=1 (if $T_D \geq 1$) and the last at time t=48 (if $T_D$=48).  Three paths are shaded as they were in Figures 6 and 7 indicating how the optimal solution changes over time under different portfolio performances.  The value of V reflects P(Ruin) at any future time point and α is the corresponding optimal asset allocation.

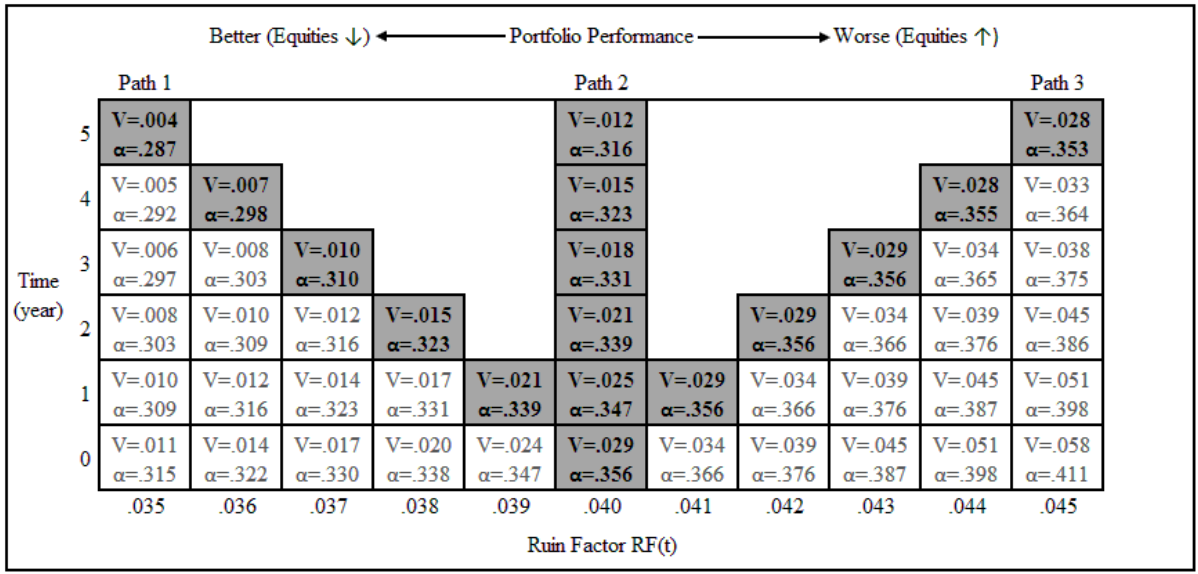

*D.  Making Adjustments over Time*

We make adjustments to the optimal strategy as follows.  Suppose retirees charting Path 1 in Figure 7 or 8 are dissatisfied with the decreasing α, and retirees charting Path 3 are dissatisfied with the non-decreasing P(Ruin).  Both prefer a shift to Path 2 after making the time t=5 withdrawal.  Table IV provides the real returns $r_{(t, \alpha)}$ that track the RF(t) bucket midpoints for each path.  Both began time t=0 with $W_R$=4% and an account balance of \$A.  In time t=0 dollars the each withdraws (4%)∗(\$A)/year.  The Path 1 retiree can shift to Path 2 by increasing the real withdrawal amount to (4%)∗(1.143)∗(\$A) for t >= 6.  The Path 3 retiree can shift by lowering the real withdrawal amount to (4%)∗(0.889)∗(\$A) for t >= 6.  The new rates are thus $W_R$= (4%)∗(1.143)=4.57% and $W_R$=(4%)∗(0.889)=3.56%, respectively, if based on \$A at time t=0.  At time t=6 the retirees begin their new strategy and withdraw (4.57%)∗(\$A)∗$\prod_{i=1}^{6} (1+I_i)$ and (3.56%)$*$($A)$*\prod_{i=1}^{6}(1+I_i)$, respectively. Essentially, they halt their original plans and begin anew with time t=5 as the new time t=0, and horizon of $T_D$=25 years in the case of Figure 7. The starting balances are (1.143)$*$($A)$*\prod_{i=1}^{5}(1+I_i)$ and (0.889)$*$($A)$*\prod_{i=1}^{5}(1+I_i)$, respectively (see Table IV). By shifting to Path 2, both use $W_R$=4%=RF(5), but it is now based on their time t=5 balances. We follow the same process when provisioning for emergencies.

**Table IV**
**Real Returns that Generate Paths 1 and 3 from Figure 7 or 8**

This figure displays the rounded returns that would track the RF(t) bucket midpoints of Paths 1 and 3 in Figures 7 and 8 (which use different α). These paths assume $W_R$=4% and $E_R$=0.0. At each time point we apply the real return then subtract the real withdrawal $(W_R)*($A)$, where $A is the time t=0 account balance. We can change $W_R$ at any time point by starting over with a new balance and consulting the grid for the optimal α. A retiree that tracks Path 1 starts time t=5 with new balance (1.143)$*$($A)$*\prod_{i=1}^{5}(1+I_i)$. This is the new time t=0 and time t=6 reflects the new time t=1 using a modified $W_R$.

| Time | Path 1 | | | Path 3 | | |
| (t) | Real Return $r_{(t, \alpha)}$ | Real Account Balance | RF(t) | Real Return $r_{(t, \alpha)}$ | Real Account Balance | RF(t) |
|---|---|---|---|---|---|---|
| 1 | 6.56% | | .039 | 1.56% | | .041 |
| 2 | 6.53% | | .038 | 1.72% | | .042 |
| 3 | 6.50% | | .037 | 1.87% | | .043 |
| 4 | 6.48% | | .036 | 2.03% | | .044 |
| 5 | 6.46% | (1.143)$*$($A) | .035 | 2.18% | (0.889)$*$($A) | .045 |

## IV. Summary and Conclusions

The discrete time models proposed in (9) and (12) yield optimal retirement decumulation strategies. Once distributional assumptions regarding asset class returns are made these models estimate the optimal strategy. As formulated they are intractable under common distributional assumptions and must be discretized. The discretized solution is thus a numerical approximation to the estimate of an optimal strategy. Since the user controls the discretization's precision, the approximation can be driven to any desired degree of accuracy. This leaves distributional assumptions as the determining factor regarding how close the user's solution is to true optimality, and different users are sure to make different distributional assumptions.

We emphasize that these decumulation strategies are based on an objective of minimizing the probability of ruin, not maximizing terminal wealth. Retirees seeking to maximize bequest wealth may use this approach by increasing $W_R$ to gain more risk exposure, then invest unspent funds aggressively outside of their retirement portfolio. However, it may be more prudent for such a retiree to employ a model that was built on a wealth maximization objective, for example that proposed by Fan, Murray, and Pittman (2013). We note that evidence suggests retirees more often bias towards loss prevention than wealth maximization. A 2012 study by ING Retirement Research Institute found that 80% of TD fund users prefer a portfolio that protects against losses, and 66% of non TD fund users agreed.

A usage scenario for an advisor who seeks to employ the models proposed here is as follows. The advisor codes or has the DP in (14) coded based on the market return assumptions they perceive as most appropriate for the coming retirement horizon using either a fixed or random $T_D$, and desired levels of precision.[3] The result is a grid similar in form to Figure 5. The advisor then customizes this grid for each retiree by shading various regions the retiree indicates they are comfortable and uncomfortable entering. A plan is then set forth at time t=0 which indicates precisely when and what type of adjustments will be made if the retiree encroaches on a region they have indicated is intolerable. The retiree thus takes comfort in the knowledge that a strategy exists, and they remain fully informed of what actions will be taken, and when they will be taken to modify that strategy based on their portfolio's performance over time.

With this research we prove that decumulation strategies whose glide-paths are fixed at the outset of retirement are suboptimal and this is formalized in Appendix G. Shifting from equities to bonds reduces volatility but also the expected return and this shift is made without regard to the retiree's withdrawal rate. It therefore increases retirement risk when defining risk as outliving one's savings. When using a safe withdrawal rate strategy, the only optimal glide-path is the one that responds to market returns as they are observed over time. We have presented a means to approximate that glide-path to any desired degree of accuracy. Lastly, we hope this research ends the perception that the safe withdrawal rate is a simplistic rule-of-thumb unjustified by academic theory. In fact, the theoretical framework underlying safe withdrawal rates is rich and exhibits a natural extension to principles of optimization that are used across a wide array of disciplines.

---

[1] These and similar figures are commonly reported in the press. See, for example, "Baby Boomers Approach 65 – Glumly" by D. Cohn and P. Taylor of the Pew Research Center. (December 20, 2010). Web URL: http://www.pewsocialtrends.org/2010/12/20/baby-boomers-approach-65-glumly/

[2] The proof is trivial. A time t=1 withdrawal amount of $(W_0)*(\$B)*(1+I_1)$ maintains the retiree's purchasing power. Replace $\$B$ with $\$A/(1-W_0)$ and $W_0$ with $W_R/(1+W_R)$ then simplify. The result is $(W_R)*(A)*(1+I_1)$ which is exactly the time t=1 withdrawal amount assumed by the model's standard form representation.

[3] A full C++ implementation is provided in Appendix H.

**Minimizing the Probability of Ruin in Retirement**

CHRISTOPHER J. ROOK*

**INTERNET APPENDIX**

* The author is a consultant statistical programmer and studies in the Department of Systems Engineering at Stevens Institute of Technology.  This document accompanies the primary research paper and includes proofs, derivations, source code, and miscellanea.

**Table of Contents**

## Appendix A.  Derivation of the Real Account Balance at Time t

LEMMA A1:  Given $Ruin^C(\leq t)$, the following equality holds:

$$\hat{r}_{(t,\alpha)} = RF(t-1)*[1 + 1/RF(t)] \tag{A.1}$$

*Proof:*  By definition,

$$RF(t) = RF(t-1)/[\hat{r}_{(t,\alpha)} - RF(t-1)] \tag{A.2a}$$

$$\rightarrow \quad [\hat{r}_{(t,\alpha)} - RF(t-1)]*RF(t) = RF(t-1) \tag{A.2b}$$

$$\rightarrow \quad [\hat{r}_{(t,\alpha)} - RF(t-1)] = RF(t-1)/RF(t) \tag{A.2c}$$

$$\rightarrow \quad \hat{r}_{(t,\alpha)} = RF(t-1)/RF(t) + RF(t-1) \tag{A.2d}$$

$$\rightarrow \quad \hat{r}_{(t,\alpha)} = RF(t-1)*[1 + 1/RF(t)] \tag{A.2e}$$

PROPOSITION A1:  Given $Ruin^C(\leq t)$, the real account balance at time t is $(\$A)*RF(0)/RF(t)$.

*Proof:*  By induction we show the proposition holds for base case times t={0, 1}.  Then assuming the proposition holds at time t=t-1 we show it must also follow at time t.

Real Account Balance (t=0):   $\$A$ (A.3a)

$$= (\$A)*(1) \tag{A.3b}$$

$$= (\$A)*RF(0)/RF(0) \tag{A.3c}$$

Real Account Balance (t=1):   $(\$A)*(\hat{r}_{(1,\alpha)}) - (\$A)*(W_R)$ (A.4a)

$$= (\$A)*(\hat{r}_{(1,\alpha)} - W_R) \quad \boxed{\text{By Lemma A1.}} \tag{A.4b}$$

$$= (\$A)*(RF(0)*[1 + 1/RF(1)] - W_R) \tag{A.4c}$$

$$= (\$A)*(RF(0)*[1 + 1/RF(1)] - RF(0)) \tag{A.4d}$$

$$= (\$A)*(RF(0)*[1 + 1/RF(1) - 1]) \tag{A.4e}$$

$$= (\$A)*RF(0)/RF(1) \tag{A.4f}$$

Real Account Balance (t=t-1):   $(\$A)*RF(0)/RF(t-1)$ $\boxed{\text{Assume}}$ (A.5)

Real Account Balance (t=t):   $[(\$A)*RF(0)/RF(t-1)]*(\hat{r}_{(t,\alpha)}) - (\$A)*W_R$ (A.6)

By Lemma A1.

$$= (\$A)*[RF(0)*RF(t-1)*[1+1/RF(t)]/RF(t-1)-W_R] \qquad (A.7a)$$

$$= (\$A)*[RF(0)*[1+1/RF(t)]-RF(0)] \qquad (A.7b)$$

$$= (\$A)*[RF(0)*(1+1/RF(t)-1)] \qquad (A.7c)$$

$$= (\$A)*RF(0)/RF(t) \qquad (A.7d)$$

We noted in Section II-C that the reciprocal of the ruin factor equals the number of real withdrawals remaining. This statement is a direct result of Proposition A1. That is, $1/RF(t) = \#$ of real withdrawals remaining at time t, and $RF(t) = 1/(\#$ of real withdrawals remaining at time t).

## Appendix B.  Criteria for Ruin(t) Given Ruin$^C$(t-1)

PROPOSITION B1:  Given Ruin$^C(\leq t-1)$, Ruin(t) occurs *iff* $\hat{r}_{(t,\alpha)} \leq RF(t-1)$.

*Proof:*

By Proposition A1.

Real Account Balance (t=t-1):      $(\$A)*RF(0)/RF(t-1)$ \hfill (A.8a)

Actual Account Balance (t=t-1):      $[(\$A)*RF(0)/RF(t-1)]*\prod_{i=1}^{t-1} (1+I_i)$ \hfill (A.8b)

Actual Account Balance (t=t)[1]:      $[(\$A)*RF(0)/RF(t-1)]*\prod_{i=1}^{t-1} (1+I_i)$ \hfill (A.8c)
$$*(1+R_{(t,\alpha)})*(1-E_R)$$

Actual Withdrawal Amount (t=t):      $(\$A)*(W_R)*\prod_{i=1}^{t} (1+I_i)$ \hfill (A.8d)

Condition Required for Ruin(t):

Account Balance (t=t)   $\leq$   Withdrawal Amount (t=t)

$$(1+R_{(t,\alpha)})*(1-E_R)*[(\$A)*RF(0)/RF(t-1)]*\prod_{i=1}^{t-1} (1+I_i) \leq (\$A)*(W_R)*\prod_{i=1}^{t} (1+I_i) \qquad (A.8e)$$

$$\leftrightarrow (1+r_{(t,\alpha)})*(1-E_R)*[1/RF(t-1)] \leq 1 \qquad (A.8f)$$

$$\leftrightarrow \hat{r}_{(t,\alpha)} \leq RF(t-1) \qquad (A.8g)$$

---

[1] This value is pre-withdrawal.

# Appendix C.  Induction for Fixed $T_D$

## C.1  Induction at Time $t=T_D$

Assume the retiree arrives at time $t=T_D$ and makes their last withdrawal.  The restricted sample space $S=\{\text{Ruin}^C(\leq T_D)\}$ includes a single event, shown at right.  The retiree need not compute $RF(T_D)$ as there are no more withdrawals, however 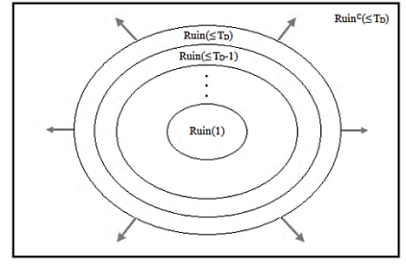 $RF(T_D) > 0$ (if computed) since $\text{Ruin}^C(\leq T_D)$ has occurred.  At $t=T_D$, $P(\text{Ruin}(>T_D))=0$, and a boundary condition (B.C.) for the value function is $V(T_D, RF(T_D)) = 0$, $\forall\ RF(T_D) > 0$.

## C.2  Induction at Time $t=T_D-1$

Assume the retiree arrives at time $t=T_D-1$, makes their 2[nd] last withdrawal and has one remaining.  The ruin factor $RF(T_D-1)$ $(> 0)$ is calculated based on the portfolio's return just observed, $\hat{r}_{(TD-1,\alpha)}$.  The retiree now faces the restricted sample 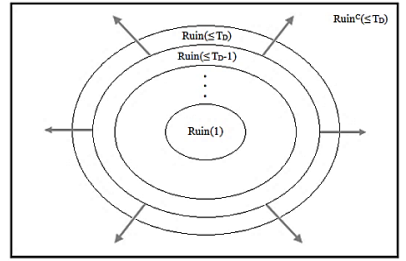 space $S=\{\text{Ruin}(T_D), \text{Ruin}^C(\leq T_D)\}$, as shown, and seeks the $\alpha$ that minimizes $P(\text{Ruin}(T_D))$.  This straight forward decision is made using the framework presented in Section II-F.  Namely, the retiree compares the tail probabilities for various asset allocations and selects the one which minimizes $P(\text{Ruin}(T_D))$.  This optimization can be expressed as:

$$V(T_D-1, RF(T_D-1)) \quad = \quad \text{Min}_{(0 \leq \alpha \leq 1)} \left\{ P(\text{Ruin}(T_D) \right\} \tag{C.1a}$$

$$\rightarrow V(T_D-1, RF(T_D-1)) \quad = \quad \text{Min}_{(0 \leq \alpha \leq 1)} \left\{ 1 - P(\text{Ruin}^C(T_D)) \right\} \tag{C.1b}$$

$$\rightarrow V(T_D-1, RF(T_D-1)) \quad = \quad \text{Min}_{(0 \leq \alpha \leq 1)} \left\{ 1 - P(\hat{r}_{(TD,\alpha)} > RF(T_D-1)) \right\} \tag{C.1c}$$

$$\rightarrow V(T_D-1, RF(T_D-1)) \quad = \quad \text{Min}_{(0 \leq \alpha \leq 1)} \left\{ 1 - (1 - F\hat{r}_{(TD,\alpha)}(RF(T_D-1))) \right\}; \tag{C.1d}$$

given the known ruin factor $RF(T_D-1)$.  Here, $F_{\hat{r}(TD,\alpha)}(\cdot)$ denotes the known/estimated CDF of $\hat{r}_{(TD,\alpha)}$.  Note that since $V(T_D, RF(T_D))=0$, we can equivalently express $V(T_D-1, RF(T_D-1))$ as:

$$V(T_D-1, RF(T_D-1)) = \text{Min}_{(0 \le \alpha \le 1)}\left\{ 1 - \underbrace{(1 - F\hat{r}_{(TD,\alpha)}(RF(T_D-1)))}_{\substack{\text{Probability of no} \\ \text{ruin at time } t=T_D.}} * \underbrace{(1 - E\hat{r}_{(TD,\alpha)}{}^+[V(T_D, RF(T_D))])}_{\substack{\text{Expected prob. of no ruin after} \\ \text{time } t=T_D, \text{ given Ruin}^C(T_D).}} \right\},$$ (C.2)

with optimal $\tilde{\alpha}=\alpha(T_D-1, RF(T_D-1)).$[2]

### C.3  Induction at Time $t=T_D-2$

Assume the retiree arrives at time $t=T_D-2$, makes their 3[rd] last withdrawal and has two remaining.  The ruin factor $RF(T_D-2)$ $(> 0)$ is calculated based on the portfolio's return just observed, $\hat{r}_{(T_D-2,\alpha)}$.  The retiree now faces the restricted sample 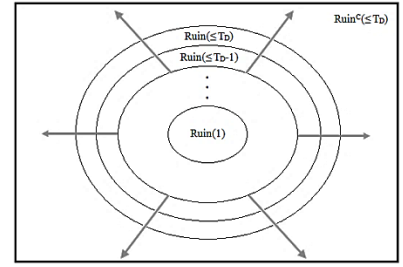 space $S=\{\text{Ruin}(T_D-1), \text{Ruin}(T_D), \text{Ruin}^C(\le T_D)\}$, as shown, and seeks to make the optimal asset allocation decision to minimize $P(\text{Ruin}(>T_D-2)) = P(\text{Ruin}(T_D-1) \cup \text{Ruin}(T_D))$, which is the probability of ruin at any future time point.  Using (6b) and (6c), we express $P(\text{Ruin}(>T_D-2))$ as:

$$P(\text{Ruin}(T_D-1) \cup \text{Ruin}(T_D)) \quad = 1 - P(\text{Ruin}^C(T_D-1) \cap \text{Ruin}^C(T_D)) \quad \text{(C.3a)}$$

$$= 1 - P(\text{Ruin}^C(T_D-1)) * \underbrace{P(\text{Ruin}^C(T_D) \mid \text{Ruin}^C(T_D-1))}. \quad \text{(C.3b)}$$

| The optimal value for this probability was just derived at time $t=T_D-1$ for all $RF(t) > 0$. |
|---|

The value function is expressed as:

$$V(T_D-2, RF(T_D-2)) = \text{Min}_{(0 \le \alpha \le 1)}\left\{ 1 - P(\text{Ruin}^C(T_D-1)) * P(\text{Ruin}^C(T_D) \mid \text{Ruin}^C(T_D-1)) \right\} \text{ (C.4a)}$$

$$\rightarrow V(T_D-2, RF(T_D-2)) = \text{Min}_{(0 \le \alpha \le 1)}\left\{ \begin{array}{l} 1 - P(\hat{r}_{(TD-1,\alpha)} > RF(T_D-2)) * \\[6pt] P(\hat{r}_{(TD,\tilde{\alpha})} > RF(T_D-1) \mid \hat{r}_{(TD-1,\alpha)} > RF(T_D-2)) \end{array} \right\} \text{ (C.4b)}$$

$$\rightarrow V(T_D-2, RF(T_D-2))^3 = \text{Min}_{(0 \le \alpha \le 1)}\left\{ \begin{array}{l} 1 - P(\hat{r}_{(TD-1,\alpha)} > RF(T_D-2)) * \\[8pt] \left[\dfrac{P(\hat{r}_{(T_D,\tilde{\alpha})} > RF(T_D-1) \cap \hat{r}_{(T_D-1,\alpha)} > RF(T_D-2))}{P(\hat{r}_{(T_D-1,\alpha)} > RF(T_D-2))}\right] \end{array} \right\} \text{ (C.4c)}$$

---

[2] The right-most term is added for notational convenience.  Recall that $V(T_D, RF(T_D)) = 0 \; \forall \; RF(T_D) > 0$ and it is therefore the expected value of zero.  Further, $\hat{r}_{(TD,\alpha)}{}^+ = (\hat{r}_{(TD,\alpha)} \mid \hat{r}_{(TD,\alpha)} > RF(T_D-1))$ as given in Section II-G.1.

[3] Our convention is to let $\tilde{\alpha}$ refer to an alpha that is optimal at a future time point, and let $\alpha$ represent one that is being optimized over at the current time point.  The optimal $\tilde{\alpha}$ is always required to minimize the current $V(\cdot)$.

The numerator of the ratio in (C.4c) reflects the probability that ruin is avoided at both times t=$T_D$-1 and $T_D$.  By definition, this is the volume of the joint PDF $f(\hat{r}_{(TD-1,\alpha)}, \hat{r}_{(TD,\tilde{\alpha})})$ over the region defined in the probability statement.  This region exists in the $\hat{r}_{(TD-1,\alpha)}$–$\hat{r}_{(TD,\tilde{\alpha})}$ plane and the joint PDF defines a 3-dimensional object that rests on the plane.  We calculate the required volume by integrating the joint PDF over the given region.  Since the integration limits for $\hat{r}_{(TD,\tilde{\alpha})}$ depend on $\hat{r}_{(TD-1,\alpha)}$ we must handle $\hat{r}_{(TD,\tilde{\alpha})}$ first, where $\hat{r}_{(TD,\tilde{\alpha})}$ ranges from RF($T_D$-1) to $\infty$.  At this induction step $\hat{r}_{(TD-1,\alpha)}$ ranges from the constant RF($T_D$-2) to $\infty$.  This region of the $\hat{r}_{(TD-1,\alpha)}$–$\hat{r}_{(TD,\tilde{\alpha})}$ plane is shown below as the cross-hatched section of Figure A1.  Assuming returns are bell-shaped, then a 3-dimensional hill object (see right) depicts 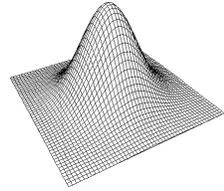 $f(\hat{r}_{(TD-1,\alpha)}, \hat{r}_{(TD,\tilde{\alpha})})$.  The required probability is the volume of this object over the region shown, and we must evaluate it across all α at time t=$T_D$-2.  Varying α changes the shape and location of the hill, and thus the probability.  The denominator of the ratio in (C.4c) is the volume of the same solid to the right of RF($T_D$-2).  We seek α that minimizes the entire expression in (C.4c) where these probabilities are two components, see (C.4d) below.

**Figure A1.  Region under $f(\hat{r}_{(TD-1,\alpha)}, \hat{r}_{(TD,\tilde{\alpha})})$ for Derivations of the Numerator and Denominator in (C.4c)**

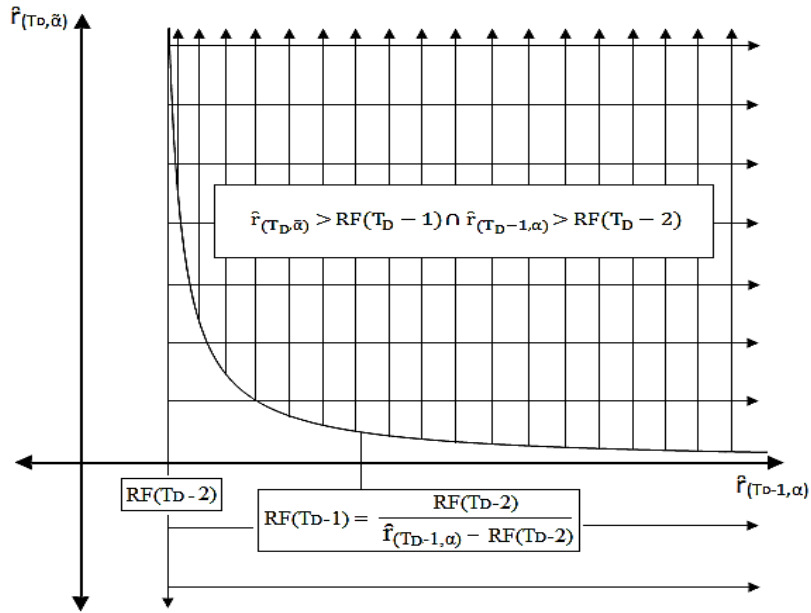

→ V(T_D-2, RF(T_D-2)) =

$$\rightarrow V(T_D\text{-}2, RF(T_D\text{-}2)) =$$

> Since RF(T_D-1) is a function of $\hat{r}_{(T_D\text{-}1,\alpha)}$, these integrals must remain nested and the ordering cannot be interchanged.

$$\text{Min}_{(0 \le \alpha \le 1)} \left\{ 1 - (1 - F\hat{r}_{(T_D\text{-}1,\alpha)}(RF(T_D\text{-}2)))* \left[ \frac{\int_{RF(T_D\text{-}2)}^{\infty} \int_{RF(T_D\text{-}1)}^{\infty} f\left(\hat{r}_{(T_D,\tilde{\alpha})}, \hat{r}_{(T_D\text{-}1,\alpha)}\right) d\left(\hat{r}_{(T_D,\tilde{\alpha})}\right) d\left(\hat{r}_{(T_D\text{-}1,\alpha)}\right)}{\int_{RF(T_D\text{-}2)}^{\infty} f\left(\hat{r}_{(T_D\text{-}1,\alpha)}\right) d\left(\hat{r}_{(T_D\text{-}1,\alpha)}\right)} \right] \right\} \quad \text{(C.4d)}$$

By conditioning and assuming returns are independent between time points we split the joint PDF in the numerator of the ratio (see Appendix G.2):

→ V(T_D-2, RF(T_D-2)) =

> This term is [1 – V(T_D-1, RF(T_D-1))] found earlier. An optimal policy must be followed at each stage of the DP.

$$\text{Min}_{(0 \le \alpha \le 1)} \left\{ 1 - (1 - F\hat{r}_{(T_D\text{-}1,\alpha)}(RF(T_D\text{-}2)))* \left[ \frac{\int_{RF(T_D\text{-}2)}^{\infty} f\left(\hat{r}_{(T_D\text{-}1,\alpha)}\right) \left[ \int_{RF(T_D\text{-}1)}^{\infty} f\left(\hat{r}_{(T_D,\tilde{\alpha})}\right) d\left(\hat{r}_{(T_D,\tilde{\alpha})}\right) \right] d\left(\hat{r}_{(T_D\text{-}1,\alpha)}\right)}{\int_{RF(T_D\text{-}2)}^{\infty} f\left(\hat{r}_{(T_D\text{-}1,\alpha)}\right) d\left(\hat{r}_{(T_D\text{-}1,\alpha)}\right)} \right] \right\} \quad \text{(C.4e)}$$

→ V(T_D-2, RF(T_D-2)) =

> By definition, this integral is the expected value of $[1 - V(\cdot)]$ over the R.V. $\hat{r}_{(T_D\text{-}1,\alpha)}{}^+$.

$$\text{Min}_{(0 \le \alpha \le 1)} \left\{ 1 - (1 - F\hat{r}_{(T_D\text{-}1,\alpha)}(RF(T_D\text{-}2)))* \left[ \frac{\int_{RF(T_D\text{-}2)}^{\infty} f\left(\hat{r}_{(T_D\text{-}1,\alpha)}\right) [1 - V(T_D\text{-}1, \ RF(T_D\text{-}1))] \ d\left(\hat{r}_{(T_D\text{-}1,\alpha)}\right)}{\int_{RF(T_D\text{-}2)}^{\infty} f\left(\hat{r}_{(T_D\text{-}1,\alpha)}\right) d\left(\hat{r}_{(T_D\text{-}1,\alpha)}\right)} \right] \right\} \quad \text{(C.4f)}$$

Now, since RF(T_D-1) is a function of $\hat{r}_{(T_D\text{-}1,\alpha)}$, namely,

$$RF(T_D\text{-}1) \quad = \quad \frac{RF(T_D\text{-}2)}{\hat{r}_{(T_D\text{-}1,\alpha)} - RF(T_D\text{-}2)} \quad \text{(C.5)}$$

the value function in (C.4f) can be written as[4],

---

[4] An intuitive explanation of the need to take expectations is that V(T_D-1, X), was already found for all positive ruin factors X, treating X as constant. At the current induction step, it is discovered that X is random with known PDF under our control via α. In the optimization over α, $E_X[V(T_D\text{-}1, X)]$ is then evaluated across the various PDFs of X.

$$V(T_D\text{-}2,\ RF(T_D\text{-}2)) =$$

$$\text{Min}_{(0\,\le\,\alpha\,\le\,1)}\left\{\begin{array}{l}1\ -\ (1-F\hat{r}_{(TD\text{-}1,\alpha)}(RF(T_D\text{-}2)))*\\[2mm](1-E\hat{r}_{(TD\text{-}1,\alpha)}{}^{+}\left[V\!\left(T_D-1,\ \dfrac{RF(T_D-2)}{\hat{r}_{(T_D-1,\alpha)}-RF(T_D-2)}\right)\right])\end{array}\right\}\quad\text{(C.6)}$$

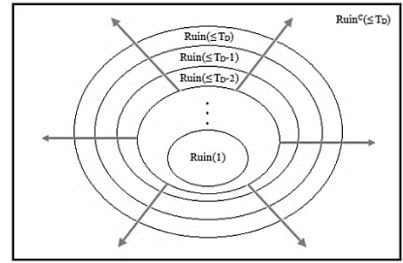

with optimality achieved at $\tilde{\alpha}=\alpha(T_D\text{-}2,\ RF(T_D\text{-}2))$ and the expectation over the conditional RV

$\hat{r}_{(TD\text{-}1,\alpha)}{}^{+}=(\hat{r}_{(TD\text{-}1,\alpha)}\mid\hat{r}_{(TD\text{-}1,\alpha)}>RF(T_D\text{-}2))$ where $\{\hat{r}_{(TD\text{-}1,\alpha)}>RF(T_D\text{-}2)\}\equiv\{RF(T_D\text{-}1)>0\}$.

### C.4 Induction at Time $t=T_D-3$

Assume the retiree arrives at time $t=T_D$-3, makes their 4th-last withdrawal and has 3 remaining.[5] The ruin factor $RF(T_D$-3$)$ $(>0)$ is calculated based on the portfolio return just observed, $\hat{r}_{(T_D\text{-}3,\alpha)}$. The retiree is facing the restricted sample space $S=\{$Ruin$(T_D$-2$)$, Ruin$(T_D$-1$)$, Ruin$(T_D)$, Ruin$^C(\le T_D)\}$ (shown at right) and seeks to make the optimal asset allocation decision to minimize P(Ruin$(T_D$-2$)$ $\cup$ Ruin$(T_D$-1$)$ $\cup$ Ruin$(T_D)$), which is the probability of ruin at any future time point. Under the restricted sample space, P(Ruin$(>T_D$-3$))$ is now defined as:

P(Ruin$(T_D$-2$)$ $\cup$ Ruin$(T_D$-1$)$ $\cup$ Ruin$(T_D)$)

$$=\ 1\ -\ P(\text{Ruin}^C(T_D\text{-}2)\ \cap\ \text{Ruin}^C(T_D\text{-}1)\ \cap\ \text{Ruin}^C(T_D))\qquad\text{(C.7a)}$$

$$=\ 1\ -\ P(\text{Ruin}^C(T_D\text{-}2))*\underbrace{P(\text{Ruin}^C(T_D\text{-}1)\ \cap\ \text{Ruin}^C(T_D)\mid\text{Ruin}^C(T_D\text{-}2))}\qquad\text{(C.7b)}$$

> Note that the optimal value for this probability was derived at time $t=T_D$-2 for all $RF(t)>0$.

The value function is thus:

$$V(T_D\text{-}3,\ RF(T_D\text{-}3))=\text{Min}_{(0\,\le\,\alpha\,\le\,1)}\left\{\begin{array}{l}1\ -\ P(\text{Ruin}^C(T_D\text{-}2))*\\[2mm]P(\text{Ruin}^C(T_D\text{-}1)\ \cap\ \text{Ruin}^C(T_D)\mid\text{Ruin}^C(T_D\text{-}2))\end{array}\right\}\quad\text{(C.8a)}$$

---

[5] Induction at time $t=T_D$-3 is nearly identical to induction at time $t=T_D$-2, and the process is generalized for time $t=T_D$-k next, then reported for any time t in Section II-G.1.

$\rightarrow V(T_D\text{-}3, RF(T_D\text{-}3)) =$

$$\text{Min}_{(0 \le \alpha \le 1)} \left\{ \begin{array}{l} 1 - P(\hat{r}_{(TD\text{-}2,\alpha)} > RF(T_D\text{-}3))_* \\[2mm] P(\hat{r}_{(T_D\text{-}1,\tilde{\alpha})} > RF(T_D\text{-}2) \cap \hat{r}_{(TD,\tilde{\alpha})} > RF(T_D\text{-}1) \mid \hat{r}_{(TD\text{-}2,\alpha)} > RF(T_D\text{-}3)) \end{array} \right\} \quad (C.8b)$$

We require an optimal policy be followed at each future stage, and these $\tilde{\alpha}$ reflect those optimal values. This $V(\cdot)$ cannot take a minimum value otherwise.

$\rightarrow V(T_D\text{-}3, RF(T_D\text{-}3)) =$

Applying the definition of a conditional probability.

$$\text{Min}_{(0 \le \alpha \le 1)} \left\{ \begin{array}{l} 1 - P(\hat{r}_{(TD\text{-}2,\alpha)} > RF(T_D\text{-}3))_* \\[3mm] \left[ \dfrac{P(\hat{r}_{(T_D\text{-}1,\tilde{\alpha})} > RF(T_D\text{-}2) \cap \hat{r}_{(T_D,\tilde{\alpha})} > RF(T_D\text{-}1) \cap \hat{r}_{(T_D\text{-}2,\alpha)} > RF(T_D\text{-}3))}{P(\hat{r}_{(T_D\text{-}2,\alpha)} > RF(T_D\text{-}3))} \right] \end{array} \right\} \quad (C.8c)$$

$\rightarrow V(T_D\text{-}3, RF(T_D\text{-}3)) = \qquad (C.8d)$

The multivariate density of the next 3 real returns integrated over the condition of no ruin.

$$\text{Min}_{(0 \le \alpha \le 1)} \left\{ \begin{array}{l} 1 - (1 - F\hat{r}_{(TD\text{-}2,\alpha)}(RF(T_D\text{-}3)))_* \\[3mm] \dfrac{\int_{RF(T_D\text{-}3)}^{\infty} \int_{RF(T_D\text{-}2)}^{\infty} \int_{RF(T_D\text{-}1)}^{\infty} f\left(\hat{r}_{(T_D,\tilde{\alpha})}, \hat{r}_{(T_D\text{-}1,\tilde{\alpha})}, \hat{r}_{(T_D\text{-}2,\alpha)}\right) d\left(\hat{r}_{(T_D,\tilde{\alpha})}\right) d\left(\hat{r}_{(T_D\text{-}1,\tilde{\alpha})}\right) d\left(\hat{r}_{(T_D\text{-}2,\alpha)}\right)}{\int_{RF(T_D\text{-}3)}^{\infty} f\left(\hat{r}_{(T_D\text{-}2,\alpha)}\right) d\left(\hat{r}_{(T_D\text{-}2,\alpha)}\right)} \end{array} \right\}$$

$\rightarrow V(T_D\text{-}3, RF(T_D\text{-}3)) = \qquad (C.8e)$

By conditioning and independence we split the joint PDF (see Appendix G.2).

$$\text{Min}_{(0 \le \alpha \le 1)} \left\{ \begin{array}{l} 1 - (1 - F\hat{r}_{(TD\text{-}2,\alpha)}(RF(T_D\text{-}3)))_* \\[3mm] \dfrac{\int_{RF(T_D\text{-}3)}^{\infty} f\left(\hat{r}_{(T_D\text{-}2,\alpha)}\right) \int_{RF(T_D\text{-}2)}^{\infty} \int_{RF(T_D\text{-}1)}^{\infty} f\left(\hat{r}_{(T_D,\tilde{\alpha})}, \hat{r}_{(T_D\text{-}1,\tilde{\alpha})}\right) d\left(\hat{r}_{(T_D,\tilde{\alpha})}\right) d\left(\hat{r}_{(T_D\text{-}1,\tilde{\alpha})}\right) d\left(\hat{r}_{(T_D\text{-}2,\alpha)}\right)}{\int_{RF(T_D\text{-}3)}^{\infty} f\left(\hat{r}_{(T_D\text{-}2,\alpha)}\right) d\left(\hat{r}_{(T_D\text{-}2,\alpha)}\right)} \end{array} \right\}$$

$\rightarrow V(T_D\text{-}3, RF(T_D\text{-}3)) = \qquad (C.8f)$

Note that the retiree has some control over the next ruin factor via choice of $\alpha$ at this time point.

$$\text{Min}_{(0 \le \alpha \le 1)} \left\{ \left[ \dfrac{\int_{RF(T_D\text{-}3)}^{\infty} f\left(\hat{r}_{(T_D\text{-}2,\alpha)}\right) [1 - V(T_D\text{-}2, \ RF(T_D\text{-}2))] \ d\left(\hat{r}_{(T_D\text{-}2,\alpha)}\right)}{\int_{RF(T_D\text{-}3)}^{\infty} f\left(\hat{r}_{(T_D\text{-}2,\alpha)}\right) d\left(\hat{r}_{(T_D\text{-}2,\alpha)}\right)} \right] \right\}$$

$1 - (1 - F\hat{r}_{(TD\text{-}2,\alpha)}(RF(T_D\text{-}3)))_*$

Since $RF(T_D\text{-}2)$ is a function of $\hat{r}_{(T_D\text{-}2,\alpha)}$, namely,

$$RF(T_D\text{-}2) \quad = \quad \frac{RF(T_D\text{-}3)}{\hat{r}_{(T_D\text{-}2,\alpha)}- RF(T_D\text{-}3)} \qquad (C.8g)$$

the expression above is by definition the expected value of $[1 - V(T_D\text{-}2, RF(T_D\text{-}2))]$ over the conditional RV $\hat{r}_{(T_D\text{-}2,\alpha)}{}^+$, and the value function can be written as:

$V(T_D\text{-}3, RF(T_D\text{-}3)) =$

$$Min_{(0 \le \alpha \le 1)}\left\{ \begin{array}{l} 1 - (1 - F\hat{r}_{(T_D\text{-}2,\alpha)}(RF(T_D\text{-}3)))* \\[2mm] (1 - E\hat{r}_{(T_D\text{-}2,\alpha)}{}^+\left[V(T_D - 2, \frac{RF(T_D\text{-}3)}{\hat{r}_{(T_D\text{-}2,\alpha)}- RF(T_D\text{-}3)})\right]) \end{array} \right\} \qquad (C.9)$$

(above: $RF(T_D\text{-}2)$ pointing to the argument)

Optimality is achieved at $\tilde{\alpha}=\alpha(T_D\text{-}3, RF(T_D\text{-}3))$ and the expectation is over the conditional RV $\hat{r}_{(T_D\text{-}2,\alpha)}{}^+ = (\hat{r}_{(T_D\text{-}2,\alpha)} \mid \hat{r}_{(T_D\text{-}2,\alpha)} > RF(T_D\text{-}3))$ where $\{\hat{r}_{(T_D\text{-}2,\alpha)} > RF(T_D\text{-}3)\} \equiv \{RF(T_D\text{-}2) > 0\}$.

### C.5 Induction at Time $t=T_D - k$

Assume the retiree arrives at time $t=T_D\text{-}k$, for $k=0,1, \ldots,$ $T_D\text{-}1$ and makes their $(k+1)$th-last withdrawal with $k$ remaining. The ruin factor $RF(T_D\text{-}k)$ $(> 0)$ is calculated based on the portfolio return just observed, $\hat{r}_{(T_D\text{-}k,\alpha)}$. The retiree is facing the

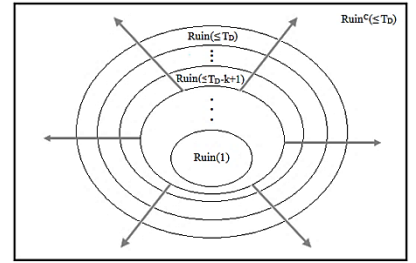

restricted sample space $S=\{Ruin(T_D\text{-}k+1), Ruin(T_D\text{-}k+2), \ldots, Ruin(T_D), Ruin^C(\le T_D)\}$ (shown at right) and seeks the optimal asset allocation to minimize $P(Ruin(>T_D\text{-}k)) = P(Ruin(T_D\text{-}k+1) \cup Ruin(T_D\text{-}k+2) \cup \ldots \cup Ruin(T_D))$, which is the probability of ruin at any future time point. Under the restricted sample space we express $P(Ruin(>T_D\text{-}k))$ as:

$P(Ruin(T_D\text{-}k+1) \cup Ruin(T_D\text{-}k+2) \cup \ldots \cup Ruin(T_D))$

$= 1 - P(Ruin^C(T_D\text{-}k+1) \cap Ruin^C(T_D\text{-}k+2 ) \cap \ldots \cap Ruin^C(T_D)) \qquad (C.10a)$

$= 1 - P(Ruin^C(T_D\text{-}k+1))*\underline{P(Ruin^C(T_D\text{-}k+2) \cap \ldots \cap Ruin^C(T_D) \mid Ruin^C(T_D\text{-}k+1))} \qquad (C.10b)$

The optimal value for this probability is derived at time $t=T_D\text{-}k+1$ for all $RF(t) > 0$.

The value function is given by:

$$V(T_D\text{-}k, RF(T_D\text{-}k)) =$$

$$\text{Min}_{(0 \leq \alpha \leq 1)} \left\{ \begin{array}{l} 1 - P(\text{Ruin}^C(T_D\text{-}k+1))* \\ \\ P(\text{Ruin}^C(T_D\text{-}k+2) \cap \ldots \cap \text{Ruin}^C(T_D) \mid \text{Ruin}^C(T_D\text{-}k+1)) \end{array} \right\} \quad \text{(C.10c)}$$

Let,

$$\hat{R}_{(t_s, t_e)} = \begin{pmatrix} \hat{r}_{(t_s, \bar{\alpha})} \\ \hat{r}_{(t_s+1, \bar{\alpha})} \\ \vdots \\ \hat{r}_{(t_e, \bar{\alpha})} \end{pmatrix}, \text{ and, } \mathbb{R}^c_{(t_s, t_e)} = \left\{ \mathbb{R}^n : \cap_{t=t_s}^{t_e} \left[ \hat{r}_{(t, \bar{\alpha})} > RF(t-1) \right] \right\}. \quad \text{(C.10d)}$$

At time $t=T_D\text{-}k$ the vector $\hat{R}_{(T_D\text{-}k+2, T_D)}$ will hold the random returns at all time points after $t=T_D\text{-}k+1$, assuming optimal asset allocations are used. The set $\mathbb{R}^c_{(T_D-k+2, T_D)}$ will represent the space in k-2 dimensions over which $\hat{R}_{(T_D\text{-}k+2, T_D)}$ satisfies the condition of $\text{Ruin}^C(>T_D\text{-}k+1)$.

$$\rightarrow V(T_D\text{-}k, RF(T_D\text{-}k)) =$$

$$\text{Min}_{(0 \leq \alpha \leq 1)} \left\{ \begin{array}{l} 1 - P(\hat{r}_{(T_D\text{-}k+1, \alpha)} > RF(T_D\text{-}k))* \\ \\ P(\hat{R}_{(T_D\text{-}k+2, T_D)} \in \mathbb{R}^c_{(T_D-k+2, T_D)} \mid \hat{r}_{(T_D\text{-}k+1, \alpha)} > RF(T_D\text{-}k)) \end{array} \right\} \quad \text{(C.10e)}$$

$$\rightarrow V(T_D\text{-}k, RF(T_D\text{-}k)) =$$

Applying the definition of a conditional probability.

$$\text{Min}_{(0 \leq \alpha \leq 1)} \left\{ \begin{array}{l} 1 - P(\hat{r}_{(T_D\text{-}k+1, \alpha)} > RF(T_D\text{-}k))* \\ \\ \left[ \dfrac{P(\hat{r}_{(T_D-k+1,\alpha)} > RF(T_D-k) \cap \hat{R}_{(T_D-k+2, T_D)} \in \mathbb{R}^c_{(T_D-k+2, T_D)})}{P(\hat{r}_{(T_D-k+1,\alpha)} > RF(T_D-k))} \right] \end{array} \right\} \quad \text{(C.10f)}$$

$$\rightarrow V(T_D\text{-}k, RF(T_D\text{-}k)) = \quad \text{(C.10g)}$$

The multivariate density of the next k real returns integrated over the condition of no ruin.

$$\text{Min}_{(0 \leq \alpha \leq 1)} \left\{ \begin{array}{l} 1 - (1 - F\hat{r}_{(TD\text{-}k+1,\alpha)}(RF(T_D\text{-}k)))* \\ \\ \dfrac{\int_{RF(T_D-k)}^{\infty} \int_{\mathbb{R}^c_{(T_D-k+2, T_D)}} f\left(\hat{r}_{(T_D-k+1,\alpha)}, \hat{R}_{(T_D-k+2,T_D)}\right) d\left(\hat{R}_{(T_D-k+2,T_D)}\right) d\left(\hat{r}_{(T_D-k+1,\alpha)}\right)}{\int_{RF(T_D-k)}^{\infty} f\left(\hat{r}_{(T_D-k+1,\alpha)}\right) d\left(\hat{r}_{(T_D-k+1,\alpha)}\right)} \end{array} \right\}$$

$\rightarrow \text{V}(T_D\text{-}k, \text{RF}(T_D\text{-}k)) =$  (C.10h)

$\text{Min}_{(0 \le \alpha \le 1)}$

By conditioning and independence
(see Appendix G.2).

$$\left\{ \begin{array}{c} 1 - (1 - F\hat{r}_{(TD\text{-}k+1,\alpha)}(\text{RF}(T_D\text{-}k)))* \\[2mm] \dfrac{\int_{\text{RF}(T_D\text{-}k)}^{\infty} f\left(\hat{r}_{(T_D\text{-}k+1,\alpha)}\right)\left[\int_{\mathbb{R}^c_{(T_D\text{-}k+2,T_D)}} f\left(\hat{R}_{(T_D\text{-}k+2,T_D)}\right) d\left(\hat{R}_{(T_D\text{-}k+2,T_D)}\right)\right] d\left(\hat{r}_{(T_D\text{-}k+1,\alpha)}\right)}{\int_{\text{RF}(T_D\text{-}k)}^{\infty} f\left(\hat{r}_{(T_D\text{-}k+1,\alpha)}\right) d\left(\hat{r}_{(T_D\text{-}k+1,\alpha)}\right)} \end{array} \right\}$$

$\rightarrow \text{V}(T_D\text{-}k, \text{RF}(T_D\text{-}k)) =$

By definition, the bracketed term
in (C.10h) is $P(\text{Ruin}^C(>T_D\text{-}k+1))$.  (C.10i)

$\text{Min}_{(0 \le \alpha \le 1)}$

$$\left\{ \begin{array}{c} 1 - (1 - F\hat{r}_{(TD\text{-}k+1,\alpha)}(\text{RF}(T_D\text{-}k)))* \\[2mm] \left[\dfrac{\int_{\text{RF}(T_D\text{-}k)}^{\infty} f\left(\hat{r}_{(T_D\text{-}k+1,\alpha)}\right) [1 - V(T_D\text{-}k+1, \ \text{RF}(T_D\text{-}k+1))] \ d\left(\hat{r}_{(T_D\text{-}k+1,\alpha)}\right)}{\int_{\text{RF}(T_D\text{-}k)}^{\infty} f\left(\hat{r}_{(T_D\text{-}k+1,\alpha)}\right) d\left(\hat{r}_{(T_D\text{-}k+1,\alpha)}\right)}\right] \end{array} \right\}$$

$\rightarrow \text{V}(T_D\text{-}k, \text{RF}(T_D\text{-}k)) =$

$\text{Min}_{(0 \le \alpha \le 1)}$

$$\left\{ \begin{array}{c} 1 - (1 - F\hat{r}_{(TD\text{-}k+1,\alpha)}(\text{RF}(T_D\text{-}k)))* \\[2mm] (1 - E\hat{r}_{(TD\text{-}k+1,\alpha)}^{+}\left[V(T_D - k + 1, \ \frac{\text{RF}(T_D\text{-}k)}{\hat{r}_{(T_D\text{-}k+1,\alpha)} - \text{RF}(T_D\text{-}k)})\right]) \end{array} \right\}$$

$\boxed{\text{RF}(T_D\text{-}k+1)}$  (C.10j)

Optimality is achieved at $\tilde{\alpha} = \alpha(T_D\text{-}k, \text{RF}(T_D\text{-}k))$ and the expectation is over the conditional RV $\hat{r}_{(TD\text{-}k+1,\alpha)}^{+} = (\hat{r}_{(TD\text{-}k+1,\alpha)} \,|\, \hat{r}_{(TD\text{-}k+1,\alpha)} > \text{RF}(T_D\text{-}k))$ where $\{\hat{r}_{(TD\text{-}k+1,\alpha)} > \text{RF}(T_D\text{-}k)\} \equiv \{\text{RF}(T_D\text{-}k+1) > 0\}$.

## Appendix D. Induction for Random $T_D$

### D.1 Induction at Time $t=S_{Max}$

Assume the retiree arrives at time $t=S_{Max}$ and makes their last withdrawal. $\text{RF}(S_{Max})$ $(> 0)$ need not be calculated since there are no more withdrawals and $\text{Ruin}^C(\le S_{Max})$ has occurred. At time $t=S_{Max}$, $P(\text{Ruin}(>S_{Max}))=0$ and a B.C. for the value function is $V_R(S_{Max}, \text{RF}(S_{Max}))=0$, $\forall$ $\text{RF}(S_{Max})>0$.

### D.2 Induction at Time $t=S_{Max}-1$

Assume the retiree arrives at time $t=S_{Max}$-1, makes their withdrawal and has at most 1 remaining. $\text{RF}(S_{Max}\text{-}1)$ $(> 0)$ is calculated based on the portfolio's return just observed, $\hat{r}_{(S_{Max}\text{-}1,\alpha)}$.

The retiree seeks $\alpha$ to minimize $P(\text{Ruin}(>S_{Max}\text{-}1)) = P(\text{Ruin}(S_{Max}))$, which we express as:

$$V_R(S_{Max}\text{-}1, RF(S_{Max}\text{-}1)) = \text{Min}_{(0 \le \alpha \le 1)}\left\{ P(\text{Ruin}(S_{Max})) \right\} \tag{D.1a}$$

$$\rightarrow V_R(S_{Max}\text{-}1, RF(S_{Max}\text{-}1)) = \text{Min}_{(0 \le \alpha \le 1)}\left\{ 1 - P(\text{Ruin}^C(S_{Max})) \right\} \tag{D.1b}$$

$$\boxed{\begin{array}{l}\text{Ruin}^C(S_{Max}) \equiv (\text{death before time } t=S_{Max} \text{ withdrawal attempt}) \\ \cup \ (\text{live until } t=S_{Max} \text{ and successfully make withdrawal})\end{array}}$$

$$\rightarrow V_R(S_{Max}\text{-}1, RF(S_{Max}\text{-}1)) =$$

$$\text{Min}_{(0 \le \alpha \le 1)}\left\{ \begin{array}{l} 1 - [P(T_D = S_{Max}\text{-}1 \mid T_D \ge S_{Max}\text{-}1)*(1) \\ + P(T_D > S_{Max}\text{-}1 \mid T_D \ge S_{Max}\text{-}1)*P(\hat{r}_{(S_{Max},\alpha)} > RF(S_{Max}\text{-}1))] \end{array} \right\} \tag{D.1c}$$

$$\boxed{\begin{array}{c}\text{This term} = \\ P(T_D = S_{Max} \mid T_D \ge S_{Max}\text{-}1)\end{array}}$$

$$\rightarrow V_R(S_{Max}\text{-}1, RF(S_{Max}\text{-}1)) =$$

$$\text{Min}_{(0 \le \alpha \le 1)}\left\{ P(T_D > S_{Max}\text{-}1 \mid T_D \ge S_{Max}\text{-}1)*[1 - P(\hat{r}_{(S_{Max},\alpha)} > RF(S_{Max}\text{-}1))] \right\} \tag{D.1d}$$

$$\rightarrow V_R(S_{Max}\text{-}1, RF(S_{Max}\text{-}1)) =$$

$$\text{Min}_{(0 \le \alpha \le 1)}\left\{ P(T_D > S_{Max}\text{-}1 \mid T_D \ge S_{Max}\text{-}1)*[1 - (1 - F_{\hat{r}(S_{Max},\alpha)}(RF(S_{Max}\text{-}1)))] \right\}, \tag{D.1e}$$

given the known ruin factor $RF(S_{Max}\text{-}1)$. We can alternatively express $V_R(S_{Max}\text{-}1, RF(S_{Max}\text{-}1))$ as:

$$V_R(S_{Max}\text{-}1, RF(S_{Max}\text{-}1)) =$$

$$\boxed{\begin{array}{c}RF(S_{Max}) = \\ RF(S_{Max}\text{-}1)/(\hat{r}_{(S_{Max},\alpha)} - RF(S_{Max}\text{-}1))\end{array}}$$

$$\text{Min}_{(0 \le \alpha \le 1)}\left\{ \begin{array}{l} P(T_D > S_{Max}\text{-}1 \mid T_D \ge S_{Max}\text{-}1)* \\ [1 - (1 - F_{\hat{r}(S_{Max},\alpha)}(RF(S_{Max}\text{-}1)))*(1 - E_{\hat{r}(S_{Max},\alpha)}{}^+[V_R(S_{Max}, RF(S_{Max}))])] \end{array} \right\} \tag{D.1f}$$

$$\boxed{\begin{array}{c}P(\text{Ruin}^C(S_{Max})) \text{ given} \\ T_D > S_{Max}\text{-}1.\end{array}} \quad \boxed{\begin{array}{c}\text{Expected prob. of no ruin after time } t=S_{Max}, \\ \text{given Ruin}^C(S_{Max}). \text{ (This expression} = 1.)\end{array}}$$

with optimal $\tilde{\alpha} = \alpha_R(S_{Max}\text{-}1, RF(S_{Max}\text{-}1))$.

### D.3 Induction at Time $t = S_{Max} - 2$

Assume the retiree arrives at time $t = S_{Max}\text{-}2$, makes their withdrawal and has at most 2 remaining. $RF(S_{Max}\text{-}2)$ ($> 0$) is calculated based on the portfolio return just observed, $\hat{r}_{(S_{Max}\text{-}2,\alpha)}$.

The retiree seeks $\alpha$ to minimize $P(\text{Ruin}(>S_{Max}\text{-}2)) = P(\text{Ruin}(S_{Max}\text{-}1) \cup \text{Ruin}(S_{Max}))$, which for a given $\alpha$ can be expressed as:

$P(\text{Ruin}(>S_{Max}\text{-}2))$

$$= P(\text{Ruin}(S_{Max}\text{-}1) \cup \text{Ruin}(S_{Max})) \tag{D.2a}$$

$$= 1 - P(\text{Ruin}^C(S_{Max}\text{-}1) \cap \text{Ruin}^C(S_{Max})) \tag{D.2b}$$

$$= 1 - \left\{ \begin{array}{l} P(T_D = S_{Max}\text{-}2 \mid T_D \geq S_{Max}\text{-}2) * (1) \\ + P(T_D > S_{Max}\text{-}2 \mid T_D \geq S_{Max}\text{-}2) * P(\hat{r}_{(SMax\text{-}1,\alpha)} > RF(S_{Max}\text{-}2) \cap \text{Ruin}^C(S_{Max})) \end{array} \right\} \tag{D.2c}$$

$$= P(T_D > S_{Max}\text{-}2 \mid T_D \geq S_{Max}\text{-}2) *$$

$$[1 - P(\hat{r}_{(SMax\text{-}1,\alpha)} > RF(S_{Max}\text{-}2)) * \underbrace{P(\text{Ruin}^C(S_{Max}) \mid \hat{r}_{(SMax\text{-}1,\alpha)} > RF(S_{Max}\text{-}2))}] \tag{D.2d}$$

$P(\text{Ruin}^C(S_{Max}) \mid \hat{r}_{(SMax\text{-}1,\alpha)} > RF(S_{Max}\text{-}2))$

$$= \left[ \frac{1}{P(\hat{r}_{(SMax\text{-}1,\alpha)} > RF(S_{Max}\text{-}2))} \right] * \underbrace{P(\hat{r}_{(SMax\text{-}1,\alpha)} > RF(S_{Max}\text{-}2) \cap \text{Ruin}^C(S_{Max}))} \tag{D.2e}$$

$P(\hat{r}_{(SMax\text{-}1,\alpha)} > RF(S_{Max}\text{-}2) \cap \text{Ruin}^C(S_{Max}))$

These 2 events are independent since $\hat{r}_{(SMax\text{-}1,\alpha)}$ and $T_D$ are independent R.V.s.

$$= P[\ (\hat{r}_{(SMax\text{-}1,\alpha)} > RF(S_{Max}\text{-}2) \cap (T_D = S_{Max}\text{-}1 \mid T_D > S_{Max}\text{-}2))$$

These 2 events are mutually exclusive.

$$\cup$$

$$(\hat{r}_{(SMax\text{-}1,\alpha)} > RF(S_{Max}\text{-}2) \cap (T_D = S_{Max} \mid T_D > S_{Max}\text{-}2) \cap \hat{r}_{(SMax,\tilde{\alpha})} > RF(S_{Max}\text{-}1))\ ] \tag{D.2f}$$

These 2 events are NOT independent since $RF(S_{Max}\text{-}1)$ is a function of $\hat{r}_{(SMax\text{-}1,\alpha)}$.

The event of making the withdrawal at $S_{Max}\text{-}1$ and also avoiding ruin at $S_{Max}$ can happen 2 ways: Make the withdrawal at $S_{Max}\text{-}1$ and experience death before $S_{Max}$, or make the withdrawal at $S_{Max}\text{-}1$, then again at $S_{Max}$. Note that all this unfolds given $T_D > S_{Max}\text{-}2 \equiv T_D \geq S_{Max}\text{-}1$, see (D.2c).

The intersection of events over $T_D$ and $(\hat{r}_{(SMax\text{-}1,\alpha)}, \hat{r}_{(SMax,\tilde{\alpha})})$ are independent since random market returns are independent of the retiree's random time of final withdrawal. Also, the union of events requiring $T_D = S_{Max}\text{-}1$ and $T_D = S_{Max}$ are mutually exclusive since both cannot occur simultaneously. Finally, the events $\hat{r}_{(S_{Max}-1,\alpha)} > RF(S_{Max}-2)$ and $\hat{r}_{(S_{Max},\tilde{\alpha})} > RF(S_{Max}-1)$ are not independent since $RF(S_{Max}-1) = RF(S_{Max}-2) / [\hat{r}_{(S_{Max}-1,\alpha)} - RF(S_{Max}-2)]$. We thus express (D.2f) as:

$P(\hat{r}_{(S_{Max}-1,\alpha)} > RF(S_{Max}-2) \cap Ruin^C(S_{Max}))$

$= P(\hat{r}_{(S_{Max}-1,\alpha)} > RF(S_{Max}-2)) * P(T_D = S_{Max}-1 \mid T_D \geq S_{Max}-1)$      (D.2g)

$\quad + P(\hat{r}_{(S_{Max}-1,\alpha)} > RF(S_{Max}-2) \cap \hat{r}_{(S_{Max},\tilde{\alpha})} > RF(S_{Max}-1)) * P(T_D = S_{Max} \mid T_D \geq S_{Max}-1)$

$= P(T_D = S_{Max}-1 \mid T_D \geq S_{Max}-1) *$      (D.2h)

$\quad\quad \int_{RF(S_{Max}-2)}^{\infty} f\big(\hat{r}_{(S_{Max}-1,\alpha)}\big)\, d\big(\hat{r}_{(S_{Max}-1,\alpha)}\big)$

$\quad + P(T_D = S_{Max} \mid T_D \geq S_{Max}-1) *$    | The joint PDF is split (see Appendix G.2). |

$\quad\quad \int_{RF(S_{Max}-2)}^{\infty} \int_{RF(S_{Max}-1)}^{\infty} f\big(\hat{r}_{(S_{Max}-1,\alpha)}, \hat{r}_{(S_{Max},\tilde{\alpha})}\big) d\big(\hat{r}_{(S_{Max},\tilde{\alpha})}\big) d\big(\hat{r}_{(S_{Max}-1,\alpha)}\big)$

$= \int_{RF(S_{Max}-2)}^{\infty} f\big(\hat{r}_{(S_{Max}-1,\alpha)}\big) * [\quad\quad\quad\quad\quad] d\big(\hat{r}_{(S_{Max}-1,\alpha)}\big)$      (D.2i)

| $P(T_D = S_{Max}-1 \mid T_D \geq S_{Max}-1) + P(T_D = S_{Max} \mid T_D \geq S_{Max}-1) * \int_{RF(S_{Max}-1)}^{\infty} f\big(\hat{r}_{(S_{Max},\tilde{\alpha})}\big)\, d\big(\hat{r}_{(S_{Max},\tilde{\alpha})}\big)$ |

| If the optimal $\alpha = \tilde{\alpha}$ is used at $t = S_{Max}-1$ then this term is precisely $Max_\alpha[P(Ruin^C(S_{Max}))]$ given the retiree arrives at $t = S_{Max}-1$ and successfully makes the withdrawal, which by definition is $1 - V_R(S_{Max}-1, RF(S_{Max}-1))$, see (D.1c). It will be shown below that $\alpha = \tilde{\alpha}$ must be used at $t = S_{Max}-1$ to minimize $V_R(S_{Max}-2, RF(S_{Max}-2))$ at $t = S_{Max}-2$. |

Substituting the term from (D.2i) back into (D.2e) reveals it is nothing more than the expectation of $[1 - V_R(S_{Max}-1, RF(S_{Max}-1))]$ over the conditional RV $\hat{r}_{(S_{Max}-1,\alpha)}^+$. If using $\alpha = \tilde{\alpha}$ at $t = S_{Max}-1$, we can express the original probability from (D.2a) for a given $\alpha$ at $t = S_{Max}-2$ with $RF(S_{Max}-2)$ as:

$P(Ruin(>S_{Max}-2))$

                                                    | $RF(S_{Max}-1)$ |

$= P(T_D > S_{Max}-2 \mid T_D \geq S_{Max}-2) *$

$\quad [1 - P(\hat{r}_{(S_{Max}-1,\alpha)} > RF(S_{Max}-2)) * (1 - E\hat{r}_{(S_{Max}-1,\alpha)}^+ [V_R(S_{Max}-1, \frac{RF(S_{Max}-2)}{\hat{r}_{(S_{Max}-1,\alpha)} - RF(S_{Max}-2)})])]$   (D.2j)

| For $P(Ruin(>S_{Max}-2))$ to be minimized over all $\alpha$ at $t = S_{Max}-2$, the function in $[\cdot]$ must take its minimum value at each $RF(S_{Max}-1)$, which occurs precisely at $V_R(S_{Max}-1, RF(S_{Max}-1))$. |

Finally, we express the value function at t=$S_{Max}$-2 as:

$V_R(S_{Max}\text{-}2, RF(S_{Max}\text{-}2)) =$

$$\text{Min}_{(0 \leq \alpha \leq 1)}\left\{ \begin{array}{l} P(T_D > S_{Max}\text{-}2 \mid T_D \geq S_{Max}\text{-}2)* \\ \\ \{1\text{-}(1\text{-}F\hat{r}_{(S_{Max}\text{-}1,\alpha)}(RF(S_{Max}\text{-}2)))*(1\text{-}E\hat{r}_{(S_{Max}\text{-}1,\alpha)}^{+}[V_R(S_{Max}\text{-}1, RF(S_{Max}\text{-}1))])\} \end{array} \right\}$$

(D.2k)

with optimality achieved at $\tilde{\alpha}=\alpha_R(S_{Max}\text{-}2, RF(S_{Max}\text{-}2))$.

### D.4 Induction at Time $t=S_{Max}-3$

Assume the retiree arrives at time t=$S_{Max}$-3, makes their withdrawal (at most 3 remain), and updates RF($S_{Max}$-3) (> 0) based on the portfolio return just observed, $\hat{r}_{(S_{Max}\text{-}3,\alpha)}$. The retiree seeks $\alpha$ to minimize P(Ruin(>$S_{Max}$-3)) = P(Ruin($S_{Max}$-2) ∪ P(Ruin($S_{Max}$-1) ∪ Ruin($S_{Max}$)), which for a given $\alpha$ can be expressed as:

P(Ruin(>$S_{Max}$-3))

= P(Ruin($S_{Max}$-2) ∪ Ruin($S_{Max}$-1) ∪ Ruin($S_{Max}$))  (D.3a)

= 1 - P(Ruin$^C$($S_{Max}$-2) ∩ Ruin$^C$($S_{Max}$-1) ∩ Ruin$^C$($S_{Max}$))  (D.3b)

$$= 1 - \left\{ \begin{array}{l} P(T_D = S_{Max}\text{-}3 \mid T_D \geq S_{Max}\text{-}3)*(1) \\ \\ \quad + P(T_D > S_{Max}\text{-}3 \mid T_D \geq S_{Max}\text{-}3)* \\ \\ \quad P(\hat{r}_{(S_{Max}\text{-}2,\alpha)} > RF(S_{Max}\text{-}3) \cap Ruin^C(S_{Max}\text{-}1) \cap Ruin^C(S_{Max})) \end{array} \right.$$

(D.3c)

= P($T_D$ > $S_{Max}$-3 | $T_D$ ≥ $S_{Max}$-3)*  (D.3d)

[ 1 - P($\hat{r}_{(S_{Max}\text{-}2,\alpha)}$ > RF($S_{Max}$-3))*P(Ruin$^C$($S_{Max}$-1) ∩ Ruin$^C$($S_{Max}$) | $\hat{r}_{(S_{Max}\text{-}2,\alpha)}$ > RF($S_{Max}$-3)) ]

P(Ruin$^C$($S_{Max}$-1) ∩ Ruin$^C$($S_{Max}$) | $\hat{r}_{(S_{Max}\text{-}2,\alpha)}$ > RF($S_{Max}$-3))

$$= \left[ \frac{1}{P(\hat{r}_{(S_{Max}-2,\alpha)} > RF(S_{Max}\text{-}3))} \right]*P(\hat{r}_{(S_{Max}\text{-}2,\alpha)} > RF(S_{Max}\text{-}3) \cap Ruin^C(S_{Max}\text{-}1) \cap Ruin^C(S_{Max}))$$  (D.3e)

$P(\hat{r}_{(S_{Max}-2,\alpha)} > RF(S_{Max}-3) \cap Ruin^C(S_{Max}-1) \cap Ruin^C(S_{Max}))$

$= P[((\hat{r}_{(S_{Max}-2,\alpha)} > RF(S_{Max}-3) \cap (T_D = S_{Max}-2 \mid T_D \geq S_{Max}-2))$

  $\cup$

  $(\hat{r}_{(S_{Max}-2,\alpha)} > RF(S_{Max}-3) \cap (T_D = S_{Max}-1 \mid T_D \geq S_{Max}-2) \cap \hat{r}_{(S_{Max}-1,\tilde{\alpha})} > RF(S_{Max}-2))$   (D.3f)

  $\cup$

  $(\hat{r}_{(S_{Max}-2,\alpha)} > RF(S_{Max}-3) \cap (T_D = S_{Max} \mid T_D \geq S_{Max}-2) \cap \hat{r}_{(S_{Max}-1,\tilde{\alpha})} > RF(S_{Max}-2) \cap \hat{r}_{(S_{Max},\tilde{\alpha})} > RF(S_{Max}-1))$

> Given $T_D > S_{Max}-3$, there are 3 mutually exclusive ways this intersection of 3 events can occur:

Invoking the independence of $T_D$ and $\hat{r}(\cdot)$, (D.3f) can be written as:

$= P(T_D = S_{Max}-2 \mid T_D \geq S_{Max}-2) * P(\hat{r}_{(S_{Max}-2,\alpha)} > RF(S_{Max}-3)$

 $+ P(T_D = S_{Max}-1 \mid T_D \geq S_{Max}-2) * P(\hat{r}_{(S_{Max}-2,\alpha)} > RF(S_{Max}-3) \cap \hat{r}_{(S_{Max}-1,\tilde{\alpha})} > RF(S_{Max}-2))$   (D.3g)

 $+ P(T_D = S_{Max} \mid T_D \geq S_{Max}-2) * P(\hat{r}_{(S_{Max}-2,\alpha)} > RF(S_{Max}-3) \cap \hat{r}_{(S_{Max}-1,\tilde{\alpha})} > RF(S_{Max}-2) \cap \hat{r}_{(S_{Max},\tilde{\alpha})} > RF(S_{Max}-1))$

$= P(T_D = S_{Max}-2 \mid T_D \geq S_{Max}-2) *$

  $\int_{RF(S_{Max}-3)}^{\infty} f(\hat{r}_{(S_{Max}-2,\alpha)}) \, d(\hat{r}_{(S_{Max}-2,\alpha)})$

 $+ P(T_D = S_{Max}-1 \mid T_D \geq S_{Max}-2) *$

> We split the joint PDF (see Appendix G.2).

  $\int_{RF(S_{Max}-3)}^{\infty} \int_{RF(S_{Max}-2)}^{\infty} f(\hat{r}_{(S_{Max}-2,\alpha)}, \hat{r}_{(S_{Max}-1,\tilde{\alpha})}) \, d(\hat{r}_{(S_{Max}-1,\tilde{\alpha})}) \, d(\hat{r}_{(S_{Max}-2,\alpha)})$   (D.3h)

 $+ P(T_D = S_{Max} \mid T_D \geq S_{Max}-2) *$

> We split the joint PDF (see Appendix G.2).

$\int_{RF(S_{Max}-3)}^{\infty} \int_{RF(S_{Max}-2)}^{\infty} \int_{RF(S_{Max}-1)}^{\infty} f(\hat{r}_{(S_{Max},\tilde{\alpha})}, \hat{r}_{(S_{Max}-1,\tilde{\alpha})}, \hat{r}_{(S_{Max}-2,\alpha)}) d(\hat{r}_{(S_{Max},\tilde{\alpha})}) d(\hat{r}_{(S_{Max}-1,\tilde{\alpha})}) d(\hat{r}_{(S_{Max}-2,\alpha)})$

Since the sum of integrals with respect to $\hat{r}_{(S_{Max}-2,\alpha)}$ is the integral of the sum, (D.3h) becomes:

  $= \int_{RF(S_{Max}-3)}^{\infty} f(\hat{r}_{(S_{Max}-2,\alpha)}) * [\qquad\qquad] \, d(\hat{r}_{(S_{Max}-2,\alpha)})$   (D.3i)

> $P(T_D = S_{Max}-2 \mid T_D \geq S_{Max}-2)$
>
> $+ P(T_D = S_{Max}-1 \mid T_D \geq S_{Max}-2) * \int_{RF(S_{Max}-2)}^{\infty} f(\hat{r}_{(S_{Max}-1,\tilde{\alpha})}) \, d(\hat{r}_{(S_{Max}-1,\tilde{\alpha})})$
>
> $+ P(T_D = S_{Max} \mid T_D \geq S_{Max}-2) *$
>
>   $\int_{RF(S_{Max}-2)}^{\infty} \int_{RF(S_{Max}-1)}^{\infty} f(\hat{r}_{(S_{Max},\tilde{\alpha})}, \hat{r}_{(S_{Max}-1,\tilde{\alpha})}) \, d(\hat{r}_{(S_{Max},\tilde{\alpha})}) \, d(\hat{r}_{(S_{Max}-1,\tilde{\alpha})})$

This entire term is recognized as $Max_\alpha[1 - P(Ruin(>S_{Max}-2))]$, (see D.2c) given that the retiree arrives at $t=S_{Max}-2$ and successfully makes the withdrawal, which is by definition $1-V_R(S_{Max}-2, RF(S_{Max}-2))$. Following an optimal policy at future time points is required to achieve a minimum for $V_R(S_{Max}-3, RF(S_{Max}-3))$ at $t=S_{Max}-3$.

Substituting (D.3i) back into the RHS of (D.3e) reveals it is the expected value of $1-V_R(S_{Max}-2, RF(S_{Max}-2))$ with respect to the RV $\hat{r}_{(S_{Max}-2,\alpha)}{}^+$. This leaves (D.3d), assuming we use optimal decumulation policies at all future time points, as:

$P(Ruin(>S_{Max}-3))$

$= P(T_D > S_{Max}-3 \mid T_D \geq S_{Max}-3) *$

$\boxed{RF(S_{Max}-2)}$

$[1- P(\hat{r}_{(S_{Max}-2,\alpha)} > RF(S_{Max}-3))*(1- E\hat{r}_{(S_{Max}-2,\alpha)}{}^+\left[V_R(S_{Max}-2, \frac{RF(S_{Max}-3)}{\hat{r}_{(S_{Max}-2,\alpha)}-RF(S_{Max}-3)})\right])]$  (D.3j)

The value function $V_R(S_{Max}-3, RF(S_{Max}-3))$ is the minimum of (D.3j) over $0 \leq \alpha \leq 1$, namely:

$V_R(S_{Max}-3, RF(S_{Max}-3)) =$  (D.3k)

$Min_{(0 \leq \alpha \leq 1)} \left\{ \begin{array}{l} P(T_D > S_{Max}-3 \mid T_D \geq S_{Max}-3) * \\ \{1 - (1 - F\hat{r}_{(S_{Max}-2,\alpha)}(RF(S_{Max}-3)))*(1 - E\hat{r}_{(S_{Max}-2,\alpha)}{}^+[V_R(S_{Max}-2, RF(S_{Max}-2))])\} \end{array} \right\}$

with optimality achieved at $\tilde{\alpha}=\alpha_R(S_{Max}-3, RF(S_{Max}-3))$.

### D.5  Induction at Time $t=S_{Max}-k$

Assume the retiree arrives at time $t=S_{Max}-k$, makes their withdrawal (at most k remain), and updates $RF(S_{Max}-k)$ (> 0) based on the portfolio return just observed, $\hat{r}_{(S_{Max}-k,\alpha)}$. The retiree seeks $\alpha$ to minimize $P(Ruin(>S_{Max}-k))=P(Ruin(S_{Max}-k+1) \cup P(Ruin(S_{Max}-k+2) \cup ... \cup Ruin(S_{Max}))$, which for a given $\alpha=\alpha_R(S_{Max}-k, RF(S_{Max}-k))$ can be expressed as:

$P(Ruin(>S_{Max}-k))$

$= P( \cup_{t=S_{Max}-k+1}^{S_{Max}} Ruin(t) )$  (D.4a)

$= 1 - P( \cap_{t=S_{Max}-k+1}^{S_{Max}} Ruin^C(t) )$  (D.4b)

$$= 1 - \left\{ \begin{array}{l} P(T_D = S_{Max}\text{-}k \mid T_D \geq S_{Max}\text{-}k) * (1) \\[2mm] + \; P(T_D > S_{Max}\text{-}k \mid T_D \geq S_{Max}\text{-}k) * \\[2mm] \quad P(\hat{r}_{(S_{Max}\text{-}k+1,\alpha)} > RF(S_{Max}\text{-}k) \cap [\, \cap_{t=S_{Max}\text{-}k+2}^{S_{Max}} Ruin^C(t) ] \,) \end{array} \right\} \qquad (D.4c)$$

$$= P(T_D > S_{Max}\text{-}k \mid T_D \geq S_{Max}\text{-}k) * \{1 - P(\hat{r}_{(S_{Max}\text{-}k+1,\alpha)} > RF(S_{Max}\text{-}k)) * \qquad (D.4d)$$

$$\underbrace{P(\cap_{t=S_{Max}\text{-}k+2}^{S_{Max}} Ruin^C(t) \mid \hat{r}_{(S_{Max}\text{-}k+1,\alpha)} > RF(S_{Max}\text{-}k)) \}}$$

$$\left[ \frac{1}{P(\hat{r}_{(S_{Max}\text{-}k+1,\alpha)} > RF(S_{Max}\text{-}k))} \right] * P\left( \underbrace{\left[ \cap_{t=S_{Max}\text{-}k+2}^{S_{Max}} Ruin^C(t) \right] \cap \hat{r}_{(S_{Max}\text{-}k+1,\alpha)} > RF(S_{Max}\text{-}k))} \right) \qquad (D.4e)$$

$$P\{ (\hat{r}_{(S_{Max}-k+1,\alpha)} > RF(S_{Max} - k)) \cap [\, (T_D = S_{Max} - k + 1 \mid T_D \geq S_{Max} - k + 1) \cup \qquad (D.4f)$$

$$\cup_{t_1=S_{Max}-k+2}^{S_{Max}} \Big\{ (T_D = t_1 \mid T_D \geq S_{Max} - k + 1) \cap \underbrace{[\cap_{t_2=S_{Max}-k+2}^{t_1}(\hat{r}_{(t_2,\tilde{\alpha})} > RF(t_2 - 1))]}\Big\} ] \}$$

$$\boxed{\text{This is: } \hat{R}_{(S_{Max}-k+2,\,t_1)} \in \mathbb{R}^c_{(S_{Max}-k+2,\,t_1)}.}$$

In (D.4e) we are operating under the assumption that $(T_D > S_{Max}\text{-}k \mid T_D \geq S_{Max}\text{-}k)$, see (D.4c). Having made it to time $t=S_{Max}\text{-}k+1$ we avoid ruin at time $t=S_{Max}\text{-}k+1$ *iff* $[\hat{r}_{(S_{Max}\text{-}k+1,\alpha)} > RF(S_{Max}\text{-}k)]$, and we avoid ruin at all times $t > S_{Max}\text{-}k+1$ *iff* $[\cap_{t=S_{Max}\text{-}k+2}^{S_{Max}} Ruin^C(t)]$, which is equivalent to:

$$(T_D = S_{Max}\text{-}k+1 \cup [(T_D > S_{Max}\text{-}k+1 \mid T_D \geq S_{Max}\text{-}k+1) \cap (Ruin^C(>S_{Max}\text{-}k+1))]). \qquad (D.4g)$$

This explains the decomposition of (D.4e) to (D.4f), where (D.4f) also expands the bracketed [·] event to all possible remaining values of $T_D$. These separate $T_D$ events form the basis of the union operator "$\cup$" and are mutually exclusive thus can be summed. The events over the intersection operator "$\cap$" cannot be multiplied because they are not independent (e.g., see Figure A1). For any events A, B, and C, the distributive property of the set operator "$\cap$" is such that $A \cap (B \cup C) \equiv (A \cap B) \cup (A \cap C)$. Using this property we distribute the event $[\hat{r}_{(S_{Max}\text{-}k+1,\alpha)} > RF(S_{Max}\text{-}k)]$ over all events separated by "$\cup$" in (D.4f) allowing it to be written as:

$P\{ \; [ \; (T_D = S_{Max} - k + 1 \mid T_D \geq S_{Max} - k + 1) \cap (\hat{r}_{(S_{Max}-k+1,\alpha)} > RF(S_{Max} - k)) \cup$  (D.4h)

$\cup_{t_1 = S_{Max}-k+2}^{S_{Max}} \left\{ (T_D = t_1 \mid T_D \geq S_{Max} - k + 1) \cap [\cap_{t_2 = S_{Max}-k+1}^{t_1} (\hat{r}_{(t_2,\tilde{\alpha})} > RF(t_2 - 1))] \right\} ] \}$

We have seen this quantity in previous induction steps. Borrowing the notation from (C.10d) that introduced $\hat{R}_{(t_s, t_e)}$ and $\mathbb{R}^c_{(t_s, t_e)}$, and using the results from Appendix G.2 to move the integration over RV $\hat{r}_{(S_{Max}-k+1,\alpha)}$ outside of the sum, we decompose (D.4h) as:

$\int_{RF(S_{Max}-k)}^{\infty} f(\hat{r}_{(S_{Max}-k+1,\alpha)}) * [ \; P[T_D = S_{Max} - k + 1 \mid T_D \geq S_{Max} - k + 1) +$  (D.4i)

$\sum_{t_1 = S_{Max}-k+2}^{S_{Max}} \left\{ P(T_D = t_1 \mid T_D \geq S_{Max} - k + 1) * \int_{\mathbb{R}^c_{(S_{Max}-k+2,t_1)}} f(\hat{R}_{(S_{Max}-k+2,t_1)}) d(\hat{R}_{(S_{Max}-k+2,t_1)}) \right\}$

$] \; d(\hat{r}_{(S_{Max}-k+1,\alpha)})$

The quantity in (D.4i) is divided by $P(\hat{r}_{(S_{Max}-k+1,\alpha)} > RF(S_{Max}-k))$ in (D.4e) leaving it as the expected value of $[\cdot]$ over the RV $\hat{r}_{(S_{Max}-k+1,\alpha)}^+$. Assuming an optimal policy is followed at all $t > S_{Max}-k+1$ we recognize $[\cdot]$ as the maximum $P(Ruin^C(>S_{Max}-k+1))$ given the retiree arrives at time $t=S_{Max}-k+1$ and makes their withdrawal, which is precisely $1 - V_R[S_{Max}-k+1, RF(S_{Max}-k+1)]$. Since our objective at $t=S_{Max}-k$ is to minimize $P(Ruin(>S_{Max}-k))$ we must use the maximum value of $[\cdot] \; \forall \; RF(S_{Max}-k)$. (This follows since for any 2 functions $h(x)>0$ and $g(x)>0$, $h(x) \geq g(x)$ $\forall \; x \rightarrow E_x[h(x)] \geq E_x[g(x)]$. The proof can proceed by contradiction and is trivial since density functions cannot take negative values.) Finally, we minimize (D.4d) over $(0 \leq \alpha \leq 1)$ expressing the value function as:

$V_R(S_{Max}-k, RF(S_{Max}-k)) =$  (D.4j)

$Min_{(0 \leq \alpha \leq 1)} \left\{ \begin{array}{l} P(T_D > S_{Max}-k \mid T_D \geq S_{Max}-k) * \{1 - (1 - F\hat{r}_{(S_{Max}-k+1,\alpha)}(RF(S_{Max}-k))) * \\ \qquad\qquad (1 - E\hat{r}_{(S_{Max}-k+1,\alpha)}^+[V_R(S_{Max}-k+1, RF(S_{Max}-k+1))]) \} \end{array} \right\}$

with optimality achieved at $\tilde{\alpha} = \alpha_R(S_{Max}-k, RF(S_{Max}-k))$.

## Appendix E.  Derivation of Ruin Factor Bucket Probabilities

Assume we are at time t and $RF(t) > 0$ is known.

$P(RF(t+1)$ in Bucket #1)

$$= \quad P(0 < RF(t+1) \leq (1.5)*(1/P_R)) \tag{E.1a}$$

$$= \quad P(0 < \frac{RF(t)}{\hat{r}_{(t+1,\alpha)} - RF(t)} \leq (1.5)*(1/P_R)) \tag{E.1b}$$

$$= \quad P(\frac{1}{0} > \frac{\hat{r}_{(t+1,\alpha)} - RF(t)}{RF(t)} \geq \frac{1}{(1.5)*\left(\frac{1}{P_R}\right)}) \tag{E.1c}$$

$$= \quad P(\infty > \hat{r}_{(t+1,\alpha)} \geq RF(t)*(1 + \frac{1}{(1.5)*\left(\frac{1}{P_R}\right)})) \tag{E.1d}$$

$$= \quad P(RF(t)*(1 + \frac{1}{(1.5)*\left(\frac{1}{P_R}\right)}) \leq \hat{r}_{(t+1,\alpha)} < \infty) \tag{E.1e}$$

$$= \quad \boxed{1 - F\hat{r}_{(t+1,\alpha)}(RF(t)*(1 + \frac{1}{(1.5)*\left(\frac{1}{P_R}\right)}))} \tag{E.1f}$$

$P(RF(t+1)$ in Bucket #i), for $(2 \leq i \leq (P_R)*RF_{Max})$

$$= \quad P(i/P_R - 1/(2P_R) < RF(t+1) \leq i/P_R + 1/(2P_R)) \tag{E.2a}$$

$$= \quad P((1/P_R)*(i - 1/2) < \frac{RF(t)}{\hat{r}_{(t+1,\alpha)} - RF(t)} \leq (1/P_R)*(i + 1/2)) \tag{E.2b}$$

$$= \quad P(RF(t)*(1 + \frac{1}{\left(\frac{1}{P_R}\right)*(i-\frac{1}{2})}) > \hat{r}_{(t+1,\alpha)} \tag{E.2c}$$
$$\geq RF(t)*(1 + \frac{1}{\left(\frac{1}{P_R}\right)*(i+\frac{1}{2})}))$$

$$= \quad P(RF(t)*(1 + \frac{1}{\left(\frac{1}{P_R}\right)*(i+\frac{1}{2})}) \leq \hat{r}_{(t+1,\alpha)} \tag{E.2d}$$
$$< RF(t)*(1 + \frac{1}{\left(\frac{1}{P_R}\right)*(i-\frac{1}{2})}))$$

$$= \quad \boxed{\begin{array}{l} F\hat{r}_{(t+1,\alpha)}(RF(t)*(1 + \frac{1}{\left(\frac{1}{P_R}\right)*(i-\frac{1}{2})})) \ - \\ \qquad F\hat{r}_{(t+1,\alpha)}(RF(t)*(1 + \frac{1}{\left(\frac{1}{P_R}\right)*(i+\frac{1}{2})})) \end{array}} \tag{E.2e}$$

P(RF(t+1) in Bucket #$(P_R)*RF_{Max}+1$)

$$= \quad P(RF_{Max} + 1/(2P_R) < RF(t+1) < \infty) \qquad \text{(E.3a)}$$

$$= \quad P(RF_{Max} + 1/(2P_R) < \frac{RF(t)}{\hat{r}_{(t+1,\alpha)} - RF(t)} < \infty) \qquad \text{(E.3b)}$$

$$= \quad P(\frac{1}{RF_{Max} + 1/(2P_R)} > \frac{\hat{r}_{(t+1,\alpha)} - RF(t)}{RF(t)} > \frac{1}{\infty}) \qquad \text{(E.3c)}$$

$$= \quad P(RF(t) * (1 + \frac{1}{RF_{Max} + 1/(2P_R)}) > \hat{r}_{(t+1,\alpha)} > RF(t)) \qquad \text{(E.3d)}$$

$$= \quad P(RF(t) < \hat{r}_{(t+1,\alpha)} < RF(t) * (1 + \frac{1}{RF_{Max} + 1/(2P_R)})) \qquad \text{(E.3e)}$$

$$= \quad \boxed{F\hat{r}_{(t+1,\alpha)}(RF(t) * (1 + \frac{1}{RF_{Max} + 1/(2P_R)})) - F\hat{r}_{(t+1,\alpha)}(RF(t))} \qquad \text{(E.3f)}$$

Lastly, in (14) these probabilities are conditional on RF(t+1) > 0 $\leftrightarrow \hat{r}_{(t+1,\alpha)} > RF(t)$, therefore each must be divided by $P(\hat{r}_{(t+1,\alpha)} > RF(t)) = 1 - F\hat{r}_{(t+1,\alpha)}(RF(t))$ for their sum to equal 1 and constitute a valid PMF. What we have derived is the conditional probability that the next ruin factor is in any of the pre-defined buckets, given that no ruin occurs at the next time point. To summarize the above, at time t the probability that RF(t+1) falls in bucket i, i = 1, 2, …, $(P_R)*RF_{Max}$, $(P_R)*RF_{Max}+1$ given that RF(t+1) > 0 is defined by the PMF:

P(RF(t+1) in Bucket i | $\hat{r}_{(t+1,\alpha)} > RF(t)$) = $\qquad\qquad$ (E.4)

$$
\begin{cases}
\dfrac{1 - F\hat{r}(t+1,\alpha)(RF(t)*(1+\frac{1}{(1.5)*\left(\frac{1}{P_R}\right)}))}{1 - F\hat{r}(t+1,\alpha)(RF(t))}, & i = 1 \\[3em]
\dfrac{F\hat{r}(t+1,\alpha)(RF(t)*(1+\frac{1}{\left(\frac{1}{P_R}\right)*(i-\frac{1}{2})})) - F\hat{r}(t+1,\alpha)(RF(t)*(1+\frac{1}{\left(\frac{1}{P_R}\right)*(i+\frac{1}{2})}))}{1 - F\hat{r}(t+1,\alpha)(RF(t))}, & i = 2, 3, …, (P_R)*RF_{Max} \\[3em]
\dfrac{F\hat{r}(t+1,\alpha)(RF(t)*(1+\frac{1}{RF_{Max} + 1/(2P_R)})) - F\hat{r}(t+1,\alpha)(RF(t))}{1 - F\hat{r}(t+1,\alpha)(RF(t))}, & i = (P_R)*RF_{Max} + 1
\end{cases}
$$

## Appendix F.  Derivation of Historical Stock and Bond Distributions

The hypothesis that real total stock (S&P 500) and bond (10 year T-Bond) returns originate from normal distributions cannot be rejected (Anderson-Darling p-values of 0.707 and 0.243, respectively).  Other distributions also provide an acceptable fit, such as the lognormal, however only using a non-zero location parameter in the case of real stock returns.  We thus assume that the underlying distributions for real total returns at year t are given by, $N(\mu, \sigma)$:

$$\text{S\&P 500 Returns} = r_{(s,t)} \sim N(0.0825, 0.2007) \qquad \text{(F.1a)}$$

$$\text{10 year T-Bond Returns} = r_{(b,t)} \sim N(0.0214, 0.0834) \qquad \text{(F.1b)}$$

In addition, a small positive correlation of $\rho = 0.04387$ was measured between real stock and bond returns at the same time point, which will be carried through the analysis.  We note that these are unconditional distributions and would roughly trace out the shape of corresponding histograms plotted using these returns.  If the returns from either investment exhibit serial correlation they cannot be treated as random samples from their unconditional distributions.  It is well known, for example, that the inflation rate exhibits strong serial correlation.

To determine the nature of serial correlation within real stock and bond returns, we assume that the average return for each is stable, or that the processes are stationary.  We estimate the overall mean for each process by the sample average, which is subtracted from the returns leaving them as deviations from their respective means.  A stationary process that is autoregressive of order p, denoted AR(p), is one that explicitly models the current observation as a linear function of the past p observations, under the assumption that they are correlated.  The general AR(p) model is defined as (Box et al. (1994)),

$$Y_t = \phi_1 Y_{t-1} + \phi_2 Y_{t-2} + \ldots + \phi_p Y_{t-p} + a_t \qquad \text{(F.2)}$$

where $Y_t$ reflects the centered observation at time t and $a_t \sim N(0, \sigma_a^2)$ are *iid* error terms.

Note that we assume the Y's are correlated despite the fact that the error terms are *iid*. This is because, for instance, $Y_t$ and $Y_{t+1}$ are both functions of the same random error term $a_t$. To determine the autoregressive order of a stationary process we examine the autocorrelations and partial autocorrelations. The autocorrelations at lag k are estimated by the sample correlations of all observations k time points apart and the partial autocorrelations at lag k reflect the correlation of observations k time points apart, after accounting for the correlation at lags < k. If an AR(k) model is fit to the data, the estimate of $\phi_k$ is also an estimate of the k-th partial autocorrelation. It is appropriate to examine both of these statistics up to lag N/4, where N reflects the number of data points collected (Box et al. (1994), here N=86). Two noteworthy findings are that both the autocorrelations and partial autocorrelations at lags > p are approximately $N(0, \sqrt{1/N})$ in an AR(p) process for large N (Box et al. (1994)). Since approximately 95% of the data in a normal distribution lies within 2 standard deviations of the mean, this result can be used to assess the lag at which serial correlation ends in a stationary process (Box et al. (1994)).

The autocorrelation estimates for inflation-adjusted stock returns are shown below on the left side of Figure F1. None of the 21 autocorrelations exceed the $2\sqrt{1/N}$ (= 0.21567) threshold, represented by the dashed line. This indicates that an AR(0) model is appropriate. The partial autocorrelations were estimated by fitting successive AR(k) models, for k=1, 2, …, 21, where the $\phi$-estimates for each model were taken as solutions to the Yule-Walker equations (Box et al. (1994)). The partial autocorrelation estimates are shown on the right side of Figure F1 and confirm the finding that real stock returns do not exhibit serial correlation. Since the autocorrelation estimates shown (not partial) and their large sample distribution apply to any linear stationary process, including a moving average MA(q), or mixed autoregressive-moving average ARMA(p, q) process, we conclude that these returns originate as completely random samples. Note that autocorrelations tail off after p in an AR(p) process, but drop off abruptly after q in an MA(q) process (Box et al. (1994)).

**Figure F1.  Analysis of Serial Correlation in Real Stock Returns**

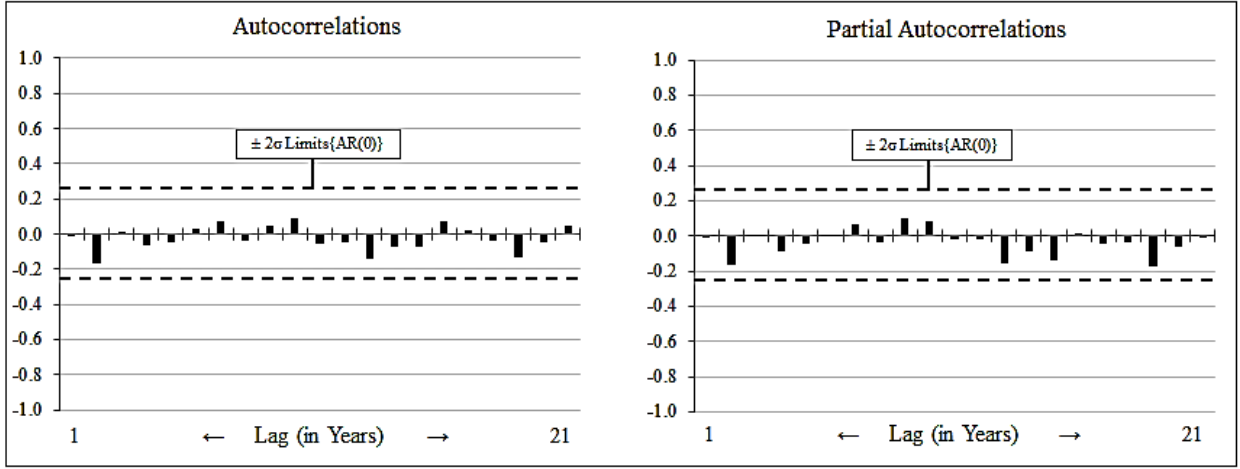

We perform a similar analysis on real bond returns and display the results in Figure F2 below. None of the 21 autocorrelations or partial autocorrelations exceed the $2\sqrt{1/N}$ (= 0.21567) tolerance limits indicating that an AR(0) process is appropriate for these returns. Again, since the autocorrelation estimates and their large sample distribution apply to any linear stationary process, we conclude that these returns also originate as completely random samples.

**Figure F2.  Analysis of Serial Correlation in Real Bond Returns**

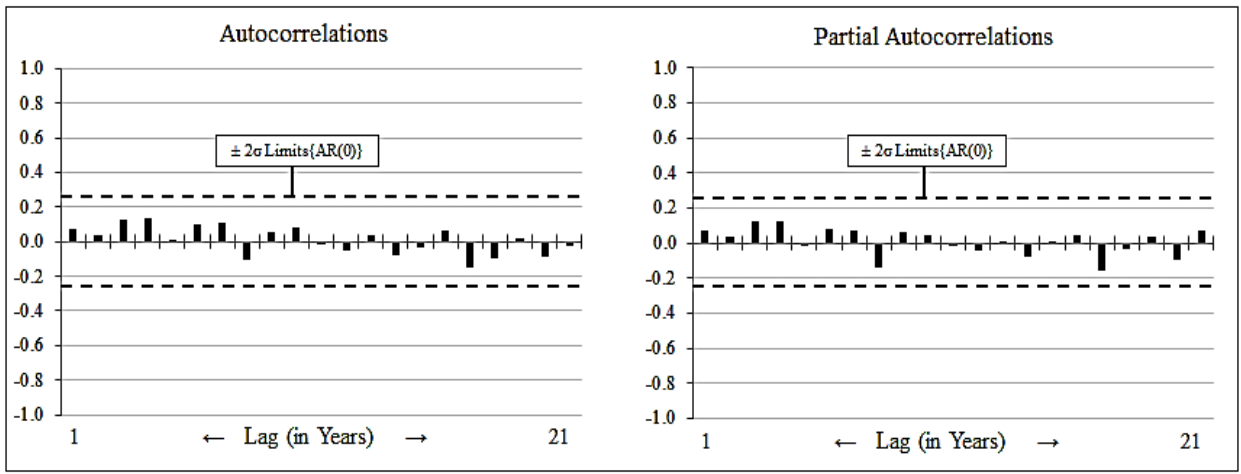

We note that these findings coincide with the theory that markets are efficient (at least in the weak sense), meaning that prices adjust for any predictive capacity inherent in historical patterns, thereby instantly removing the predictive information. Lastly, we assume that real total stock and bond returns jointly follow a bivariate normal distribution.

We can test the stationarity assumption after fitting an AR(p) model to the data. The fitted model defines a characteristic polynomial in the backshift operator B (where $B^k[Y_t] = Y_{t-k}$) using the estimated coefficients $\hat{\phi}$ and satisfying $\phi(B) * Y_t = a_t$ from (F.2). We immediately see that $\phi(B) = 1 - \hat{\phi}_1 B - \hat{\phi}_2 B^2 - \ldots - \hat{\phi}_p B^p$ is such a polynomial and stationarity is confirmed if all roots have magnitude > 1 (Box et al. (1994)). (Complex roots would appear as conjugate pairs, a $\pm$ b$i$, with magnitude $\sqrt{a^2 + b^2}$.) Since an AR(0) process was found appropriate for both returns, the stationarity test was not performed. Note that MA(q) models are stationary by construct, and ARMA(p, q) models are stationary when their autoregressive component meets the condition just stated (Box et al. (1994)). Regression models were also fit and indicate that the average real return for stocks and bonds does not change as a linear function of time.

Due to the nature of compounding, investors generally do not experience the arithmetic mean return of a security over time. The geometric mean is a more accurate reflection of the compounded average performance. Based on the 86 years of historical data used here, the geometric means of inflation-adjusted stock and bond returns are 0.063 and 0.018, respectively. Eighty-six years of data were simulated (N=1 million times each) using the distributions for stocks and bonds given in (F.1a) and (F.1b) and the average geometric mean values were 0.063 and 0.018, respectively, matching their historical counterparts.[6]

---

[6] Both negative and positive outliers were removed in the same relative proportions to avoid square roots of negative numbers. (About 1 outlier per 70,000 simulated histories was removed.)

At each stage/state of the discrete DP (see Figure 5) the minimum probability of ruin is found over all asset allocations in the set $\alpha_{\{\cdot\}}$. Let $R_{(s,t)}$ and $R_{(b,t)}$ denote the actual total returns (not inflation-adjusted) at year t, for stocks and bonds respectively. A portfolio with stock and bond proportions $\alpha$, and $(1-\alpha)$ thus has a total return represented by the RV,

$$R_{(t,\alpha)} = \alpha * R_{(s,t)} + (1-\alpha) * R_{(b,t)}, \tag{F.3a}$$

with compounded return:

$$1 + R_{(t,\alpha)} = 1 + [\alpha * R_{(s,t)} + (1-\alpha) * R_{(b,t)}]. \tag{F.3b}$$

Adjusting for inflation, the compounded real return for this portfolio at year t is given by the RV:

$$1 + r_{(t,\alpha)} \mid I_t, R_{(t,\alpha)} = [1 + R_{(t,\alpha)}]/(1+I_t) = \{1 + [\alpha * R_{(s,t)} + (1-\alpha) * R_{(b,t)}]\}/(1+I_t) \tag{F.4a}$$

$$\rightarrow \quad 1 + r_{(t,\alpha)} \mid I_t, R_{(t,\alpha)} = \{\alpha * [1 + R_{(s,t)}] + (1-\alpha) * [1 + R_{(b,t)}]\}/(1+I_t) \tag{F.4b}$$

$$\rightarrow \quad 1 + r_{(t,\alpha)} \mid I_t, R_{(t,\alpha)} = \alpha * [1 + R_{(s,t)}]/(1+I_t) + (1-\alpha) * [1 + R_{(b,t)}]/(1+I_t) \tag{F.4c}$$

$$\rightarrow \quad 1 + r_{(t,\alpha)} \mid I_t, R_{(t,\alpha)} = \alpha * (1 + r_{(s,t)}) + (1-\alpha) * (1 + r_{(b,t)}) \tag{F.4d}$$

$$\rightarrow \quad 1 + r_{(t,\alpha)} \mid I_t, R_{(t,\alpha)} = \alpha * N(1.0825, 0.2007) + (1-\alpha) * N(1.0214, 0.0834) \tag{F.4e}$$

$$\rightarrow \quad 1 + r_{(t,\alpha)} \mid I_t, R_{(t,\alpha)} \sim N(\alpha * (1.0825) + (1-\alpha) * (1.0214), \tag{F.4f}$$

$$\sqrt{\alpha^2 * (0.2007)^2 + (1-\alpha)^2 * (0.0834)^2 + 2 * \alpha * (1-\alpha) * (.00073)} \; ),$$

where in general,

$$\text{Var}(\alpha X_1 + (1-\alpha) X_2) = \alpha^2 * \text{Var}(X_1) + (1-\alpha)^2 * \text{Var}(X_2) + 2 * \alpha * (1-\alpha) * \text{Cov}(X_1, X_2) \tag{F.4g}$$

and,

$$\text{Cov}(X_1, X_2) = \rho * \sqrt{\text{Var}(X_1) * \text{Var}(X_2)}, \tag{F.4h}$$

holds for any 2 random variables $X_1$ and $X_2$. Finally, assuming an expense ratio of $E_R$ is paid to the financial institution yearly, the inflation/expense-adjusted return at year t is given by the RV:

$$\hat{r}_{(t,\alpha)} = (1 + r_{(t,\alpha)})(1 - E_R) \sim N((1-E_R) * [\alpha * (1.0825) + (1-\alpha) * (1.0214)], \tag{F.5}$$

$$(1-E_R) * \sqrt{\alpha^2 * (0.2007)^2 + (1-\alpha)^2 * (0.0834)^2 + 2 * \alpha * (1-\alpha) * (.00073)} \; ).$$

# Appendix G.  Miscellanea

## G.1  The Equity Glide-Path Debate

Much of the pracitioner-based literature on retirement research has been devoted to finding the best equity glide-path to use in decumulation.  We have summarized the conclusions from several such studies within the literature review section of this paper.  In fact a spirited debate on the topic has ensued and a natural question to ask is whether or not our model can ultimately settle the issue.  Let $\alpha_t \in [0, 1]$ denote the equity ratio used at time t for t=0, 1, …, $T_D$-1 (fixed).  The inflation/expense-adjusted random return between times t=0 and t=1 is $\hat{r}_{(1,\alpha_0)}$ which has the asset allocation set at time t=0 and is observed at time t=1.  We define an equity glide-path as any set $\alpha_{\{gp\}} = \{\alpha_0, \alpha_1, …, \alpha_{T_{D}-1}\}$ which is fixed at time t=0.  Common equity glide-paths would then satisfy the conditions:  declining ($\alpha_i > \alpha_j$), rising ($\alpha_i < \alpha_j$), and constant ($\alpha_i = \alpha_j$) ∀ i<j = 0, 1, …, $T_D$-1.  Applying our notation, the probability of ruin at any time point during decumulation using equity glide-path $\alpha_{\{gp\}}$ is:

$P(\text{Ruin}(> 0))$

$= P(\text{Ruin}(1) \cup \text{Ruin}(2) \cup … \cup \text{Ruin}(T_D))$ 

(G.1a)

$= 1 - P(\text{Ruin}^C(1) \cap \text{Ruin}^C(2) \cap … \cap \text{Ruin}^C(T_D))$ 

(G.1b)

$= 1 - P([\hat{r}_{(1,\alpha_0)} > RF(0)] \cap [\hat{r}_{(2,\alpha_1)} > RF(1)] \cap … \cap [\hat{r}_{(T_D,\alpha_{T_{D}-1})} > RF(T_D-1)])$ 

(G.1c)

(G.1d)

Independent RVs for any equity glide-path $\alpha_{\{gp\}}$.

$= 1 - \int_{RF(0)}^{\infty} \int_{RF(1)}^{\infty} … \int_{RF(T_D-1)}^{\infty} f\left(\hat{r}_{(1,\alpha_0)}, \hat{r}_{(2,\alpha_1)}, …, \hat{r}_{(T_D,\alpha_{T_{D}-1})}\right) d\left(\hat{r}_{(1,\alpha_0)}\right) d\left(\hat{r}_{(2,\alpha_1)}\right) … d\left(\hat{r}_{(T_D,\alpha_{T_{D}-1})}\right)$

The joint PDF of all future returns using equity glide-path $\alpha_{\{gp\}}$.

The expression in (G.1d) is the probability of ruin at any time point during decumulation given equity glide-path $\alpha_{\{gp\}}$.  For any given $\alpha_{\{gp\}}$ the code that implements our models (see Appendix

H) can be used to derive this probability with minimal modification (by fixing $\alpha_t$ to its known value at each time point). Since we are evaluating a given strategy, simulation can also be used. To enter the equity glide-path debate we turn this problem around and seek the set of values $\{\alpha_0, \alpha_1, \ldots, \alpha_{T_D-1}\}$ that minimizes the expression in (G.1d). We recognize this as the optimization problem addressed in Section II-G but with the added constraint $\alpha(t, RF_1(t)) = \alpha(t, RF_2(t)) \ \forall \ t = 0, 1, \ldots, T_D-1$ and $0 < RF_1(t) < RF_2(t)$. This constraint ensures that the asset allocation $\alpha_t$ at time t is constant. Proceeding by induction as before the step at time $t = T_D$ does not change, but we encounter a problem at time $t = T_D-1$. Assuming $RF(T_D-1)$ is known there is no single $\alpha_{T_D-1}$ that minimizes $P(Ruin(T_D)) \ \forall \ RF(T_D-1) > 0$. Without imposing new conditions or rules we cannot make the decision at time $t = T_D-1$ and therefore cannot proceed to time $t = T_D-2$.

While our model cannot help settle the equity glide-path debate, we see that framing the retirement decumulation problem this way is flawed. The retiree is under no obligation to fix $\alpha_{\{gp\}}$ at time $t = 0$ and the models we have proposed in (9) and (12) do not. We delay the asset allocation decision until each time point t when RF(t) is known and express the optimal $\alpha$ to use between times t and t+1 as a function of (t, RF(t)), namely $\tilde{\alpha} = \alpha(t, RF(t))$. We have also shown that the ruin factor, RF(t), encapsulates into a single value all prior information needed to make this decision. Further, since adding constraints to an optimization problem cannot improve the objective, our model will perform at least as well as any based on a predetermined equity glide-path $\alpha_{\{gp\}}$. To achieve optimality with a predetermined $\alpha_{\{gp\}}$ one must somehow guess in advance which glide-path will be right. Simply assessing all of them is not sufficient. (Note that with $T_D = 30$ and $P_\alpha = 1,000$ there are $1,000^{30}$ possible equity glide-paths.) We conclude that all solutions based on preset equity glide-paths, including those used in target-date funds, are suboptimal when implemented in decumulation with a safe withdrawal rate $W_R$.

*G.2 Splitting the Joint PDF during Backward Induction*

In Appendix G.1 we highlighted a difference between this research and much of the literature. Namely, we delay the asset allocation decision until the last moment, which allows us to compute and use the ruin factor RF(t) when making the decision. The optimal α at each time point is denoted by α̃=α(t, RF(t)) and our goal is to determine this function for each t and RF(t)>0. When the retiree sets their asset allocation at time t the observed return at time t+1 is assumed independent of all returns observed at times ≤ t. During backward induction, however, we split the PDF of the time t+1 unobserved return from the joint PDF of all time > t+1 unobserved returns. At time t we compute RF(t) then optimize to find α̃=α(t, RF(t)). This asset allocation defines the PDF of the RV $\hat{r}_{(t+1,\tilde{\alpha})}$ which is the inflation/expense-adjusted return for the next time point. The subsequent time point's return is $\hat{r}_{(t+2,\tilde{\alpha})}$ where α̃=α(t+1, RF(t+1)), but since RF(t+1) = RF(t)/[$\hat{r}_{(t+1,\tilde{\alpha})}$ – RF(t)], the PDF of the RV $\hat{r}_{(t+2,\tilde{\alpha})}$ is a function of the RV $\hat{r}_{(t+1,\tilde{\alpha})}$, raising the question: Can the joint PDF be separated? We can separate the joint PDF by conditioning on $\hat{r}_{(t+1,\tilde{\alpha})}$ and we know the marginal PDF of $\hat{r}_{(t+1,\tilde{\alpha})}$ from independence. This will be made clear by first deriving the value function using a different approach and then returning to our original method and demonstrating precisely the conditioning that occurs, allowing separation of the joint PDF. Without loss of generality, we will focus on the induction from Appendix C.4, where $T_D$ is fixed and t=$T_D$-3. We observe that for any event and random variable X ~ $f_X(x)$ (see, Ross (2007)):

$$P(\text{Event}) = \int_{-\infty}^{\infty} P(\text{Event} \mid X = x) * f_X(x) \, dx \tag{G.2a}$$

During induction at time t=$T_D$-3 from Appendix C.4 we are interested in:

$$P(\text{Ruin}(> T_D\text{-3})) = P(\text{Ruin}(T_D\text{-2}) \cup \text{Ruin}(T_D\text{-1}) \cup \text{Ruin}(T_D)) \tag{G.2b}$$

$$= 1 - P(\text{Ruin}^C(T_D\text{-2}) \cap \text{Ruin}^C(T_D\text{-1}) \cap \text{Ruin}^C(T_D)) \tag{G.2c}$$

$$= 1 - P(\text{Ruin}^C(T_D\text{-2})) * P(\text{Ruin}^C(T_D\text{-1}) \cap \text{Ruin}^C(T_D) \mid \text{Ruin}^C(T_D\text{-2})) \tag{G.2d}$$

The value function is:

$$V(T_D\text{-}3, RF(T_D\text{-}3)) = Min_{(0 \leq \alpha \leq 1)} \left\{ \begin{array}{l} 1 - P(Ruin^C(T_D\text{-}2))* \\ \\ P(Ruin^C(T_D\text{-}1) \cap Ruin^C(T_D) \mid Ruin^C(T_D\text{-}2)) \end{array} \right\} \quad \text{(G.2e)}$$

$\rightarrow V(T_D\text{-}3, RF(T_D\text{-}3))$

> Applying the definition of a conditional probability.

$$= Min_{(0 \leq \alpha \leq 1)} \left\{ \begin{array}{l} 1 - P(Ruin^C(T_D\text{-}2))* \\ \\ \left[ \frac{P(Ruin^C(T_D-1) \cap Ruin^C(T_D) \cap Ruin^C(T_D-2))}{P(Ruin^C(T_D-2))} \right] \end{array} \right\} \quad \text{(G.2f)}$$

$\rightarrow V(T_D\text{-}3, RF(T_D\text{-}3))$

> Since $Ruin^C(T_D\text{-}2) \equiv \hat{r}_{(T_D-2,\alpha)} > RF(T_D\text{-}3)$.

$$= Min_{(0 \leq \alpha \leq 1)} \left\{ \begin{array}{l} 1 - P(\hat{r}_{(T_D-2,\alpha)} > RF(T_D\text{-}3))* \\ \\ \left[ \frac{P(Ruin^C(T_D-1) \cap Ruin^C(T_D) \cap \hat{r}(T_D-2,\alpha) > RF(T_D-3))}{P(\hat{r}(T_D-2,\alpha) > RF(T_D-3))} \right] \end{array} \right\} \quad \text{(G.2G)}$$

We now apply the observation in (G.2a) to the numerator in (G.2G), which yields:

$$P(Ruin^C(T_D\text{-}1) \cap Ruin^C(T_D) \cap \hat{r}_{(T_D-2,\alpha)} > RF(T_D\text{-}3)) \quad \text{(G.2H)}$$

$$= \int_{-\infty}^{\infty} P(Ruin^C(T_D-1) \cap Ruin^C(T_D) \cap \hat{r}(T_D-2,\alpha) > RF(T_D-3) \mid \hat{r}(T_D-2,\alpha) = r) * f_{\hat{r}(T_D-2,\alpha)}(r) \, dr$$

We separate this integral into 2 components, where the first vanishes and the second simplifies:

> This probability is zero when $\hat{r}_{(T_D-2,\alpha)} \leq RF(T_D\text{-}3)$.

$$= \int_{-\infty}^{RF(T_D-3)} P(Ruin^C(T_D-1) \cap Ruin^C(T_D) \cap \hat{r}(T_D-2,\alpha) > RF(T_D-3) \mid \hat{r}(T_D-2,\alpha) = r) * f_{\hat{r}(T_D-2,\alpha)}(r) dr \quad \text{(G.2I)}$$

$$+ \int_{RF(T_D-3)}^{\infty} P(Ruin^C(T_D-1) \cap Ruin^C(T_D) \cap \hat{r}(T_D-2,\alpha) > RF(T_D-3) \mid \hat{r}(T_D-2,\alpha) = r) * f_{\hat{r}(T_D-2,\alpha)}(r) dr$$

> This event will always occur when $\hat{r}_{(T_D-2,\alpha)} > RF(T_D\text{-}3)$ and can be dropped.

In (G.2G), (G.2I) was the numerator of an expression [ · ], and including the denominator yields:

$$[\cdot] = \frac{\int_{RF(T_D-3)}^{\infty} P(Ruin^C(T_D-1) \cap Ruin^C(T_D) \mid \hat{r}(T_D-2,\alpha)=r) * f_{\hat{r}(T_D-2,\alpha)}(r) dr}{\int_{RF(T_D-3)}^{\infty} f_{\hat{r}(T_D-2,\alpha)}(r) \, dr} \quad \text{(G.2J)}$$

$$\boxed{\text{This is the PDF of the RV } \hat{r}_{(TD-2,\alpha)}{}^+ = (\hat{r}_{(TD-2,\alpha)} \mid \hat{r}_{(TD-2,\alpha)} > RF(T_D-3)).}$$

(G.2K)

$$= \int_{RF(T_D-3)}^{\infty} \left[ \frac{f_{\hat{r}(TD-2,\alpha)}(r)}{\int_{RF(T_D-3)}^{\infty} f_{\hat{r}(TD-2,\alpha)}(r)\,dr} \right] * P(\text{Ruin}^C(T_D-1) \cap \text{Ruin}^C(T_D) \mid \hat{r}(T_D-2,\alpha) = r)\,dr$$

$$= E\hat{r}_{(TD-2,\alpha)}{}^+ [\,P(\text{Ruin}^C(T_D-1) \cap P(\text{Ruin}^C(T_D) \mid \hat{r}_{(TD-2,\alpha)}]$$

(G.2L)

Substituting (G.2L) back into (G.2G) clarifies our task, which is to choose an α that minimizes the bracketed expression $\{\cdot\}$ and assign it to the function V(T_D-3, RF(T_D-3)).

$\rightarrow$ V(T_D-3, RF(T_D-3)) (G.2M)

$$= \text{Min}_{(0 \le \alpha \le 1)} \left\{ \begin{array}{l} 1 - P(\hat{r}_{(TD-2,\alpha)} > RF(T_D-3))* \\[2mm] E\hat{r}_{(TD-2,\alpha)}{}^+ [P(\text{Ruin}^C(T_D-1) \cap P(\text{Ruin}^C(T_D) \mid \hat{r}_{(TD-2,\alpha)}] \end{array} \right\}$$

The events Ruin$^C$(T_D-1) and Ruin$^C$(T_D) are determined by future returns, namely those observed at times t=T_D-1 and t=T_D. Since they are based on future asset allocation decisions we are free to consider any values. However, to minimize the expression we must use the largest possible value of $[P(\text{Ruin}^C(T_D-1) \cap P(\text{Ruin}^C(T_D) \mid \hat{r}_{(TD-2,\alpha)}]$ for each α. If we do not, the bracketed term $\{\cdot\}$ cannot be a minimum. We recognize that $[P(\text{Ruin}^C(T_D-1) \cap P(\text{Ruin}^C(T_D) \mid \hat{r}_{(TD-2,\alpha)}]$ is maximized at $[1 - V(T_D-2, RF(T_D-2))]$ by definition of the value function and (G.2M) becomes:

$\rightarrow$ V(T_D-3, RF(T_D-3)) (G.2N)

$$= \text{Min}_{(0 \le \alpha \le 1)} \left\{ \begin{array}{l} 1 - (1 - F\hat{r}_{(TD-2,\alpha)}(RF(T_D-3)))* \\[2mm] (1 - E\hat{r}_{(TD-2,\alpha)}{}^+ [V(T_D-2, RF(T_D-2))]) \end{array} \right\}$$

This is identical to what was derived in (C.9). In (C.8c), during induction for fixed T_D at time t=T_D-3, we express a probability statement as the integration over a corresponding joint PDF $f(\hat{r}_{(TD-2,\alpha)}, \hat{r}_{(TD-1,\tilde{\alpha})}, \hat{r}_{(TD,\tilde{\alpha})})$, then split $f(\hat{r}_{(TD-2,\alpha)})$ from $f(\hat{r}_{(TD-1,\tilde{\alpha})}, \hat{r}_{(TD,\tilde{\alpha})})$ in (C.8e). Since $\hat{r}_{(TD-1,\tilde{\alpha})}$ uses $\tilde{\alpha}=\alpha(T_D-2, RF(T_D-2))$ and $RF(T_D-2) = RF(T_D-3)/[\hat{r}_{(TD-2,\alpha)} - RF(T_D-3)]$ some may question the validity of this operation. The same is true for RVs $\hat{r}_{(TD-2,\alpha)}$ and $\hat{r}_{(TD,\alpha)}$. The numerator of the ratio in (C.8c) is $P[\hat{r}_{(TD-1,\tilde{\alpha})} > RF(T_D\text{-}2) \cap \hat{r}_{(TD,\tilde{\alpha})} > RF(T_D\text{-}1) \cap \hat{r}_{(TD-2,\alpha)} > RF(T_D\text{-}3)]$. Using (G.2a) we can express this probability as:

$$P[\hat{r}_{(TD-1,\tilde{\alpha})} > RF(T_D\text{-}2) \cap \hat{r}_{(TD,\tilde{\alpha})} > RF(T_D\text{-}1) \cap \hat{r}_{(TD-2,\alpha)} > RF(T_D\text{-}3)] \qquad \text{(G.2O)}$$

$$= \int_{-\infty}^{\infty} P(\hat{r}_{(TD-1,\tilde{\alpha})} > RF(T_D\text{-}2) \cap \hat{r}_{(TD,\tilde{\alpha})} > RF(T_D\text{-}1) \cap \hat{r}_{(TD-2,\alpha)} > RF(T_D\text{-}3) | \hat{r}_{(TD-2,\alpha)} = r) * f_{\hat{r}(T_D-2,\alpha)}(r)\, dr$$

$$\boxed{\text{This probability is zero when } \hat{r}_{(TD-2,\alpha)} \leq RF(T_D\text{-}3).}$$

$$\qquad \text{(G.2P)}$$

$$= \int_{-\infty}^{RF(T_D\text{-}3)} P(\hat{r}_{(TD-1,\tilde{\alpha})} > RF(T_D\text{-}2) \cap \hat{r}_{(TD,\tilde{\alpha})} > RF(T_D\text{-}1) \cap \hat{r}_{(TD-2,\alpha)} > RF(T_D\text{-}3) | \hat{r}_{(TD-2,\alpha)} = r) * f_{\hat{r}(T_D-2,\alpha)}(r)\, dr$$

$$+ \int_{RF(T_D\text{-}3)}^{\infty} P(\hat{r}_{(TD-1,\tilde{\alpha})} > RF(T_D\text{-}2) \cap \hat{r}_{(TD,\tilde{\alpha})} > RF(T_D\text{-}1) \cap \hat{r}_{(TD-2,\alpha)} > RF(T_D\text{-}3) | \hat{r}_{(TD-2,\alpha)} = r) * f_{\hat{r}(T_D-2,\alpha)}(r)\, dr$$

$$\boxed{\text{This event will always occur when } \hat{r}_{(TD-2,\alpha)} > RF(T_D\text{-}3) \text{ and can be dropped.}}$$

$$= \int_{RF(T_D\text{-}3)}^{\infty} f_{\hat{r}(T_D-2,\alpha)}(r) * P(\hat{r}_{(TD-1,\tilde{\alpha})} > RF(T_D\text{-}2) \cap \hat{r}_{(TD,\tilde{\alpha})} > RF(T_D\text{-}1) | \hat{r}_{(TD-2,\alpha)} = r)\, dr \qquad \text{(G.2Q)}$$

$$\boxed{\text{This probability is replaced by its integral expression.}}$$

$$\qquad \text{(G.2R)}$$

$$= \int_{RF(T_D\text{-}3)}^{\infty} f(\hat{r}_{(T_D-2,\alpha)}) \int_{RF(T_D\text{-}2)}^{\infty} \int_{RF(T_D\text{-}1)}^{\infty} f(\hat{r}_{(T_D,\tilde{\alpha})}, \hat{r}_{(T_D-1,\tilde{\alpha})}) d(\hat{r}_{(T_D,\tilde{\alpha})}) d(\hat{r}_{(T_D-1,\tilde{\alpha})}) d(\hat{r}_{(T_D-2,\alpha)})$$

$$\boxed{\begin{array}{l}\text{The joint PDF has been separated which} \\ \text{validates the operation in (C.8e).}\end{array}}$$

We conclude by emphasizing some subtle points from above. At time t we set the $\alpha$ that defines the RV $\hat{r}_{(t+1)}$ which is observed at time t+1. At time t+1 we set the $\alpha$ that defines the RV $\hat{r}_{(t+2)}$ which is observed at time t+2. Regardless of the $\hat{r}_{(t+2)}$ we choose to represent our portfolio between times t+1 and t+2 it will be independent of $\hat{r}_{(t+1)}$ via market efficiency. However, the choice of $\hat{r}_{(t+2)}$ does depend on $\hat{r}_{(t+1)}$. Further, since we observe returns sequentially in time we can always condition on knowledge of what has happened previously if it helps to simplify expressions. This is true because backward induction assumes the prior RF(t) is known, and it is also true when we use the model since we start at time t=0 and proceed sequentially in time.

## Appendix H.  Full C++ Implementation

The implementation consists of 5 functions and one header that are compiled into a 64-bit console application (executable).  The application is launched by either double-clicking the executable or by invoking it from the command line or via a batch file.  A single parameter is optionally accepted which is the number of tasks the user wants to process concurrently as it runs.  If no value is supplied the code determines the maximum number of concurrent processing units on the machine running it and uses this value.  The maximum number of concurrent processing units is generally the number of CPU cores.  If an Intel® processor with Hyper-Threading® (HT) is used the capacity of each core doubles.  For example, the runtime statistics given next are from a Dell® Precision® workstation with dual 10-core 2.5GHz Intel® Xeon® processors.  Since all Xeon® processors use HT this PC can execute 2x10x2=40 concurrent jobs.  The application is multi-threaded and the task of processing each time point is split as follows.  First the code approximates the RF(t) bucket at which heavy algorithm pruning begins.  It then splits the remaining buckets into equal-sized collections and launches jobs to process all buckets simultaneously in separate threads.  Once all threads for a given time point finish the results are aggregated and processing for the next time point begins.

When launched the application queries the user for the location of a directory that contains a control file and optionally a file of age probabilities (required for random $T_D$).  The control file defaults to filename "control.txt" and the age probabilities file to "ageprobs.txt".  Both filenames can be changed in the header if desired.  Examples of 6 control files (including runtime statistics) and the age probability file (derived from SSA.gov) are shown next.  This application uses the C++ boost© library which will need to be installed (including binaries) on the machine creating the executable and it was optimized for speed when compiled using Microsoft® Visual Studio® (a project level setting).

---

Contents: (A 3-line text file of space-delimited values as shown below.)

```
StockMean StockVar BondMean BondVar StockBondCov RF_Max E_R PrunePwr
P_R P_α
NumRand Details
```

Example 1: (Fixed $T_D$=50, $P_R$=1,000, $P_\alpha$=100, $RF_{Max}$=2.75, $E_R$=0.0%, Runtime=00:00:43)

```
0.082509 0.0402696529 0.021409 0.0069605649 0.0007344180 2.75 0.000 4.00
1000 100
0 50
```

---

Example 2: (Fixed $T_D$=50, $P_R$=5,000, $P_\alpha$=1,000, $RF_{Max}$=2.75, $E_R$=0.5%, Runtime=00:28:02)

```
0.082509 0.0402696529 0.021409 0.0069605649 0.0007344180 2.75 0.005 4.00
5000 1000
0 50
```

---

Example 3: (Fixed $T_D$=50, $P_R$=10,000, $P_\alpha$=1,000, $RF_{Max}$=2.75, $E_R$=1.0%, Runtime=01:42:36)

```
0.082509 0.0402696529 0.021409 0.0069605649 0.0007344180 2.75 0.010 4.00
10000 1000
0 50
```

---

Example 4: (Random $T_D$/N=2, $S_{Max}$=46, $P_R$=1,000, $P_\alpha$=100, $RF_{Max}$=2.75, $E_R$=0.0%, Runtime=00:01:17)

```
0.082509 0.0402696529 0.021409 0.0069605649 0.0007344180 2.75 0.000 4.00
1000 100
2 M 65 F 67
```

---

Example 5: (Random $T_D$/N=9, $S_{Max}$=50, $P_R$=5,000, $P_\alpha$=1,000, $RF_{Max}$=2.75, $E_R$=0.5%, Runtime=02:22:08)

```
0.082509 0.0402696529 0.021409 0.0069605649 0.0007344180 2.75 0.005 4.00
5000 1000
9 M 62 F 63 M 66 F 67 F 70 F 72 M 74 M 75 F 76
```

---

Example 6: (Random $T_D$/N=1, $S_{Max}$=48, $P_R$=10,000, $P_\alpha$=1,000, $RF_{Max}$=2.75, $E_R$=1.0%, Runtime=11:01:24)

```
0.082509 0.0402696529 0.021409 0.0069605649 0.0007344180 2.75 0.010 4.00
10000 1000
1 F 65
```

---

Note: *Runtime units are HH:MM:SS. The means/variances match the historical values that were derived in Appendix F. A 50-year fixed $T_D$ solution contains all solutions for fixed $T_D$ under 50 years. The parameter PrunePwr=4.0 indicates that heavy algorithm pruning begins when P(Ruin) equals the largest possible P(Ruin) for that time point to within 4 decimal places. Once ruin becomes a near certainty $\alpha$=1 and it is not necessary to evaluate all $\alpha$-values. A higher PrunePwr value has a longer runtime and the user can confirm an equivalent solution by adjusting this value. The # years processed for random $T_D$ is the larger of (111-min male age) and (113-min female age) where 111 and 113 are the maximum possible male/female ages, respectively, as derived from SSA.gov life-tables. (See the file "ageprobs.txt" below.) When $T_D$ is random the last possible withdrawal is at $t=S_{Max}$ but the last decision point is at $t=S_{Max}-1$.*

---------------------------------------------------------------------------------------------------------

| | | |
|---|---|---|
| 50 | 0.00534626517475267O | 0.00328338544226365O |
| 51 | 0.0058106882101352­20 | 0.0035448014806604­40 |
| 52 | 0.006264310709811210 | 0.003795760877521360 |
| 53 | 0.006717933209487190 | 0.004015350349774660 |
| 54 | 0.007171555709163170 | 0.004234939822002960 |
| 55 | 0.007657579815958870 | 0.004475442577353010 |
| 56 | 0.008176005529874280 | 0.004747315257288520 |
| 57 | 0.008683630708083120 | 0.005071471144897680 |
| 58 | 0.009180455350585390 | 0.005458366881724930 |
| 59 | 0.009677279993087660 | 0.005897545826231530 |
| 60 | 0.010228107314122800 | 0.006399464619953360 |
| 61 | 0.010822136777984200 | 0.006943209979818680 |
| 62 | 0.011513371063204700 | 0.007518325264291610 |
| 63 | 0.012269408562664700 | 0.008103897190300420 |
| 64 | 0.013111850347777200 | 0.008741752323988580 |
| 65 | 0.014029895882835800 | 0.009452803948427840 |
| 66 | 0.015034345703546900 | 0.010247508705154100 |
| 67 | 0.016038795524258000 | 0.011104955331095500 |
| 68 | 0.017064846416382300 | 0.012035594407788100 |
| 69 | 0.018112498379919600 | 0.013049888636767600 |
| 70 | 0.019278956236229300 | 0.014168749281105900 |
| 71 | 0.020531818378191600 | 0.015392176340802900 |
| 72 | 0.021827882662980100 | 0.016667886608179200 |
| 73 | 0.023134747483475200 | 0.017943596875555500 |
| 74 | 0.024463213375383400 | 0.019282046992147100 |
| 75 | 0.025899684624357400 | 0.020787803373312600 |
| 76 | 0.027422560158983900 | 0.022429496094444400 |
| 77 | 0.028891433015077500 | 0.024050275532504500 |
| 78 | 0.030273901585518600 | 0.025629228404421100 |
| 79 | 0.031591566941720300 | 0.027208181276337700 |
| 80 | 0.032887631226508800 | 0.028891700563613000 |
| 81 | 0.034140493368471100 | 0.030721612832390500 |
| 82 | 0.035166544260595300 | 0.032561981742703900 |
| 83 | 0.035922581760055300 | 0.034339610803802000 |
| 84 | 0.036365403724024700 | 0.036033586732613200 |
| 85 | 0.036516611223916700 | 0.037539343113778700 |
| 86 | 0.036300600509785300 | 0.038804596739619200 |
| 87 | 0.035663368903097600 | 0.039693411270168200 |
| 88 | 0.034572514796733900 | 0.040132590214647900 |
| 89 | 0.032963234976454800 | 0.040028023799316100 |
| 90 | 0.030857130513673500 | 0.039285602250269300 |
| 91 | 0.028265001944096400 | 0.037894868925998300 |
| 92 | 0.025305655160496000 | 0.035845367184967500 |
| 93 | 0.022076294984231200 | 0.033189380234856200 |
| 94 | 0.018717328379487600 | 0.030021017849487100 |
| 95 | 0.015347561239037500 | 0.026371649953467900 |
| 96 | 0.012129001598479300 | 0.022481779302123700 |
| 97 | 0.009277660171944530 | 0.018591908650779500 |
| 98 | 0.006825938566552900 | 0.014890257547081000 |
| 99 | 0.004881842139370110 | 0.011585958821745600 |
| 100 | 0.003391368211863310 | 0.008804492173203810 |
| 101 | 0.002300514105499630 | 0.006545857601455560 |
| 102 | 0.001522875534626520 | 0.004726401974213920 |
| 103 | 0.000972048213591394 | 0.003335668649943010 |
| 104 | 0.000604829999567979 | 0.002269091213284120 |
| 105 | 0.000356417678316845 | 0.001495299739629630 |
| 106 | 0.000216010714131421 | 0.000951554379764307 |
| 107 | 0.000108005357065710 | 0.000575115284472933 |
| 108 | 0.000064803214239426 | 0.000345069170683760 |
| 109 | 0.000032401607119713 | 0.000188219547645687 |
| 110 | 0.000010800535706571 | 0.000094109773822844 |
| 111 | 0.000010800535706571 | 0.000052283207679358 |
| 112 | 0.000000000000000000 | 0.000020913283071743 |
| 113 | 0.000000000000000000 | 0.000010456641535872 |

<u>Note:</u>  *Columns in this file represent: age, male death probability, and female death probability for a 50-year old. They are computed from life-tables published at SSA.gov and each column sums to 1. For example, a male aged 50 has a 0.00534626517475267O probability of death before age 51. As seen, the maximum possible male age is 111 and the maximum possible female age is 113. There are no blank lines in this file and each field is delimited by a single space.*

```
/*
/ Copyright (c) 2014 Chris Rook, All Rights Reserved.
/
/    Redistribution and use in source and binary forms, with or without modification, are permitted provided that the following conditions are met:
/
/       1. Redistributions of source code must retain the above copyright notice, this list of conditions and the following disclaimer.
/       2. Redistributions in binary form must reproduce the above copyright notice, this list of conditions and the following disclaimer in the
/          documentation and/or other materials provided with the distribution.
/       3. Neither the name of the copyright holder nor the names of its contributors may be used to endorse or promote products derived from this
/          software without specific prior written permission.
/
/       THIS SOFTWARE IS PROVIDED BY THE COPYRIGHT HOLDERS AND CONTRIBUTORS "AS IS" AND ANY EXPRESS OR IMPLIED WARRANTIES, INCLUDING, BUT NOT LIMITED
/       TO, THE IMPLIED WARRANTIES OF MERCHANTABILITY AND FITNESS FOR A PARTICULAR PURPOSE ARE DISCLAIMED. IN NO EVENT SHALL THE COPYRIGHT HOLDER OR
/       CONTRIBUTORS BE LIABLE FOR ANY DIRECT, INDIRECT, INCIDENTAL, SPECIAL, EXEMPLARY, OR CONSEQUENTIAL DAMAGES (INCLUDING, BUT NOT LIMITED TO,
/       PROCUREMENT OF SUBSTITUTE GOODS OR SERVICES; LOSS OF USE, DATA, OR PROFITS; OR BUSINESS INTERRUPTION) HOWEVER CAUSED AND ON ANY THEORY OF
/       LIABILITY, WHETHER IN CONTRACT, STRICT LIABILITY, OR TORT (INCLUDING NEGLIGENCE OR OTHERWISE) ARISING IN ANY WAY OUT OF THE USE OF THIS SOFTWARE,
/       EVEN IF ADVISED OF THE POSSIBILITY OF SUCH DAMAGE.  (Source: http://opensource.org/licenses/BSD-3-Clause.)
/
/ Filename:  minpruin.cpp  (Defines the entry point for the console application.)
/
/ Function:  main()
/
/ Summary:
/
/       Compiling the code will generate a single executable file that can be invoked with a double click or via a batch file with one parameter.  When invoked
/       the directory location where input files reside and output files are written to is queried, stored as a string vbl and passed to other functions.
/       The control file (control.txt) is then read.  If TD is specified as random the function derhrates() is invoked which reads the file of male/female
/       death probabilities (ageprobs.txt) and derives the hazard rates that apply to the group listed in the last line of the control file.  The hazard rates
/       are stored locally and written to the file specified by the constant hrfile for reference.  When TD is fixed, hazard rates are not needed or derived.
/       When invoked, the executable may have either zero or one argument(s).  (See below.)  The arrangement specified by the control file determines the
/       total number of years to process.  Iteration always begins at the terminal year (i.e., total # of years minus 1) and ends at year 0.  This reflects the
/       fact that the first asset allocation (alpha) decision point is at time t=0 and the last is at time t=(total # years - 1).  The first withdrawal attempt
/       (assuming survival) is at time t=1 and the last withdrawal attempt (assuming survival) is at time t=(total # years).  For example, if TD is fixed
/       and 50 years are specified in the control file then processing starts at year=49 and ends at year=0.  (This results in 50 alpha decisions made by the
/       retiree, the first at year=0 (i.e., t=0) and the last at year=49 (i.e., t=49).  The total # of years to process for random TD depends on the specific
/       ages listed for the group (MPU).  For example, if random TD for a 65 year old male and female couple is specified in the control file then a total of 48
/       years are processed since the maximum age allowed for a male is 111 and for a female is 113 (see the file ageprobs.txt).  Processing 48 years of data
/       implies that 48 alpha decisions are made, here the first is at year=0 (both M/F at 65) and the last at year=47 (F is age 112).  (See the paper for a
/       warning about dated hazard risk.  Rerun the optimization if the hazard probabilities differ notably over time from those written to hrates.txt.)  Using a
/       different ageprobs.txt file can change total # of years that need to be processed which, in general is: Max((Max male age - Min male age),(Max female
/       age - Min female age)).  Here maximums are allowable and minimums are actual as specified in the control file.  When processing each year the function
/       getprprobs() is invoked and reads the probabilities derived for the previous year.  These are then passed to the function optimize() which derives the
/       probabilities and alphas for the current year and writes the results to a file for reference.  Once complete, the function combfiles() is invoked which
/       combines all separate files into a single file to be read in as the prior probabilities during processing of the next year/time point.  When at year=0,
/       combfiles() also concatenates all data from all prior years into both a vertical text file and 2 horizontal csv files, where the csv files can be read
/       into a spreadsheet.  This is helpful when simulating the optimal solution to verify the result and only the csv file of optimal alphas is needed.
/
/       Notes:  1.) The terms "year" and "time point" are used interchangably throughout this application and mean the same thing.
/               2.) The file "ageprobs.txt" was derived from life tables at SSA.gov.
/
/ Parameters:
/
/       The executable takes zero or one argument(s) when invoked.  If an argument is specified it reflects the maximum number of jobs to process concurrently
/       by the application as it executes.  If no argument is specified this value defaults to the number of independent processing units that exist on the
/       machine running the application.  (This application was developed and tested on the Windows 7® operating system.)
/
/ Input Files:
/
/              1.) Parameter control file of specific form whose name is defined by the string constant paramfile (defined in the header and located in the directory
```

```
/                        specified by the user).
/                   2.) Age probability file of specific form whose name is set as the string constant ageprfile (defined in the header file and located in the directory
/                        specified by the user.  This file is read and processed by the function derhrates()).
/
/ Output Files:
/
/       1.) The derived hazard rates for the random arrangement specified in the control file.  The hazard rate file is written by function derhrates() using
/            filename specified by string constant hrfile defined in the header to the directory specified by the user.
/       2.) FinalResults_V.txt  (Vertical text file of all results from all years processed, written by combfiles() after year=0 finishes.)
/       3.) FinalAlphaResults_H.csv  (Horizontal csv file of optimal alphas, written by combfiles() after year=0 finishes.  Use this file to simulate the result.)
/       4.) FinalProbResults_H.csv  (Horizontal csv file of optimal probabilities, written by combfiles() after year=0 finishes.)
/
/       Note:  A number of temporary data files are written to the directory specified by the user during the processing of each year and across years.  These
/            are removed by the application as it runs.  They are created to avoid carrying large amounts of data inside arrays that are not needed.  This was
/            a design decision made to speed up runtime.  When processing a given year/time point, only the prior year's data are needed.  Data from all other
/            years/time points is not used (by nature of DPs).
/
/ Return Value:
/
/                   This function returns the integer value of 0 (success) or an error code (failure).
/-------------------------------------------------------------------------------------------------------------------------------------------------------------------*/

#include "stdafx.h"

int main(int argc, char *argv[])
{
        // Declare variables.
        int precisions[2], numrand, numyears, p=0, rc, pllproc=0;
        long buckets[2], * * bktarys, bktsprun;
        double params[8];
        long double * hr, * prV;
        string rootdir, trbin;
        vector<long> unqbkts;
        boost::thread * t;

        // Get the directory location where setup files are stored, strip any leading/trailing blanks.
        cout << "Enter the directory where the setup files reside (eg, c:\\mprsetup\\): " << endl;
        cin >> rootdir; cin.get();
        boost::algorithm::trim(rootdir);
        cout << endl;

        // Read in parameter control file.  Load first line into array params[] and 2nd line into array precisions[].
        // (3rd line is processed conditionally based on the arrangement specified by the user i.e., fixed vs random TD.)
        ifstream getparams(rootdir+paramfile);
        if (getparams.is_open())
        {
                getparams >> params[0] >> params[1] >> params[2] >> params[3] >> params[4] >> params[5] >> params[6] >> params[7];
                getparams >> precisions[0] >> precisions[1];
                getparams >> numrand;
                if (numrand > 0)
                {
                        char * genders = new char[numrand];
                        int * ages = new int[numrand];
                        if (numrand == 1)
                                getparams >> genders[0] >> ages[0];        // Gender and age of single person with random TD.
                        else                                               // Random TD for group.  Load arrays with genders and ages.
                        {
                                string people;
                                std::getline(getparams, people);
                                stringstream ss(people);
                                for (int i=1; i <= numrand && ss.good(); ++i)
                                {
```

```cpp
                                ss >> genders[i-1] >> ages[i-1];
                                genders[i-1] = toupper(genders[i-1]);

                                // Data check for invalid gender.  (Invalid ages are checked in derhrates().)
                                if (genders[i-1] != 'M' && genders[i-1] != 'F')
                                {
                                        cout << "ERROR: Invalid gender in file " + rootdir + paramfile + " for person #" << i << ": " << genders[i-1] << endl;
                                        cout<< "EXITING...main()..." << endl; cin.get();
                                        exit (EXIT_FAILURE);
                                }
                        }
                }

                // Build the hazard rates if random TD is specified and no file exists in the directory. (Don't leave an old one sitting there.)
                ifstream gethrates(rootdir+hrfile);
                numyears = derhrates(rootdir, numrand, genders, ages, gethrates.is_open());
                hr = new long double[numyears+1];
                delete [] genders;  genders = nullptr;
                delete [] ages;  ages = nullptr;

                // File of hazard rates has now been built and written for reference, load it to the hr[] array.
                if (!gethrates.is_open())
                        gethrates.open(rootdir+hrfile);

                // Read in survival probabilities that have been written to the file hrates.txt for this arrangement.
                // (Doing it this way is useful when needing to run a single year for debugging purposes.)
                if (gethrates.is_open())
                {
                        while (!gethrates.eof() && p < numyears+1)
                        {
                                gethrates >> hr[p] >> trbin;
                                p=p+1;
                        }
                        gethrates.close();
                }
                else
                {
                        cout << "ERROR: Could not read file: " << rootdir + hrfile << endl;
                        cout << "EXITING...main()..." << endl; cin.get();
                        exit (EXIT_FAILURE);
                }
        }
        else
        {
                // For fixed TD load hr[] with zeros (hrates.txt is not created).
                // Using a fixed TD is similar to using a random TD with all discrete hazard probabilities = 0 except the one at TD, which equals 1.
                // That is, P(TD=t|TD>=t) = 0 for t=0,1,2,...,TD-1 and P(TD=t|TD>=t)=1 for t=TD. (Pr[death before TD] = 0 and Pr[death at TD]=1.)
                getparams >> numyears;                // Account for TD using HR=1.00.
                hr = new long double[numyears+1];
                for (int j=0; j<numyears; ++j)
                        hr[j]=0.0;
                hr[numyears]=1.00;
        }
        getparams.close();
}
else
{
        cout << "ERROR: Could not read file: " << rootdir + paramfile << endl;
        cout << "EXITING...main()..." << endl; cin.get();
        exit (EXIT_FAILURE);
}
```

```cpp
// Parse arguments to main().  Pllproc is optional, default is optimal # of threads on the machine running the application.
if (argc == 2)
        pllproc = stoi(argv[1]);
else if (argc != 1)
{
        cout << "ERROR: Parameter misspecification.  Incorrect # of parameters to the executable (expecting zero or one...)." << endl;
        cout << "EXITING...main()..." << endl; cin.get();
        exit (EXIT_FAILURE);
}

// When pllproc==0, replace with the # of independent processing units on the computer running the application.
// Note that main() runs in its own thread but that is not accounted for here as it sits idle.
if (pllproc == 0)
        pllproc = boost::thread::hardware_concurrency();
if (pllproc == 1)
        pllproc = pllproc++;      // If just one processing unit use 2 threads.

// Iterate over all years, launching separate threads to proccess optimal size collections of buckets concurrently.
for (int yr=numyears-1; yr>=0; yr--)
{
        cout << endl << "Processing for year " << yr << " has begun ..." << endl;

        // Retrieve the prior year's probabilities which have been written to a file and load into an array.
        prV = new long double[(long) (params[5]*precisions[0] + 1.5)];
        unqbkts=getprprobs(rootdir, yr, numyears, params[5], precisions[0], prV, hr[yr+1]);

        // Function call to determine pruning/parallel processing parameters for the current year.
        buckets[0]=0;
        buckets[1]=0;
        rc = optimize(rootdir, yr, params, precisions, buckets, hr, prV, unqbkts);

        // Derive the optimal # of buckets to process per run.
        bktsprun = (long) (rc / (pllproc-1));
        cout << "-->  # Buckets processed per thread (excluding last one): " << bktsprun << endl;

        // Build thread array of optimal size then launch calls to optimize() concurrently.
        bktarys = new long * [pllproc];
        t = new boost::thread[pllproc];
        for (int i=1; i<=pllproc; ++i)
        {
                // Define buckets then process them in concurrent threads.
                bktarys[i-1] = new long[2];
                bktarys[i-1][0]=bktsprun*(i-1) + 1;
                bktarys[i-1][1]=(i < pllproc)*bktsprun*(i) + (i==pllproc)*((long) (params[5]*precisions[0] + 0.5));
                cout << "-->  Begin concurrent processing of buckets " << bktarys[i-1][0] << " through " << bktarys[i-1][1] << " ..." << endl;
                t[i-1] = boost::thread(optimize, rootdir, yr, boost::cref(params), boost::cref(precisions), boost::cref(bktarys[i-1]), boost::cref(hr),
                boost::cref(prV), boost::cref(unqbkts));
        }

        // Wait for all threads to finish, then proceed.
        for (int i=0; i<pllproc; ++i)
                t[i].join();

        // Free dynamic memory allocations.
        for (int i=0; i<pllproc; ++i)
        {
                delete [] bktarys[i];  bktarys[i] = nullptr;
        }
        delete [] bktarys; bktarys = nullptr;
        delete [] t;   t = nullptr;
        delete [] prV; prV = nullptr;
```

```
                    // Clear contents of unqbkts vector and reset capacity.  (This container is reused within the loop.)
                    unqbkts.clear();
                    unqbkts.shrink_to_fit();

                    // Combine all files for each year processed, and if at year=0 concatenate data files from all years.
                    combfiles(rootdir, yr, pllproc, bktsprun, params[5], precisions[0], numyears);

                    cout << "Processing for year " << yr << " has finished." << endl;
            }

            // Free memory of hazard rate array and exit.
            delete [] hr; hr = nullptr;
            return 0;
}
```

<div align="center">

### Code File:  derhrates.cpp

</div>

```
/ Filename:  derhrates.cpp
/
/ Function:  derhrates()
/
/ Summary:
/
/       Derive the discrete hazard rates for the random TD arrangement specified by the last line of the control file.  This function is not called for
/       fixed TD.  The derived hazard rates are written to a file whose name is specified by the string constant hrfile (defined in the header file), and
/       the total # of years that need to be processed are determined by this function and returned by the call.
/
/ Parameters:
/
/       1.) Root directory where input files reside and output files are written.
/       2.) # of persons specified by this arrangement in the control file.
/       3.) Array of genders for the random persons involved in this arrangement.
/       4.) Array of ages for the random persons involved in this arrangement.
/       5.) True/false indicator of whether or not the hazard rates file already exists.  If it exists, this function only returns the total number of
/            years that need to be processed by this arrangement and the hazard rates are not re-derived.  (Do not leave an old hazard rates file in the
/            root directory.  If one is there it will be used but an alert message is given to the user.  Kill the job with task manager if necessary.)
/
/ Input Files:
/
/       1.) Age probability file of specific form whose name/location is set as the string constant ageprfile which is defined in the header file and
/            resides in the directory specified by the user.
/
/ Output Files:
```

```
/
/       1.) The derived hazard rates for the random arragement specified in the control file using filename specified by hrfile which is defined in the
/           header file and resides in the directory specified by the user.
/
/ Return Value:
/
/       This function returns the # of years that need to be processed for this arrangement, representing the maximum # of alpha decisions to make.
/----------------------------------------------------------------------------------------------------------------------------------------------------------*/

#include "stdafx.h"

int derhrates(const string root, const int npersons, const char * gndrs, const int * ags, const bool hrexst)
{
        int age, prvage, remyrs, strtage=0, minmage=999, maxmage=0, maxamage, minfage=999, maxfage=0, maxafage;
        long double prob1, prob2, * * perspmf, * * perscdf, sumprobs, mchksum=0, fchksum=0, * tdcdf, * hrates;
        vector<long double> mprobs, fprobs;

        // Read age probabilities for M/F into corresponding arrays.  These are TD probabilities starting at a given age and will need to be adjusted for each person
        // based on their age at retirement start.  Also, use the file to derive the maximum possible M/F ages, which are ages that have non-zero probabilities.
        ifstream getprobs(root+ageprfile);
        if (getprobs.is_open())
        {
                while (!getprobs.eof())
                {
                        getprobs >> age >> prob1 >> prob2;
                        if (strtage == 0)
                        {
                                strtage = age;
                                prvage = age-1;
                        }
                        if (age == prvage+1)
                        {
                                if (prob1 >= 0.0)
                                        mprobs.push_back(prob1);
                                if (prob2 >= 0.0)
                                        fprobs.push_back(prob2);
                                if (prob1 > 0.0)
                                        maxamage=age;
                                if (prob2 > 0.0)
                                        maxafage=age;
                                prvage = age;
                        }
                }
                getprobs.close();
        }
        else
        {
                cout << "ERROR: Could not read file: " << root + ageprfile << endl;
                cout << "EXITING...derhrates()..." << endl; cin.get();
                exit (EXIT_FAILURE);
        }

        // Confirm that the probabilities read sum to 1.00 for each gender (allow for floating pt precision diffs).
        for (int i=(int) mprobs.size()-1; i>= 0; --i)
                mchksum=mchksum+mprobs[i];
        for (int i=(int) fprobs.size()-1; i>= 0; --i)
                fchksum=fchksum+fprobs[i];
        if (min(mchksum,fchksum) < (1.00 - pow(0.1,15)) || max(mchksum,fchksum) > (1.00 + pow(0.1,15)))
        {
                cout.setf(ios_base::fixed, ios_base::floatfield); cout.precision(25);
                cout << "ERROR: Probabilities in " << root + ageprfile << " do not sum to 1 for one or both genders.  See below..." << endl;
                cout << "--> Male probability sum   = " << mchksum << endl;
```

```cpp
                cout << "--> Female probability sum = " << fchksum << endl;
                cout << "EXITING...derhrates()..." << endl; cin.get();
                exit (EXIT_FAILURE);
}

// Find the minimum male/female ages specified in the control file at retirement start.
for (int i=0; i<npersons; ++i)
{
                if (gndrs[i] == 'M' && ags[i] < minmage)
                        minmage = ags[i];
                else if (gndrs[i] == 'F' && ags[i] < minfage)
                        minfage = ags[i];
}

// Find the maximum male/female ages specified in the control file at retirement start.
for (int i=0; i<npersons; ++i)
{
                if (gndrs[i] == 'M' && ags[i] > maxmage)
                        maxmage = ags[i];
                else if (gndrs[i] == 'F' && ags[i] > maxfage)
                        maxfage = ags[i];
}

// Check that no age specified on the last line of the control file is less than the start age in the probabilities file.
if (minmage < strtage || minfage < strtage)
{
                cout << "ERROR: An age in " << root + paramfile << " is less than the minimum age in " << root + ageprfile << "." << endl;
                cout << "--> Smallest male age (" << paramfile << ") = " << minmage << endl;
                cout << "--> Smallest female age (" << paramfile << ") = " << minfage << endl;
                cout << "--> Smallest allowable age (" << ageprfile << ") = " << strtage << endl;
                cout << "EXITING...derhrates()..." << endl; cin.get();
                exit (EXIT_FAILURE);
}

// Check that no male age specified on the last line of the control file is greater than the last (>0 prob) male age in the age probs file.
if (maxmage > maxmage)
{
                cout << "ERROR: A male age in " << root + paramfile << " is greater than the maximum allowed in " << root + ageprfile << "." << endl;
                cout << "--> Largest male age (" << paramfile << ") = " << maxmage << endl;
                cout << "--> Largest allowable male age (" << ageprfile << ") = " << maxamage << endl;
                cout << "EXITING...derhrates()..." << endl; cin.get();
                exit (EXIT_FAILURE);
}

// Check that no female age specified on the last line of the control file is greater than the last (>0 prob) female age in the age probs file.
if (maxfage > maxafage)
{
                cout << "ERROR: A female age in " << root + paramfile << " is greater than the maximum allowed in " << root + ageprfile << "." << endl;
                cout << "--> Largest female age (" << paramfile << ") = " << maxfage << endl;
                cout << "--> Largest allowable female age (" << ageprfile << ") = " << maxafage << endl;
                cout << "EXITING...derhrates()..." << endl; cin.get();
                exit (EXIT_FAILURE);
}

// Derive total # years to process for this arrangement (plus 1). (Includes SMax, but no alpha decision is made at SMax.)
remyrs = max(maxamage-minmage+1, maxafage-minfage+1);

// If the hazard rates file exists use it.  Otherwise derive it and write to file (using the details on the last line of the control file).
if (hrexst)
                cout << endl << "ALERT: Hazard rates file: " << root + hrfile << " already exists and will be used.  (Last line of the control file is ignored.)" << endl;
else
{
```

```
// Construct the person-specific probability arrays (pmfs and cdfs), shifting each to begin at t=0 which represents the start of retirement.
// Build array of pointers to individual TD probability arrays (start at time t=0 for each person).  Then build inner arrays that will
// apply to each individual person retiring, containing remyrs elements. (Do this for both the individual person pmfs and cdfs.)
perspmf = new long double * [npersons];
perscdf = new long double * [npersons];
tdcdf = new long double [remyrs];
hrates = new long double [remyrs];
for (int i=0; i<npersons; ++i)
{
        // Check that age values in parameter file are valid.  (Invalid gender values are checked in main().)
        if (ags[i] < strtage || gndrs[i] == 'M' && ags[i] > maxamage || gndrs[i] == 'F' && ags[i] > maxafage)
        {
                cout << "ERROR: Invalid age in file control.txt for person #" << i << " of gender=" << gndrs[i] << ": " << ags[i] << endl;
                cout << "EXITING...derhrates()..." << endl; cin.get();
                exit (EXIT_FAILURE);
        }

        // Create the individual person PMF and CDF arrays.
        perspmf[i] = new long double [remyrs];
        perscdf[i] = new long double [remyrs];

        // Populate the individual arrays for each person.  These are probabilities of death that
        // will sum to one and a value exists for each of the remyrs processed by this arrangement.
        sumprobs = 0.0;
        for (int j=((int) mprobs.size()-1); j>=(ags[i]-strtage); j--)
        {
                if (gndrs[i] == 'M')
                        sumprobs = sumprobs + mprobs[j];
                else if (gndrs[i] == 'F')
                        sumprobs = sumprobs + fprobs[j];
        }

        // Load the individual person PMF array with the applicable conditional PMF value. (Initialize with zeros.)
        for (int j=0; j<remyrs; ++j)
                perspmf[i][j] = 0.0;
        for (int j=0; (gndrs[i] == 'M' && j<(maxamage-(ags[i]-1))) || (gndrs[i] == 'F' && j<(maxafage-(ags[i]-1))); ++j)
        {
                if (gndrs[i] == 'M')
                        perspmf[i][j] = mprobs[ags[i]-strtage+j] / sumprobs;
                else if (gndrs[i] == 'F')
                        perspmf[i][j] = fprobs[ags[i]-strtage+j] / sumprobs;
        }

        // Check that sum of probabilities equals 1 (within floating point precision). If not put out alert.
        sumprobs = 0.0;
        for (int j=0; j<remyrs; ++j)
                sumprobs = sumprobs + perspmf[i][j];
        if (sumprobs < (1.00 - pow(0.1,15)) || sumprobs > (1.00 + pow(0.1,15)))
        {
                cout.setf(ios_base::fixed, ios_base::floatfield); cout.precision(50);
                cout << "Alert: Sum of probabilities for Person " << i+1 << " is not 1.00." << endl << "Sum is = " << sumprobs << endl;
        }

        // Load the individual person cdf array with the applicable cumulative probability.
        // And load the TD cdf array with the applicable cumulative probability.
        for (int j=0; j<remyrs; ++j)
        {
                perscdf[i][j] = 0;
                for (int k=0; k<=j; ++k)
                        perscdf[i][j] = perscdf[i][j] + perspmf[i][k];
                if (i==0)
                        tdcdf[j] = perscdf[0][j];
```
43

```cpp
                            else
                                    tdcdf[j] = tdcdf[j]*perscdf[i][j];
                    }
            }

            // Build the hazard rates and write to file.
            hrates[0] = tdcdf[0];
            for (int j=1; j<remyrs; ++j)
                    hrates[j] = min((tdcdf[j] - tdcdf[j-1]) / (1.0 - tdcdf[j-1]), 1.00);

            // Write hazard rate probabilities to file for reference, this file will be read back in if processing years in separate calls.
            ofstream hrout(root+hrfile);
            hrout.setf(ios_base::fixed, ios_base::floatfield);
            for (int j=0; j<remyrs; ++j)
            {
                    hrout.precision(50); hrout << hrates[j] << " "; hrout.precision(0); hrout << "(t=" << j << ")" << endl;
            }
            hrout.close();

            // Free dynamic memory allocations.
            for (int i=0; i<npersons; ++i)
            {
                    delete [] perspmf[i];  perspmf[i] = nullptr;
                    delete [] perscdf[i];  perscdf[i] = nullptr;
            }
            delete [] perspmf;  perspmf = nullptr;
            delete [] perscdf;  perscdf = nullptr;
            delete [] tdcdf; tdcdf = nullptr;
            delete [] hrates; hrates = nullptr;
        }

        // Return # years to process for this arrangement, which is 1 less than the # written to the hr file (since SMax is written).
        return (remyrs-1);
}
```

<u>Code File:</u>  getprprobs.cpp

```
/        This function reads the prior time point's probability file and populates an array with these values.  It also returns a vector containing only the
/        bucket #'s from the prior time point that have unique probabilities assigned to them.  When deriving the expected probability of ruin for any time
/        point after the next time point it is not necessary to process every bucket if sequential buckets hold the same probability.  Therefore the optimize()
/        function only processes those buckets with unique probabilities.  Note that the probability of falling into a new larger bucket that was formed from
/        multiple sequential buckets is higher.  When at time t=TD-1 or t=SMax-1 (i.e., the first decision point processed via induction) we set V(t,RF(t))=0
/        for all RF(t) > 0.  There is no external file to read when processing times t=TD-1 or t=SMax-1.
/
/ Parameters:
/
/        1.) Root directory where input files reside and output files are written.
/        2.) Current year/time point being processed.  (These start at 0 and end at one less than the total # being processed by the given arrangement specified
/             in the control file.  When we proceed using induction, we start at the end and finish at the beginning.)
/        3.) Total number of years processed by this control file arrangement (i.e., total # of alpha-decisions required).
/        4.) Maximum ruin factor value specified in the control file (i.e., RFMax).
/        5.) Precision value for the ruin factor discretization as specified in the control file (i.e., PR).
/        6.) An empty array of appropriate size (i.e., total # of buckets + 1) long doubles that will be populated by this function.
/        7.) Value of the prior time point's hazard rate (0.0 for fixed TD and t < TD, 1 at TD).  This is to determine when the highest possible probability has
/             been reached and all buckets from this time point onward can be treated as one collection.
/
/ Input Files:
/
/        1.) File of probabilities from the prior time point that has already been derived.  This file has the form "Year_X_All_Buckets.txt", where X refers to
/             actual time point.  Time points start at 0 and end at 1 less than the total # being processed by the given arrangement.  Each time point reflects
/             an alpha-decision that needs to be made.  Time t=0 reflects the first asset allocation decision at the start of retirement and time T=SMax-1 reflects
/             the last possible asset allocation decision that needs to be made the year before SMax when TD is random.  When TD is fixed, TD-1 is the last time
/             point requiring an asset allocation decision.  Note that the prior year/time point probabilities being read are from the current value + 1.
/
/ Output Files:
/
/                None.
/
/ Return Value:
/
/        This function returns a vector of bucket #'s with unique probabilities (from the prior timepoint which has already been processed).  (It also populates
/        the array Vp whose location is passed to this function.)
/-------------------------------------------------------------------------------------------------------------------------------------------------------------------*/

#include "stdafx.h"

vector<long> getprprobs(const string root, const int curyear, const int nyrs, const double rfmax, const int pprec, long double * Vp, const long double prhrate)
{
        int inyr, prMax=0;
        long * bktPrn, inb;
        double inrf, inalpha;
        long double inprob, prevprob;
        const long nbuckets = (long) (rfmax*pprec + 0.5);
        vector<long> PrB;
        string ifname;
        ifstream fin;
        ifname = root + "Year_" + to_string((long long) (curyear+1)) + "_All_Buckets.txt";

        // Array to indicate (0/1) where unique probabilities from prior time point reside.  By only processing unique probabilities at the prior
        // time point we can reduce processing time.  (We will always process the first and last buckets.)
        bktPrn = new long [nbuckets];

        if (curyear == (nyrs-1))                          // If at last decision point (either t=TD-1 or t=SMax-1) then V(t+1,RF(t+1))=0 for all RF(t+1)>0.
        {
                for (int i=0; i<nbuckets; ++i)
                        Vp[i]=0.00;
                Vp[nbuckets]=0.00;                        // Pruin for RF(t+1) region > RFMax at TD-1/SMax-1.
        }
        else {
```

```cpp
                // Open and read the prior time point's probability file and load the values into an array.
                fin.open(ifname);
                if (fin.is_open())
                {
                        while (!fin.eof())
                        {
                                fin >> inyr >> inrf >> inprob >> inalpha;
                                if (inyr == (curyear + 1))
                                {
                                        inb = (long) (inrf*pprec + 0.5);
                                        Vp[inb-1]=inprob;
                                }
                        }
                        fin.close();
                        Vp[nbuckets]=(1.00 - prhrate);    // PRuin for RF(t+1) region > RFMax at all time points except TD-1/SMax-1.
                }
                else
                {
                        cout << "ERROR: Could not read prior year probability file: " << ifname << endl;
                        cout << "EXITING...getprprobs()..." << endl; cin.get();
                        exit (EXIT_FAILURE);
                }
        }

        // Check the data just read in for any issues and load pruning array.
        for (long b=1; b <= nbuckets; ++b)
        {
                // Populate the pruning array with 1's and 0's, where 1 indicates unique probability (last bucket in the sequence).
                if (b > 1 && b < nbuckets && Vp[b-1] != Vp[b] && prMax == 0)  // Always process first and last buckets individually.
                {
                        if (Vp[b] >= (1.00 - prhrate))                        // The next bucket holds the maximum possible probability, stop scanning.
                                prMax = 1;
                        bktPrn[b-1] = 1;                                      // This bucket will account for a unique probability.
                }
                else if (b > 1 && b < nbuckets)
                        bktPrn[b-1] = 0;                                      // This bucket will not account for a unique probability.
                else
                        bktPrn[b-1] = 1;                                      // First and last buckets will always be processed.

                if (b==1)
                        prevprob=0;

                // Check for and report issues found in the prior time point's probability file.
                if ( (Vp[b-1] < 0) || (Vp[b-1] > (1.00 - prhrate) + (2.0*pow(0.1,16))) || (Vp[b-1] < (prevprob - pow(0.1,15))) )
                {
                        cout.setf(ios_base::fixed, ios_base::floatfield);
                        cout.precision(50);
                        cout << "There is an issue with the previous years data (see below):" << endl;
                        cout << "Vp[" << (b-2) << "] = " << Vp[b-2] << endl;
                        cout << "Vp[" << (b-1) << "] = " << Vp[b-1] << endl;
                        cout << "EXITING...getprprobs()..." << endl; cin.get();
                        exit(EXIT_FAILURE);
                }
                prevprob = Vp[b-1];
        }

        // Build an array with values that are the bucket #'s from the prior time point that have a
        // unique probability and thus will be processed during the expected value computation.
        for (long b=1; b<=nbuckets; ++b)
                if (bktPrn[b-1] == 1)
                        PrB.push_back(b);
```

```cpp
          // Free memory of dynamically sized arrays.
          delete [] bktPrn;  bktPrn = nullptr;

          // For informational purposes.
          cout << "-->  # Unique bucket probs at prior time point reported by getprprobs(): " << PrB.size() << endl;

          // Return the vector PrB.
          return PrB;
}
```

## Code File:  optimize.cpp

```
/ Filename:  optimize.cpp
/
/ Function:  optimize()
/
/ Summary:
/
/       This function derives the optimal alpha and probability values for a single year/time point and bucket range specified by the call.  When the first
/       and last buckets are set to 0 this function makes a single pass over the given year/time point and approximates the bucket number where heavy algorithm
/       pruning begins.  Heavy pruning sets alpha=pa which is the alpha value that takes effect when the optimal probability of ruin exceeds the first k
/       decimal places of the maximum possible probability of ruin for this year/arrangment, where k is the last parameter specified on the first line of the
/       control file.  Thus a value of 4.00 indicates that pruning begins when the probability of ruin exceeds the first 4 decimal places of the highest possible
/       probability of ruin for this year/arrangement.  Note that this value is stored locally in the variable prnpwr and only has an impact when TD is random.
/       The initial pass approximates this value by only considering alpha=pa.  Optimal bucket sizes are then formed based off of this approximation and then
/       the actual calls to this function are made where the point of heavy pruning is determined exactly.  The approximation is only used to estimate optimal
/       bucket collection sizes for threaded processing.  For all other cases the specific buckets listed are processed and written to a file with the bucket
/       boundaries contained in the file name.  The function combfiles() is invoked by main() after all buckets for a year/time point have been processed and
/       written to a file and it aggregates all individual files then deletes them leaving a single file containing all relevant optimal values for each year/
/       time point.  Once the final year/time point has been processed (i.e., t=0) the individual files for each year/time point are aggregated into final
/       (horizontal & vertical) files for the specific arrangement being modeled and the year-specific files are deleted.
/
/ Parameters:
/
/       1.) Root directory where input files reside and output files are written.
/       2.) Year value to process.  (This is the timepoint value which always starts at zero and ends at # time points to process for this arrangement less 1.)
/       3.) Array containing all values from the first line of the control file.
/       4.) Array containing the precision values from the 2nd line of the control file.
/       5.) Array containing the bucket limits for this call.
/       6.) Array containing the hazard rates derived for this arrangement.
/       7.) Array of optimal probabilities derived for the prior year/time point (i.e., current year/time point + 1)
/       8.) Vector of unique bucket endpoints for prior year where collections of buckets with equal probabilities are treated as a single bucket in the
/            expected value calculation.
```

```
/
/ Input Files:
/
/        None.  (The file of prior probabilities is read by the function getprprobs() and passed to this function via the parameter Vp.)
/
/ Output Files:
/
/        1.) The file "Year_Y_Buckets_S_thru_E.txt" is written to directory root.  (Where Y is the time point, S is the start bucket #, and E is the end bucket #.
/            Once processing for this year has ended these files are aggregated by combfiles() into a single file named "Year_Y_All_Buckets.txt".)
/
/ Return Value:
/
/        The approximate bucket # where heavy algorithm pruning begins when invoking this function with start and end buckets equal to 0.  Otherwise return
/        the value 0.
/----------------------------------------------------------------------------------------------------------------------------------------------------*/

#include "stdafx.h"

int optimize(const string root, const int y, const double prms[8], const int prc[2], const long bkts[2], const long double * sprobs, const long double * Vp, const
vector<long> & PrB)
{
        const int pr=prc[0], pa=prc[1];
        const double smn=prms[0], svr=prms[1], bmn=prms[2], bvr=prms[3], cv=prms[4], rfmax=prms[5], er=prms[6], prnpwr=prms[7];
        const long nbuckets=(long) (rfmax*pr + 0.5), nuqbkts=(long) PrB.size();
        int prunepnt = 0, ties, stalpha=0;
        long stbkt=bkts[0], endbkt=bkts[1];
        double rf, alpha, OPT_alpha, mean, std, * Ac;
        long double OPT_pruin = 1.0, cdfval, eprob, * Vc, rhs_cdf, lhs_cdf, pruin, tiethresh, pruneprob;
        boost::math::normal normdist;
        ofstream fout;

        // Create the arrays to hold current year optimal alphas and probabilities.
        Ac = new double [nbuckets];
        Vc = new long double [nbuckets];

        // Before processing the current year, run through all buckets using only alpha=100% to find the pruning point.
        if (stbkt == 0 && endbkt == 0)
        {
                stalpha = pa;
                stbkt = 1;
                endbkt = nbuckets;
        }

        // One-half of the maximum possible PRuin is the tie threshold, in general.
        tiethresh = (0.5)*((1.00 - sprobs[y]));

        // Prune point #2 begins at set number of decimals after max probability for this arrangement.
        pruneprob = floor(pow(10.0,prnpwr)*(1.00 - sprobs[y])) / pow(10.0,prnpwr);

        if (bkts[0] == 0 && bkts[1] == 0)
        {
                cout.setf(ios_base::fixed, ios_base::floatfield);
                cout.precision(10); cout << "-->  Value of pruneprob reported by optimize() is: " << pruneprob << endl;
        }
        for (long b = stbkt; b <= endbkt; ++b)        // PRuin at any future time point.
        {
                ties = 0;                             // Initialize for le/gt comparisons.
                rf = (double) b / pr;                 // Derive ruin factor.
                OPT_alpha = 99.00;                    // Initialize to unrealistic value.
                OPT_pruin = 99.00;                    // Initialize to unrealistic value.
                for (int a=stalpha; a <= pa; ++a)
                {
```

```cpp
            if ((prunepnt == 0 && (a == stalpha || OPT_pruin > 0.00)) || (prunepnt == 1 && a == pa))
            {
                    alpha = (double) a / pa;
                    mean = (1.00-er)*(1.00 + alpha*smn + (1.00-alpha)*bmn);
                    std = (1.00-er)*sqrt(pow(alpha,2)*svr + pow(1.00-alpha,2)*bvr + 2.00*(alpha)*(1.00-alpha)*(cv));
                    normdist = boost::math::normal(mean,std);
                    cdfval = cdf(normdist, rf);
                    if (cdfval == 1.00)
                            eprob = Vp[nbuckets];
                    else
                    {
                            //------------------------------------------------------------------------------------------------------------------//
                            rhs_cdf = 1.00;
                            lhs_cdf = cdf(normdist, rf*(1 + (pr/1.5)));
                            eprob = (rhs_cdf - lhs_cdf)*Vp[0];                         // First bucket, unique processing.
                            rhs_cdf = lhs_cdf;
                            for (long pb=2; pb <= nuqbkts; ++pb)                       // All others but last bucket, std proc for unq probs only.
                            {
                                    lhs_cdf = cdf(normdist, rf*(1.0 + pr/(PrB[pb-1] + 0.5)));
                                    eprob = eprob + (rhs_cdf - lhs_cdf)*Vp[PrB[pb-1]-1];
                                    rhs_cdf = lhs_cdf;
                            }
                            eprob = eprob + (rhs_cdf - cdfval)*Vp[nbuckets];          // Last bucket, unique processing.
                            eprob = eprob / (1.00 - cdfval);                          // Make the probability conditional.
                            //------------------------------------------------------------------------------------------------------------------//
                    }

                    // Deal with numerical instability near zero.
                    if (ties == 0)
                    {
                            pruin = (1.00 - sprobs[y])*(cdfval + eprob - (cdfval*eprob));
                            if (pruin > tiethresh)
                                    ties = 1;
                    }

                    // Deal with numerical instability near one.
                    if (ties == 1)
                            pruin = (1.00) - (sprobs[y] + (1.00-cdfval)*(1.00-eprob) - sprobs[y]*(1.00-cdfval)*(1.00-eprob));

                    // Update optimal values.
                    if ((a == 0) || (ties == 0 && pruin < OPT_pruin) || (ties == 1 && pruin <= OPT_pruin))
                    {
                            OPT_pruin = pruin;
                            OPT_alpha = alpha;
                    }
            }
    }
    // Load optimal values into correspnding arrays.
    Ac[b-1] = OPT_alpha;
    Vc[b-1] = OPT_pruin;

    // Set pruning for this timepoints processing, allow for floating point precision diffs.
    if (prunepnt == 0 && Vc[b-1] >= (pruneprob - (pow(0.1,16) + pow(0.1,17))))
    {
            prunepnt = 1;

            // Initial run finds the (approximate) bucket where pruning starts then uses this information to split the remaining buckets by
            // sizes that will run in shortest time.  For this run, end the call as soon as the bucket number has been found.
            if (bkts[0] == 0 && bkts[1] == 0)
            {
                    cout << "-->  Pruning RC returned by optimize() is: " << b << endl;
```

```cpp
                                return b;
                        }
                }
        }

        // Write current year to file with bucket boundaries contained in the name.
        string outfile = root + "Year_" + to_string((long long) y) + "_Buckets_" + to_string((long long) stbkt) + "_thru_" + to_string((long long) endbkt) + ".txt";

        fout.open(outfile);
        for (long b = stbkt; b <= endbkt; ++b)
        {
                fout.setf(ios_base::fixed, ios_base::floatfield);
                fout << y << " ";
                fout.precision(10); fout << (double) b / pr << " ";
                fout.precision(50); fout << Vc[b-1] << " ";
                fout.precision(10); fout << Ac[b-1] << endl;
        }
        fout.close();

        // Free dynamic memory allocations and exit.
        delete [] Ac;  Ac = nullptr;
        delete [] Vc;  Vc = nullptr;
        return 0;
}
```

## Code File: combfiles.cpp

```
/ Filename:  combfiles.cpp
/
/ Function:  combfiles()
/
/ Summary:
/
/       Function to aggregate the individual bucket-specific output files written by optimize() into single files for each time point.  These files are
/       subsequently read by getprprobs() when processing the next time point.  At the last time point (i.e., year=0) all files for the individual time
/       points are aggregated into 3 files then deleted.  The first is a (space-delimited) vertical file containing all data for all time points.  This
/       file is named FinalResults_V.txt.  The remaining 2 files are comma-separated horizontal files for import into a spreadsheet, and these files are
/       named FinalAlphaResults_H.csv and FinalProbResults_H.csv.
/
/ Parameters:
/
/       1.) Root directory where input files reside and output files are written.
/       2.) Current year/time point number being processed.  (Years start at 0 and end at one less than the total # of years being processed by the given
```

```
/           arrangement specified in the control file.)
/       3.) Number of concurrent processes used to derive the optimal values for each time point.
/       4.) Number of buckets processed per thread within each time point.  (Except the last one.  This is specific to each time point.)
/       5.) Maximum ruin factor value specified in the control file (i.e., RFMax).
/       6.) Precision value for the ruin factor discretization as specified in the control file (i.e., PR).
/       7.) Total number of years processed by the arrangement specified in the control file.
/
/ Input Files:
/
/       1.) For a given time point, input files of the form "Year_X_Buckets_S_thru_E.txt" are read from the root directory.  (Where X is the time point,
/           S is the start bucket #, and E is the end bucket #.  This function aggregates these individual files into a single file named:  Year_X_All_Buckets.txt.
/       2.) Year_X_All_buckets.txt  (Written by this function for each time point and all such files are read by this function and aggregated at the last time point,
/           which is year=0.)
/
/ Output Files:
/
/       1.) Year_X_All_Buckets.txt, where X reflects the year/time point being processed.
/       2.) FinalResults_V.txt  (Vertical text file of all results from all years processed.)
/       3.) FinalAlphaResults_H.csv  (Horizontal csv file of just optimal alphas.  Use this file to simulate and confirm the result.)
/       4.) FinalProbResults_H.csv  (Horizontal csv file of optimal probabilities.)
/
/       Note:  The last 3 output files above are only written at the last time point when all processing ends (i.e., year=0).  The first file is written at every
/              time point.
/
/ Return Value:
/
/       None.
/-------------------------------------------------------------------------------------------------------------------------------------------------------------*/

#include "stdafx.h"

void combfiles(const string root, const int curyear, const int pllruns, const long bcktsprun, const double rfmax, const int pprec, const int nyrs)
{
        int numfiles=0, fnlyear;
        const long nbkts=(long) (rfmax*pprec + 0.5);
        long fnlbkt;
        double fnlrf, fnlalpha;
        long double fnlprob;
        string fndfname, fndfile, concatyr, concatall;
        ifstream chk4file, normin;
        ofstream fnloutp, fnlouta;

        numfiles = 0;
        for (int j=1; j<=pllruns; ++j)
        {
                if (j <= (pllruns-1))
                        fndfname = "Year_" + to_string((long long) curyear) + "_Buckets_" + to_string((long long) bcktsprun*(j-1) + 1) + "_thru_"
                        + to_string((long long) bcktsprun*(j)) + ".txt";
                else
                        fndfname = "Year_" + to_string((long long) (curyear)) + "_Buckets_" + to_string((long long) bcktsprun*(j-1) + 1) + "_thru_"
                        + to_string((long long) nbkts) + ".txt";
                fndfile = root + fndfname;

                // Construct string for concatenation.
                concatyr=concatyr + " " + fndfile;

                // Attempt to open file and update file counter if successful.
                chk4file.open(fndfile);
                if (chk4file.is_open())
                        numfiles = numfiles + 1;
                chk4file.close();
        }
```

```cpp
if (numfiles == pllruns)
{
        // Concatenate files into 1 and remove individual files.
        concatyr="type" + concatyr + " > " + root + "Year_" + to_string((long long) curyear) + "_All_Buckets.txt 2>nul & del " + root + "Year_"
                + to_string((long long) (curyear)) + "_Buckets*.txt";
        system(concatyr.c_str());
}
else
{
        cout << "ERROR: Attempt to combine files after year=" << curyear << " has failed." << endl;
        cout << "EXITING...combfiles()..." << endl; cin.get();
        exit (EXIT_FAILURE);
}

// If at last time point then build the normalized and transposed final data files and delete all intermediate data files.
if (curyear == 0)
{
        // Build normalized data file.
        concatall="type " + root + "Year_" + to_string((long long) curyear) + "_All_Buckets.txt";
        for (int j=curyear+1; j<nyrs; ++j)
                concatall = concatall + " " + root + "Year_" + to_string((long long) j) + "_All_Buckets.txt";
        concatall=concatall + " > " + root + "FinalResults_V.txt 2>nul & del " + root + "Year_*.txt";
        system(concatall.c_str());

        // Build transposed probability/alpha data files.
        long double * * fnlprobs;                               // Pointer to long double pointer for probabilities.
        double * * fnlalphas;                                   // Pointer to double pointer for alphas.
        fnlprobs = new long double * [nyrs];                    // Use long double pointer to hold array of long double pointers.
        fnlalphas = new double * [nyrs];                        // Use double pointer to hold array of double pointers.
        for (int j=0; j<nyrs; ++j)                              // (A pointer to any type can hold an array of that type.)
        {
                fnlprobs[j] = new long double [nbkts];          // Each array value now points to array of long doubles (2D array).
                fnlalphas[j] = new double [nbkts];              // Each array value now points to array of doubles (2D array).
        }

        // Open normalized file just created, transpose it and write to csv file.
        normin.open(root+"FinalResults_V.txt");
        if (normin.is_open())
        {
                while (!normin.eof())
                {
                        normin >> fnlyear >> fnlrf >> fnlprob >> fnlalpha;
                        fnlbkt = (long) (fnlrf*pprec + 0.5);
                        if (fnlbkt >= 1 && fnlbkt <= (long) (rfmax*pprec + 0.5))
                        {
                                fnlprobs[fnlyear][fnlbkt-1] = fnlprob;
                                fnlalphas[fnlyear][fnlbkt-1] = fnlalpha;
                        }
                }
                normin.close();
        }
        else
        {
                cout << "ERROR: Could not read file: " << root+"FinalResults_V.txt" << endl;
                cout << "EXITING...combfiles()..." << endl; cin.get();
                exit (EXIT_FAILURE);
        }

        // Transpose and write to file.
        fnloutp.open(root+"FinalProbResults_H.csv");  fnloutp.setf(ios_base::fixed, ios_base::floatfield);  fnloutp << "RF";
        fnlouta.open(root+"FinalAlphaResults_H.csv");  fnlouta.setf(ios_base::fixed, ios_base::floatfield);  fnlouta << "RF";
```

```cpp
                    for (long i=0; i<=nbkts; ++i)              // The first line (i=0) of the csv files holds column headers.
                    {
                            if (i > 0)
                            {
                                    float _rf_;  _rf_ = (float) i/pprec;
                                    fnloutp.precision(10); fnloutp << _rf_;
                                    fnlouta.precision(10); fnlouta << _rf_;
                            }
                            for (int j=0; j<nyrs; ++j)
                            {
                                    if (i == 0)
                                    {
                                            fnloutp << ", Time (t=" << j << ")";  fnlouta << ", Time (t=" << j << ")";
                                    }
                                    else
                                    {
                                            fnloutp.precision(50); fnloutp << "," << fnlprobs[j][i-1];
                                            fnlouta << "," << fnlalphas[j][i-1];
                                    }
                            }
                            fnloutp << endl;  fnlouta << endl;
                    }
                    fnloutp.close();
                    fnlouta.close();

                    // Free dynamic memory allocations.
                    for (int j=0; j<(nyrs); ++j)
                    {
                            delete [] fnlprobs[j];  fnlprobs[j] = nullptr;
                            delete [] fnlalphas[j];  fnlalphas[j] = nullptr;
                    }
                    delete [] fnlprobs;  fnlprobs = nullptr;
                    delete [] fnlalphas;  fnlalphas = nullptr;
          }
}
```

## Code File:  stdafx.h

```cpp
/*
/ Copyright (c) 2014 Chris Rook, All Rights Reserved.
/
/ Filename:  stdafx.h  (header file to include)
/ ----------------------------------------------------*/

#pragma once

// Header files

#include "targetver.h"
#include <stdio.h>
#include <tchar.h>
#include <iostream>
#include <fstream>
#include <string>
#include <math.h>
#include <stdlib.h>
#include <boost/math/distributions.hpp>
#include <boost/algorithm/string.hpp>
#include <boost/thread/thread.hpp>
#include <windows.h>
#include <dos.h>
#include <vector>
```

```cpp
using namespace std;

// Define constants.

const string paramfile="control.txt";       // Name of parameter control input file.
const string ageprfile="ageprobs.txt";      // Name of age probability input file.
const string hrfile="hrates.txt";           // Name of hazard rate probability output file.

// Define Function prototypes.

int optimize(const string root, const int strtyr, const double prms[8], const int prc[2], const long bkts[2], const long double * sprobs, const long double * Vp, const
          vector<long> & PrB);
int derhrates(const string root, const int numpersons, const char * ppl, const int * ags, const bool hrexst);
void combfiles(const string root, const int curyear, const int pllruns, const long bcktsprun, const double rfmax, const int pprec, const int nyears);
vector<long> getprprobs(const string root, const int curyear, const int nyrs, const double rfmax, const int pprec, long double * Vp, const long double prhrate);
```

-----------------------------------------------------------------------------------------------------------------------
```
0.0000000000000003119716336171020000000000000000000  (t=0)
0.0000000000000021922536698186529000000000000000000  (t=1)
0.0000000000113691519087302570000000000000000000000  (t=2)
0.0000000020003863625408732000000000000000000000000  (t=3)
0.0000000019508109036756375000000000000000000000000  (t=4)
0.0000001312223201754271500000000000000000000000000  (t=5)
0.0000000682760046549956060000000000000000000000000  (t=6)
0.0000029370323648036120000000000000000000000000000  (t=7)
0.0000010922010790464607000000000000000000000000000  (t=8)
0.0000036145327414987219000000000000000000000000000  (t=9)
0.0000108401675325234710000000000000000000000000000  (t=10)
0.0000298458435402582570000000000000000000000000000  (t=11)
0.0000762090969603561550000000000000000000000000000  (t=12)
0.0001818864733465654200000000000000000000000000000  (t=13)
0.0004076082101761658600000000000000000000000000000  (t=14)
0.0008590424421066637900000000000000000000000000000  (t=15)
0.0017048901456105734000000000000000000000000000000  (t=16)
0.0031935943913145057000000000000000000000000000000  (t=17)
0.0056589746749071438000000000000000000000000000000  (t=18)
0.0094993249668862551000000000000000000000000000000  (t=19)
0.0151181890915877470000000000000000000000000000000  (t=20)
0.0228750816382969940000000000000000000000000000000  (t=21)
0.0330159228244574080000000000000000000000000000000  (t=22)
0.0456376517854389030000000000000000000000000000000  (t=23)
0.0606917461091673600000000000000000000000000000000  (t=24)
0.0779638873690633450000000000000000000000000000000  (t=25)
0.0971373962552448960000000000000000000000000000000  (t=26)
0.1178406631868527300000000000000000000000000000000  (t=27)
0.1395759506103998500000000000000000000000000000000  (t=28)
0.1619001022525754000000000000000000000000000000000  (t=29)
0.1844673511176513600000000000000000000000000000000  (t=30)
0.2071355813553053800000000000000000000000000000000  (t=31)
0.2291719682928737000000000000000000000000000000000  (t=32)
0.2505744072042152000000000000000000000000000000000  (t=33)
0.2709771954632418400000000000000000000000000000000  (t=34)
0.2900855319019218800000000000000000000000000000000  (t=35)
0.3080096158184624200000000000000000000000000000000  (t=36)
0.3250776200154276200000000000000000000000000000000  (t=37)
0.3424647541089663700000000000000000000000000000000  (t=38)
0.3601044468071747900000000000000000000000000000000  (t=39)
0.3795533934696363280000000000000000000000000000000  (t=40)
0.3993070818458502300000000000000000000000000000000  (t=41)
0.4202378046266770400000000000000000000000000000000  (t=42)
0.4410484111686362500000000000000000000000000000000  (t=43)
0.4660883229270871800000000000000000000000000000000  (t=44)
0.4952102261932866400000000000000000000000000000000  (t=45)
0.5324950209634413700000000000000000000000000000000  (t=46)
0.5462672776065660200000000000000000000000000000000  (t=47)
0.5986628734659862300000000000000000000000000000000  (t=48)
0.7541499167480122400000000000000000000000000000000  (t=49)
0.9999999999970230480000000000000000000000000000000  (t=50)
```

***Note:*** *"hrates.txt" is a file derived by the function derhrates() and is written for random $T_D$ arrangements. It contains the discrete hazard rates for the person/MPU defined on the last line of the control file. (See Section 4.6.2 for reference.) The probabilities shown above are from the $5^{th}$ "control.txt" example (MPU of size N=9). All random arrangements have a single set of discrete hazard rates which is the only random-specific input required by the random $T_D$ model. Because an MPU is defined by a single hazard rate file the MPU size does not impact runtime. In the example above, $S_{Max}=50$ and a maximum of 50 decisions need to be made, the first at t=0 and the last at $t=S_{Max}-1=49$. No decision is made at time $t=S_{Max}$ since $P(T_D > S_{Max}) = 0$.*

---------------------------------------------------------------------------------------------------------------------------

```
0  0.0016000000  0.00000000000000000192720969208370620000000000000000  0.1980000000
0  0.0017000000  0.00000000000000000057845452863663540000000000000000  0.1990000000
0  0.0018000000  0.00000000000000000162135290408941400000000000000000  0.1990000000
0  0.0019000000  0.00000000000000000427354233369064750000000000000000  0.2000000000
                                              .
                                              .
                                              .
10 0.0108000000  0.00000000222864883182142450000000000000000000000000  0.2140000000
10 0.0109000000  0.00000000256035079632214650000000000000000000000000  0.2140000000
10 0.0110000000  0.00000000293658568468094670000000000000000000000000  0.2150000000
10 0.0111000000  0.00000000336267466389999980000000000000000000000000  0.2150000000
                                              .
                                              .
                                              .
20 0.0296000000  0.00001079481991831277300000000000000000000000000000  0.2360000000
20 0.0297000000  0.00001125172852740063400000000000000000000000000000  0.2360000000
20 0.0298000000  0.00001172534597344354000000000000000000000000000000  0.2370000000
20 0.0299000000  0.00001221616101523663900000000000000000000000000000  0.2370000000
                                              .
                                              .
                                              .
30 0.0934000000  0.00775593991107923190000000000000000000000000000000  0.3640000000
30 0.0935000000  0.00780733692967890570000000000000000000000000000000  0.3640000000
30 0.0936000000  0.00785894808515004080000000000000000000000000000000  0.3650000000
30 0.0937000000  0.00791077179182503140000000000000000000000000000000  0.3650000000
                                              .
                                              .
                                              .
40 0.0001000000  0.00000000000000000000000000000000000085482880000000  0.1440000000
40 0.0002000000  0.00000000000000000000000000000000000088469110000000  0.1440000000
40 0.0003000000  0.00000000000000000000000000000000000091611610000000  0.1440000000
40 0.0004000000  0.00000000000000000000000000000000000094839560000000  0.1440000000
                                              .
                                              .
                                              .
45 0.4978000000  0.08876385008462925300000000000000000000000000000000  0.4170000000
45 0.4979000000  0.08883038694792594300000000000000000000000000000000  0.4180000000
45 0.4980000000  0.08889690950186190900000000000000000000000000000000  0.4200000000
45 0.4981000000  0.08896342425482126100000000000000000000000000000000  0.4210000000
                                              .
                                              .
                                              .
46 0.5223000000  0.04775335689522756900000000000000000000000000000000  0.8510000000
46 0.5224000000  0.04780431578767197600000000000000000000000000000000  0.8530000000
46 0.5225000000  0.04785521417275828700000000000000000000000000000000  0.8550000000
46 0.5226000000  0.04790604639994011600000000000000000000000000000000  0.8560000000
                                              .
                                              .
                                              .
47 0.8091000000  0.00104883991561901090000000000000000000000000000000  0.1740000000
47 0.8092000000  0.00105293787325091270000000000000000000000000000000  0.1740000000
47 0.8093000000  0.00105704928401912120000000000000000000000000000000  0.1750000000
47 0.8094000000  0.00106117491595328910000000000000000000000000000000  0.1750000000
```

*__Note:__ Select rows from the file "FinalResults_V.txt" are displayed. This is a vertical file written by the application containing all model output. The values shown are from the 6[th] "control.txt" example. Column definitions are: time point [t], ruin factor RF(t), minimum probability of ruin $V_R(t, RF(t))$, and optimal asset allocation $\alpha_R(t, RF(t))$. In addition, there are 2 horizontal files which together contain the same information as the single vertical file and these are named: "FinalAlphaResults_H.csv" & "FinalProbResults_H.csv". To simulate and confirm the result use the file "FinalAlphaResults_H.csv". A method for doing this is discussed in the next section.*

Additional Notes:

- The source code provided has not been independently validated. The user should conduct full code validation before using it within a production environment.[7]

- To simulate a specific solution load the file "FinalAlphaResults_H.csv" into Microsoft® Excel® and use a simulator such as Oracle® Crystal Ball®. A vlookup table is used to retrieve the optimal alpha at each time point.[8] (Recall that each RF(t) centers its bucket.)

- The basic approach for coding the DP suggested in Figure 5 was altered to avoid unacceptable runtimes. Most coding effort was spent modifying the implementation to reduce runtimes which also has the undesirable effect of complicating the code.

- This implementation has been tested on the Windows 7 Professional and Windows 7 Home Edition operating systems. Some file management operations in the code are Windows-specific and would need to be updated for use on other operating systems.

- The code provided will produce a similar but slightly better solution than was presented in Figures 6, 7, and 8. Runtimes are now such that there is no longer a need to partition the RF(t) dimension when discretizing it as was done in those solutions.

- Increasing the RF(t) discretization precision $P_R$ results in a more accurate numerical approximation whereas increasing the alpha precision $P_\alpha$ lowers the probability of ruin $V(t, RF(t))$.

***

---

[7] The author would greatly appreciate being informed of code errors/bugs as well as techniques to reduce runtimes.
[8] A template for simulating the optimal solution will be supplied upon request.